\documentclass[AMA,Times1COL]{WileyNJDv5} 

\articletype{RESEARCH ARTICLE}%

\received{Date Month Year}
\revised{Date Month Year}
\accepted{Date Month Year}
\journal{Journal}
\volume{00}
\copyyear{2023}
\startpage{1}

\usepackage{subfigure}

\hypersetup{
	colorlinks  = true,
	linkcolor   = red,
	citecolor   = blue,
	urlcolor    = black,
	pdfborder   = {0 0 0}
}

\begin{document}

\title{Matched Asymptotic Expansions-Based Transferable Neural Networks for  Singular Perturbation Problems}

\author[1]{Zhequan Shen}

\author[2]{Lili Ju}

\author[1,3]{Liyong Zhu}

\authormark{Zhequan Shen, Lili Ju, Liyong Zhu}
\titlemark{Matched Asymptotic Expansions-Based Transferable Neural Networks for  Singular Perturbation Problems}

\address[1]{\orgdiv{School of Mathematical Sciences}, \orgname{Beihang University}, \orgaddress{\state{Beijing, 100191}, \country{China}}}

\address[2]{\orgdiv{Department of Mathematics}, \orgname{University of South Carolina}, \orgaddress{\state{Columbia,  South Carolina,  29063}, \country{USA}}}

\address[3]{\orgdiv{LMIB(Beihang University)}, \orgname{Ministry of Education}, \orgaddress{\state{Beijing, 100191}, \country{China}}}

\corres{Corresponding author: Liyong Zhu. \email{liyongzhu@buaa.edu.cn}}



\abstract[Abstract]{
	In this paper, by utilizing the theory of matched asymptotic expansions, an efficient and accurate  neural network method, named as  ``MAE-TransNet", is developed for solving singular perturbation problems in general dimensions, whose solutions usually change drastically in some narrow boundary layers.
	The TransNet is a two-layer neural network with specially pre-trained hidden-layer neurons. 
	In the proposed MAE-TransNet, the inner and outer solutions produced from the matched asymptotic expansions are first approximated by a TransNet with nonuniform hidden-layer neurons  and a TransNet with uniform hidden-layer neurons, respectively. Then, these two solutions are combined with a matching term to obtain the composite solution, which approximates the asymptotic expansion solution of the singular perturbation problem.
	This process enables the MAE-TransNet method to retain the precision of the matched asymptotic expansions while maintaining the efficiency and accuracy of TransNet.  Meanwhile, the rescaling of the sharp region allows the same pre-trained network parameters to be applied to boundary layers with various thicknesses, thereby improving the transferability of the method.  Notably, for  coupled boundary layer problems,  a computational framework based on  MAE-TransNet is also constructed to effectively address issues resulting from the lack of  relevant matched asymptotic expansion theory in such problems.
	Our  MAE-TransNet is thoroughly  compared with  TransNet, PINN, and Boundary-Layer PINN (BL-PINN) on various benchmark  problems including  1D linear and nonlinear problems with boundary layers, the 2D Couette flow problem, a 2D coupled boundary layer problem, and the 3D Burgers vortex problem. 
	Numerical results demonstrate that MAE-TransNet significantly outperforms other neural network methods in capturing the characteristics of boundary layers, improving the accuracy, and reducing the computational cost.
}



\maketitle



\section{Introduction}
\label{sec1}

A singular perturbation generally occurs when a small parameter multiplies the highest derivative, leading to the formation of thin boundary layers in the solution. These boundary layers reveal a wide range of natural phenomena such as high Reynolds number flows \cite{COUPEZ201365} and high Peclet number transfer \cite{RIOU2012302} in fluid dynamics, oxygen concentration transport \cite{steinruck2012asymptotic} in mathematical biology, and heat transfer simulation \cite{LI2023116107} in manufacturing technologies. In this paper, we consider the following singular perturbation problems with boundary layers:
\begin{equation}\label{equ: SPP}
	\left\{
	\begin{aligned}
		& \mathcal{L}(\boldsymbol{u}; \varepsilon) = f, && \boldsymbol{x} \in \Omega,\\
		& \mathcal{B}(\boldsymbol{u}) =  g, && \boldsymbol{x} \in \partial \Omega,
	\end{aligned}
	\right.
\end{equation}
where $\mathcal{L}$ denotes a differential operator defined on the domain (an open bounded set) $\Omega\in{\mathbb R}^d$, $\mathcal{B}$ denotes a boundary operator defined on the domain boundary $\partial \Omega$, and $0<\varepsilon \ll 1$ is a small thickness parameter which is usually multiplied to the highest order derivative term (e.g., a small diffusion coefficient). 

Accurately solving the singular perturbation problem (\ref{equ: SPP}) is quite challenging due to the large gradients within the boundary layers. Traditional numerical methods require extremely dense uniform meshes, leading to significant wastage of computational resources. Therefore, nonuniform mesh methods are developed to reduce computational costs by refining the meshes within the boundary layer \cite{roos2008robust, clavero2003uniformly, ge2011multigrid, du2018adaptive, kumar2015adaptive}. Despite this, most classic numerical methods, such as finite difference methods (FDM) \cite{kumar2022new, hsieh2018robust, singh2020parameter, patidar2006uniformly} and finite element methods (FEM) \cite{TOPRAKSEVEN2024130, zhang2021high, lin2018weak, gharibi2021convergence, fitzsimons1985petrov, ZHANG2017549}, still could be impractical when dealing with the case of very small singular perturbation parameters \cite{cao2023physics}, especially for multidimensional multilayer problems.

In recent years, numerous neural network methods have been developed to incorporate PDE information into the loss function and solve the  corresponding problem through optimization techniques, such as Deep Ritz method (DRM) \cite{yu2018deep}, Deep Galerkin method (DGM) \cite{sirignano2018dgm} and Physics-informed neural networks (PINN) \cite{raissi2019physics, 19M1274067}. Although these methods have presented good performance on many types of PDEs, it is challenging for them to capture the sharp changes within the boundary layers. Several  PINN-based methods have been proposed to solve singular perturbation problems in \cite{cao2023physics, CAO2024117222, HUANG2024100496, wang2024general}. The method proposed in \cite{aldirany2024multi} includes computing an initial approximation to the problem using a simple neural network and estimating a finer correction in an iterative manner by solving the problem for the residual with a new network of increasing complexity. In the work of Boundary-Layer PINN (BL-PINN) \cite{arzani2023theory}, the solution to boundary layer problems in the singular perturbation theory is divided into solutions in inner and outer regions, and multiple asymptotic expansion basis functions in both regions are approximated by PINN. BL-PINN integrates perturbation methods with PINN, successfully approximating solutions for singular perturbation problems with boundary layers. 
However, the above-mentioned neural network methods usually have deep network architectures, which require a significant amount of computational cost due to the multitude of parameters in nonlinear and nonconvex optimization. Moreover, in addressing singular perturbation problems with boundary layers, these approaches often employ hyperparameters that are dependent on $\varepsilon$. Consequently, as $\varepsilon$ changes, these hyperparameters must be readjusted, leading to additional computational costs for training.

To overcome the challenge of high computational costs associated with deep neural networks, some researchers introduce shallow neural network (i.e, very few number of network layers) methods along with pretrained parameters for the hidden layers, which only require solving a simple least squares problem concerning the parameters of the output layer, achieving greater accuracy for various PDEs when compared with the conventional neural network methods \cite{dong2021local, chen2022bridging, dong2022computing, chen2023random}.  Deep least-squares method proposed in \cite{cai2020deep} makes use of the neural network to approximate the solution of singularly perturbed reaction-diffusion equations through the compositional construction and employ first-order system least-squares functional as a loss function. For boundary layer problems in the singular perturbation theory, a numerical scheme based on the concept of Extreme Learning Machines (ELMs) has been proposed in \cite{calabro2021extreme}. Although these methods have demonstrated advantages in handling certain specific singular perturbation problems with boundary layers, they do not guarantee effectiveness for sufficiently small values of $\varepsilon$. Moreover, in practice, the hyperparameters in these methods usually lack interpretability, which increases the cost of selecting these parameters. 
To reduce the cost of parameter tuning, a novel neural network method incorporating a scaling layer, together with an adaptive parameter tuning strategy, was proposed in \cite{XU2025114129}.

Matched asymptotic expansions (MAEs) \cite{nayfeh2024perturbation, bender2013advanced, 21M1436087, 1017007} are used when the problem solution experiences a rapid change over a short interval (e.g., the solution to singular perturbation problems with boundary layers). MAE usually involves the construction of an expansion valid in the region of rapid change, called the inner solution, and an expansion valid outside this region, called the outer solution. The composite solution is consequently expressed as a sum of the inner solution, the outer solution, and a matching term which ensures the validity across the entire domain. Taking the case of Prandtl's MAEs, the error between the composite solution and the analytical solution achieves $O(\varepsilon)$ over the entire computational domain \cite{nayfeh2024perturbation}, which implies that the error decreases as $\varepsilon$ diminishes. However, it is quite challenging to analytically solve the outer and inner solutions for the majority of problems \cite{faria2017equation}. 

Transferable neural network (TransNet) \cite{zhang2024transferable} is a two-layer neural network (i.e., one hidden layer and one output layer) with specially pretrained hidden-layer neurons, and its key point is to construct a transferable neural feature space based on reparameterization of the hidden-layer neurons and approximation properties without using information from PDEs. It is noteworthy that the parameters in the neural feature space have explicit geometric meanings, which can guide parameter selection and thereby significantly reduce the cost of parameter tuning. The uniform  neuron distribution theorem ensures the transferability of neurons in the hidden layer. The experimental results in \cite{zhang2024transferable} verify the excellent performance of TransNet, specifically, various problems with different types of PDEs, domains, or boundary conditions are solved with much greater accuracy than the state-of-the-art methods. Moreover, due to the utilization of pretrained hidden layers, TransNet significantly outperforms PINN in terms of computational efficiency. However, when applied to singular perturbation problems, it remains difficult to effectively handle the sharp changes in the boundary layers. 

Inspired by the work of BL-PINN, MAE and TransNet, in this paper, we propose a MAEs-based transferable neural network (named as ``MAE-TransNet") method for solving singular perturbation problems  with boundary layers in general dimensions. Furthermore,  we also develop a novel  computational framework  for solving coupled boundary layer problems based on MAE-TransNet, which addresses issues of lacking relevant MAE theory in those problems.
Our method integrates the traditional perturbation methods into neural network frameworks to enable the new network method to inherit the advantages of both approaches simultaneously. The proposed MAE-TransNet, rooted in the efficiency and accuracy of TransNet, achieves higher accuracy as the parameter $\varepsilon$ decreases, consistent with the theory of MAEs.

Our method distinguishes itself from existing neural network approaches, such as the popular BL-PINN method, in the following ways: first, we compute both the inner and outer solutions across the entire computational domain. This contrasts with domain decomposition methods that separately compute inner expansions within boundary layer regions and outer expansions outside. Second, unlike approaches that simultaneously solve the solutions in both the inner and outer regions of the boundary layers, we first solve the outer solution over the entire computational domain, and then compute the inner solution with the matching principle. Third, instead of relying on PINNs, we employ the TransNet method, which utilizes a two-layer neural network with predetermined hidden-layer neuron parameters to solve the simplified boundary value problems for inner and outer solutions. Fourth, diverging from methods that use identical networks for both the inner and outer solutions, we utilize nonuniformly and uniformly distributed hidden-layer neurons for the inner and outer solutions, respectively, to capture their distinct characteristics.

The rest of the paper is organized as follows. Section \ref{sec2} briefly reviews the theory of MAE for singular perturbation problems and the TransNet method for solving PDEs. In Section \ref{sec3}, the MAE-TransNet method is developed with corresponding algorithms for both one-dimensional and multidimensional cases. Section \ref{sec4} presents numerical results and comparisons on a series of benchmark experiments in various dimensional spaces, highlighting the superior performance of MAE-TransNet compared to TransNet, PINN and BL-PINN. Finally, Section \ref{sec5} concludes this work and discusses some future research directions.

\section{Related Work}
\label{sec2}
In this section, we present some related work, including Prandtl's MAEs approach \cite{nayfeh2024perturbation} and the TransNet method developed in \cite{zhang2024transferable}, which collectively form the foundation of our methods.

\subsection{Prandtl's Matched Asymptotic Expansions}
\label{sec2.1}
Prandtl's MAEs consist of a first-order outer expansion (\textit{outer solution}) and a first-order inner expansion (\textit{inner solution}). The outer expansion is valid outside the boundary layer and the inner expansion is valid in the sharp-change region, and  then an approximate solution of the original singular perturbation problem (\textit{composite solution}) can be  written as a sum of the outer solution, the inner solution, and a matching term which cancels the outer expansion in the inner region and the inner expansion in the outer region. 

\begin{figure*}[!htb]
	\centering
	\includegraphics[width=.9\textwidth]{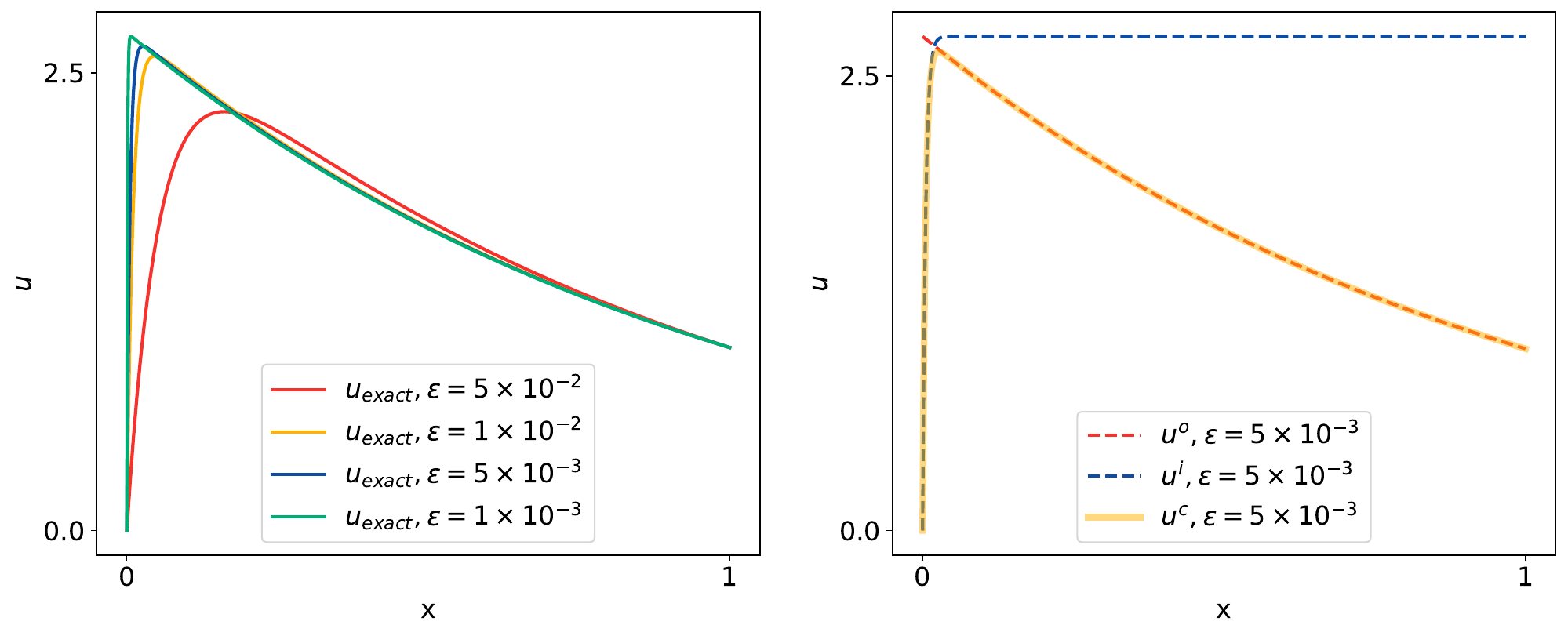}
	\caption{Left: The exact solutions of the 1D singular perturbation problem (\ref{equ: MAE}) with different values of $\varepsilon$. Right: The outer, inner and composite solutions for the problem (\ref{equ: MAE}) with $\varepsilon=5\times10^{-3}$.}
	\label{fig: MAE-example}
\end{figure*}

For purpose of illustration, let us consider the following 1D singular perturbation problem in the domain $\Omega = (0,1)$: 
\begin{equation}\label{equ: MAE}
	\left\{
	\begin{aligned}
		&\varepsilon \frac{\mathrm{d}^2 u(x)}{\mathrm{d} x^2} + \frac{\mathrm{d} u(x)}{\mathrm{d} x} + u(x) = 0, \quad x \in  (0, 1), \\ 
		&u(0)=0, \quad u(1)=1.
	\end{aligned}
	\right.
\end{equation}
It is well known that one boundary layer occurs at $x=0$  with the thickness $\varepsilon$ for this problem.
The outer, inner, composite, and exact solutions are denoted as $u^o(x), u^i(x), u^c(x)$, and $u_{exact}(x)$, respectively. As shown in Figure \ref{fig: MAE-example}, the left plot presents the exact solutions of the problem (\ref{equ: MAE}) with different values of $\varepsilon$, while the right plot presents the outer, inner, and composite solutions for the problem (\ref{equ: MAE}) with $\varepsilon=5\times10^{-3}$.

We first focus on the boundary value problem  satisfied by the outer solution $u^o(x)$. Through analysis in \cite{nayfeh2024perturbation}, the boundary condition $u(0)=0$ is to be dropped, then as $\varepsilon \rightarrow 0$ the problem (\ref{equ: MAE}) reduces to the following boundary value problem:
\begin{equation}\label{equ: MAE_outer}
	\left\{
	\begin{aligned}
		&\frac{\mathrm{d} u^o(x)}{\mathrm{d} x} + u^o(x) = 0, \quad x \in (0, 1), \\ 
		&u^o(1)=1.
	\end{aligned}
	\right.
\end{equation}
It is easy to find that the solution of the problem (\ref{equ: MAE_outer}) is given by 
\begin{equation}\label{equ: MAE_outer_solution}
	u^o(x)=e^{1-x}. \nonumber
\end{equation}

To obtain the inner solution $u^i(x)$, let us introduce the following scaling transformation to magnify the boundary layer
\begin{equation}\label{equ: MAE_inner_scaling}
	\zeta=\frac{x}{\delta(\varepsilon)}:=\frac{x}{\varepsilon}, 
\end{equation}
where $\delta(\varepsilon)\approx  O(\varepsilon)$ is usually determined by the settings of problem \cite{nayfeh2024perturbation} and here we take $\delta=\varepsilon$. Then we have $u^i(x)=u^i(\varepsilon\zeta)$ and denote $\bar{u}^i(\zeta)=u^i(\varepsilon\zeta)$ with $\zeta \in [0, \frac{1}{\varepsilon}]$. Based on the problems (\ref{equ: MAE}) and (\ref{equ: MAE_outer}), we know the boundary condition $\bar{u}^i(0)=0$ should be imposed. In order to determine the other boundary condition, the \textit{outer limit} (denoted by $u^{out}(x)$) and \textit{inner limit} (denoted by  $\bar{u}^{in}(\zeta)$) of a function  $u(x;\varepsilon)$ are introduced:
\begin{align} 
	u^{out}(x) = \lim_{\substack{\varepsilon \rightarrow 0,\text{ } x \text{ fixed}}} u(x; \varepsilon), \nonumber \\
	\bar{u}^{in}(\zeta) = \lim_{\substack{\varepsilon \rightarrow 0,\text{ } \zeta \text{ fixed}}} u(\varepsilon\zeta; \varepsilon), \nonumber
\end{align}
and the \textit{matching principle} is defined by
\begin{equation}\label{equ: MAE_matching_principle}
	\lim_{x \rightarrow 0} u^{out}(x)=\lim_{\zeta \rightarrow \infty} \bar{u}^{in}(\zeta).
\end{equation}
From the matching principle (\ref{equ: MAE_matching_principle}), the inner limit of the outer solution should be equal to the outer limit of the inner solution and we get the second boundary condition $$\lim_{\varepsilon\rightarrow 0}\bar{u}^i(\frac{1}{\varepsilon})={u}^o(0) = e.$$ Therefore, as $\varepsilon \rightarrow 0$, the problem (\ref{equ: MAE}) after the scaling transformation by $1/\varepsilon$ reduces to the following boundary value problem:
\begin{equation}\label{equ: MAE_inner}
	\left\{
	\begin{aligned}
		&\frac{\mathrm{d}^2 \bar{u}^i(\zeta)}{\mathrm{d} \zeta^2} + \frac{\mathrm{d} \bar{u}^i(\zeta)}{\mathrm{d} \zeta} = 0, \quad \zeta \in (0, \frac{1}{\varepsilon}),\\ 
		&\bar{u}^i(0)=0, \quad \bar{u}^i(\frac{1}{\varepsilon})=e.
	\end{aligned}
	\right.
\end{equation}
Note the computational domain for $\bar{u}^i(\zeta)$ is $\Omega^{\zeta} = (0, \frac{1}{\varepsilon})$.
It can be found that  the solution of problem (\ref{equ: MAE_inner}) is given by
\begin{equation}\label{equ: MAE_inner_solution}
	\begin{aligned}
		\bar{u}^i(\zeta)&=e - e^{1-\zeta}= e - e^{1-x/\varepsilon}=u^i(x). 
	\end{aligned}
\end{equation}

Finally, a uniformly valid solution called composite solution $u^c(x)$ is obtained:
\begin{equation}\label{equ: MAE_composite}
	u^c(x) = u^o(x) + (u^i(x) - (u^o)^{in}(x)),\quad x\in [0,1],
\end{equation}
which is regarded as an approximate solution of the exact solution $u_{exact}(x)$. 
For the problem (\ref{equ: MAE}), its composite solution is given by $u^c(x)= e^{1-x} - e^{1-x/\varepsilon}$ since $(u^o)^{in}(x) = e$ for any $x \in [0,1]$. 
The error between   $u^c(x)$ and $u_{exact}(x)$ achieves a magnitude of $O(\varepsilon)$ over the entire computational domain \cite{nayfeh2024perturbation}, which implies that the accuracy  improves as $\varepsilon$ decreases.

MAEs overcome the difficulty of straightforward expansions in simultaneously addressing sharp and gentle gradients by employing a magnified scale to capture steep changes and the original scale to characterize features outside the sharp-change regions. Thus, the MAEs have been used to solve various complex problems involving sharp gradients in solutions \cite{tang2022asymptotic, bressloff2021asymptotic}. However, the major drawback of the MAEs method lies in the necessity to explicitly solve both the outer and inner solutions, which can be impossible for a majority of problems \cite{faria2017equation}. Transferable neural network is one of the efficient numerical methods for addressing these issues. 

\subsection{Transferable Neural Networks}
\label{sec2.2}
\begin{figure*}[!htb]
	\centering
	\includegraphics[width=.9\textwidth]{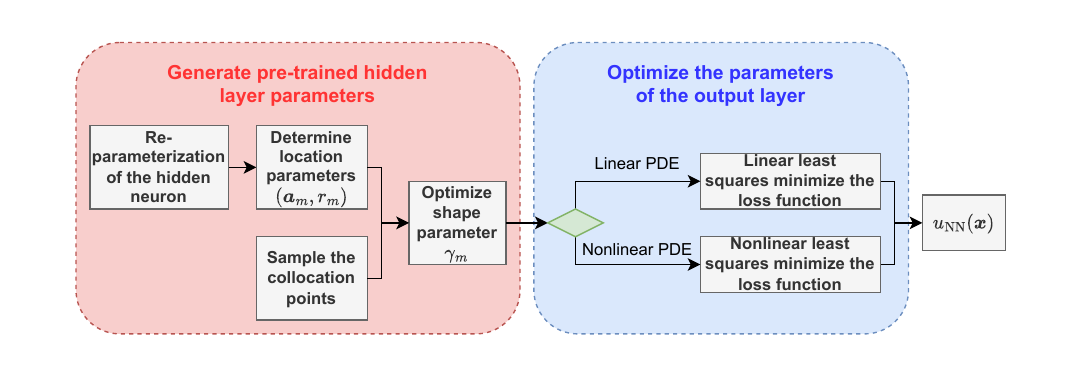}
	\caption{The flowchart of the transferable neural network method for solving PDEs.}
	\label{fig: TransNet}
\end{figure*}

The TransNet \cite{zhang2024transferable} is a fully-connected two-layer neural network (one hidden layer and one output layer) with pretrained hidden-layer neurons for solving PDEs of the following general formulation:
\begin{equation}\label{equ: TransNet_PDE}
	\left\{
	\begin{aligned}
		& \mathcal{L}(\boldsymbol{u}) = f, \quad \boldsymbol{x} \in \Omega,\\
		& \mathcal{B}(\boldsymbol{u}) =  g, \quad \boldsymbol{x} \in \partial \Omega,
	\end{aligned}
	\right.
\end{equation}
where $\Omega$ denotes the spatial domain in  $\mathbb{R}^d$ (or the spatial-temporal domain in $\mathbb{R}^d\times (0,T]$  for time-dependent PDE problems). 
As shown in Figure \ref{fig: TransNet}, The computational procedure of the TransNet method consists of two stages: generating pretrained hidden layer parameters, and optimizing the parameters of the output layer.

In the first stage, the pretrained model is the neural feature space $\mathcal{P}_{\mathrm{NN}}$, expanded by hidden-layer neurons (or called neural basis functions)  $\left\{\sigma\left(\boldsymbol{w}_m^{\top} \boldsymbol{x}+b_m\right)\right\}_{m=1}^M$:
\begin{equation}\label{equ: TransNet_NFS}
	\mathcal{P}_{\mathrm{NN}}=\operatorname{span}\left\{1, \sigma\left(\boldsymbol{w}_1^{\top} \boldsymbol{x}+b_1\right), \ldots, \sigma\left(\boldsymbol{w}_M^{\top} \boldsymbol{x}+b_M\right)\right\},
\end{equation} 
where $M$ is the number of hidden-layer neurons, $\sigma(\cdot)$ is the activation function, and $\boldsymbol{w}_m\in \mathbb{R}^d, b_m\in \mathbb{R}$ is the weight and bias of the $m$-th hidden-layer neuron. 
To achieve the uniform distribution of neurons in $\mathcal{P}_{\mathrm{NN}}$, the initial step involves the reparameterization of the hidden-layer neuron as:
\begin{equation}
	\sigma\left(\boldsymbol{w}_m^{\top} \boldsymbol{x}+b_m\right)
	= \sigma\left(\gamma_m\left(\boldsymbol{a}_m^{\top} \boldsymbol{x}+r_m\right)\right), \nonumber
\end{equation}
where $\boldsymbol{a}_m \in \mathbb{R}^d, \|\boldsymbol{a}_m\|_2=1$ is the unit normal vector of the partition hyperplane corresponding to the $m$-th hidden-layer neuron, that is,
\begin{equation}\label{equ: partition hyperplane}
	\boldsymbol{w}_m^{\top} \boldsymbol{x}+b_m=0. \nonumber
\end{equation}
Here, $r_m \in \mathbb{R}$, $|r_m|$ represents the distance from the origin to the partition hyperplane, $(\boldsymbol{a}_m, r_m)$ are named as location parameters, while $\gamma_m \in \mathbb{R}_+$ is the shape parameter. Let us take $\sigma(\cdot)=\tanh(\cdot)$ throughout this paper, which is widely used for solving PDEs. The shape parameter $\gamma_m$ controls the transition layer width of the activation function near the partition hyperplane. 
{\em Location parameter $\left\{\boldsymbol{a}_m\right\}_{m=1}^M$ are i.i.d., uniformly distributed on the $d$-dimensional unit sphere and $\left\{r_m\right\}_{m=1}^M$ are i.i.d., uniformly distributed in $[0,1]$.}
For simplicity, it is usually assumed that  all hidden-layer neurons  share the same shape parameter, that is, $\gamma_m=\gamma$. The next step is then to optimize the shape parameter $\gamma$ through the residuals at the collocation points for approximating Gaussian random fields, which can be achieved by some simple line search  algorithms (see \cite{zhang2024transferable} for details). 
At this stage, the construction of $\mathcal{P}_{\mathrm{NN}}$ is completed. 
To simplify notation, we assume that solution $\boldsymbol{u}$ in (\ref{equ: TransNet_PDE}) is a scalar function, then the network solution $u_\mathrm{NN}\in \mathcal{P}_\mathrm{NN}$ is expressed as
\begin{equation}\label{equ: u_NN}
	u_{\mathrm{NN}}(\boldsymbol{x})=\sum_{m=1}^M \alpha_m \sigma\left(\boldsymbol{w}_m^{\top} \boldsymbol{x}+b_m\right)+\alpha_0,
\end{equation}
where $\alpha_1, \alpha_2, \dots, \alpha_M$ are the weights of the output layer, and $\alpha_0$ is the bias of that.

In the second stage, the parameters of the hidden layer are frozen, the weights and bias of the output layer, that is, $\alpha_0, \alpha_1, \dots, \alpha_M$ in (\ref{equ: u_NN}) can be obtained through least squares solvers, which minimizes the PDE residual loss function: 
\begin{equation}\label{equ: TransNet_loss}
	{\mathrm{Loss}} = \lambda_1\|\mathcal{L}(u(\boldsymbol{x})) - \mathcal{L}(u_{\mathrm{NN}}(\boldsymbol{x}))\|_2^2 + \lambda_2\|\mathcal{B}(u(\boldsymbol{x})) - \mathcal{B}(u_{\mathrm{NN}}(\boldsymbol{x})) \|_2^2,
\end{equation}
where $\lambda_1$ and $\lambda_2$ are two positive weighting hyperparameters for balancing the loss terms from the interior domain and the boundary.
In all experiments of this paper, we simply set  $\lambda_1=\lambda_2=1$.

Algorithm \ref{alg: TransNet} summarizes the main steps of the TransNet method.  In this paper, the optimal $\gamma$ for each example can be efficiently determined via the golden section search with low computational cost as done in \cite{lu2025multiple}, and thus for simplicity, the value of $\gamma$ is directly provided as an input.

\begin{algorithm}[!htb]
	\caption{\enskip TransNet method for solving PDE problems \cite{zhang2024transferable}.}
	\label{alg: TransNet}
	\begin{algorithmic}[1]
		\Require The shape parameter $\gamma$ and the number of hidden-layer neurons $M$
		
		\Ensure The TransNet solution $u_{\mathrm{NN}}$

		\State Generate the location parameters $\left\{(\boldsymbol{a}_m, r_m)\right\}_{m=1}^{M}$ on $\Omega$ as the process described in \cite{zhang2024transferable} and form the set of hidden-layer neurons $\left\{\sigma\left(\gamma\left(\boldsymbol{a}_m^{\top} \boldsymbol{x}+r_m\right)\right)\right\}_{m=1}^{M}$ (uniformly distributed in $\Omega$).
		
		\State Generate collocation points in $\Omega$ and on $\partial \Omega$ by uniform sampling.
		
		\State Optimize the parameters of the output layer $\boldsymbol{\alpha}$ by minimizing the loss function (\ref{equ: TransNet_loss}), which is associated with the PDE problem \eqref{equ: TransNet_PDE},  over the collocation points using a certain least squares solver.
		
		\State Return $u_{\mathrm{NN}}$ based on (\ref{equ: u_NN}).
	\end{algorithmic}
\end{algorithm}

Wide ranges of numerical experiments  indicate that TransNet has much higher accuracy (numerically exponential-type convergence with respect to the number of neurons) and lower computing cost compared to PINN for various PDE problems with relatively smooth solutions \cite{zhang2024transferable}. However, numerical experiments in Section \ref{sec4} reveal TransNet's frequent inability to capture the highly sharp variations occurring within boundary layers of singular perturbation problems. In this paper, we address this limitation by employing MAEs to handle singular perturbation problems with boundary layers, and TransNet is then subsequently used to solve the corresponding simplified problems for the inner and outer solutions.

\section{The Proposed Method -- Matched Asymptotic Expansions-Based Transferable Neural Network}
\label{sec3}
The TransNet method meets significant challenges when dealing with singular perturbation problems, as it struggles to capture large gradient information within the boundary layer. In this section, we present the MAE-TransNet method for both one-dimensional and multidimensional singular perturbation problems with boundary layers, which could effectively resolve such issues.

\subsection{1D Singular Perturbation Problems With Boundary Layers}
\label{sec3.1}
In this subsection, for 1D singular perturbation problems, we first present how the MAE-TransNet method captures a single boundary layer and subsequently extend the method to multiple boundary layers.

\subsubsection{The Case of Single Boundary Layer}
\label{sec3.1.1}

\begin{figure*}[!htb]
	\centering
	\includegraphics[width=1\textwidth]{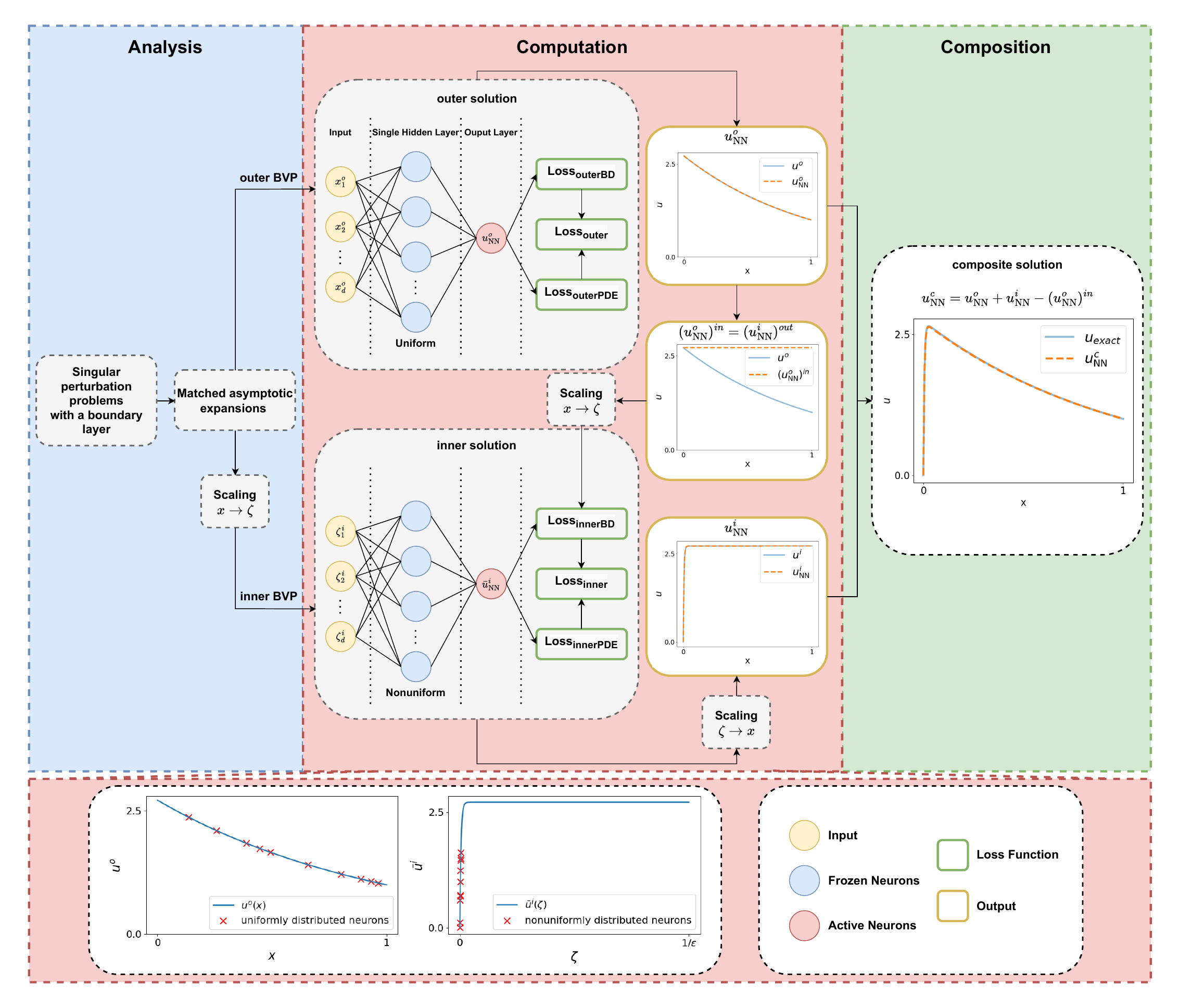}
	\vspace{-0.5cm}
	\caption{A schematic diagram of MAE-TransNet for 1D singular perturbation problems with single boundary layer.}
	\label{fig: MAE-TransNet_single}
\end{figure*}

For the purpose of illustration, let us again  consider the 1D singular perturbation problem (\ref{equ: MAE}), whose solution  has a boundary layer  with width $O(\varepsilon)$ close to the point $x=0$. As shown in Figure \ref{fig: MAE-TransNet_single}, the schematic diagram of MAE-TransNet is divided into three modules: analysis, computation, and composition. 

In the analysis module, the boundary value problem (\ref{equ: MAE_outer}) for the outer solution $u^o(x)$ (defined in $\overline\Omega = [0,1]$) is first derived, while the scaling transformation  $\zeta  = x/\delta(\varepsilon)$  (see \eqref{equ: MAE_inner_scaling} with $\delta(\varepsilon)=\varepsilon$) is employed to magnify the boundary layer and construct the  corresponding boundary value problem  (\ref{equ: MAE_inner})  (defined in the scaled domain $\overline{\Omega^{\zeta}}=[0,1/\varepsilon]$) for the inner solution $\bar{u}^i(\zeta)$.
In the computation module, the variation of the $u^o(x)$ in $\Omega$ is relatively gentle, thus the TransNet with uniformly distributed hidden-layer neurons in $\Omega$ is able to yield an accurate network outer solution $u_{\mathrm{NN}}^o(x)$. For $\bar{u}^i(\zeta)$, leveraging the existing $u_{\mathrm{NN}}^o$ and the matching principle outlined in (\ref{equ: MAE_matching_principle}), the inner limit of the outer solution $(u_{\mathrm{NN}}^o)^{in}$ is integrated to the problem (\ref{equ: MAE_inner}) as a supplementary boundary condition. 
It is noteworthy that since the gradient information of $\bar{u}^i(\zeta)$ concentrates within the scaled boundary layer, inspired by \cite{pieper2024nonuniform}, the nonuniformly distributed hidden-layer neurons are employed to compute the network inner solution  $\bar{u}_{\mathrm{NN}}^i(\zeta)$, in which the hidden-layer neurons are uniformly distributed within a small spherical region covering only  the scaled boundary layer region (such as $(0,1)\subset\Omega^{\zeta}$ in this case) rather than the entire scaled computational domain $\Omega^{\zeta}$.  
In the composite module,  the network composite solution $u_{\mathrm{NN}}^c$ is finally obtained through the combination of $u_{\mathrm{NN}}^o$, $u_{\mathrm{NN}}^i$ and the matching term $(u_{\mathrm{NN}}^o)^{in}$ according to $(\ref{equ: MAE_composite})$.

\begin{figure*}[!htb]
	\centering
	\includegraphics[width=0.9\textwidth]{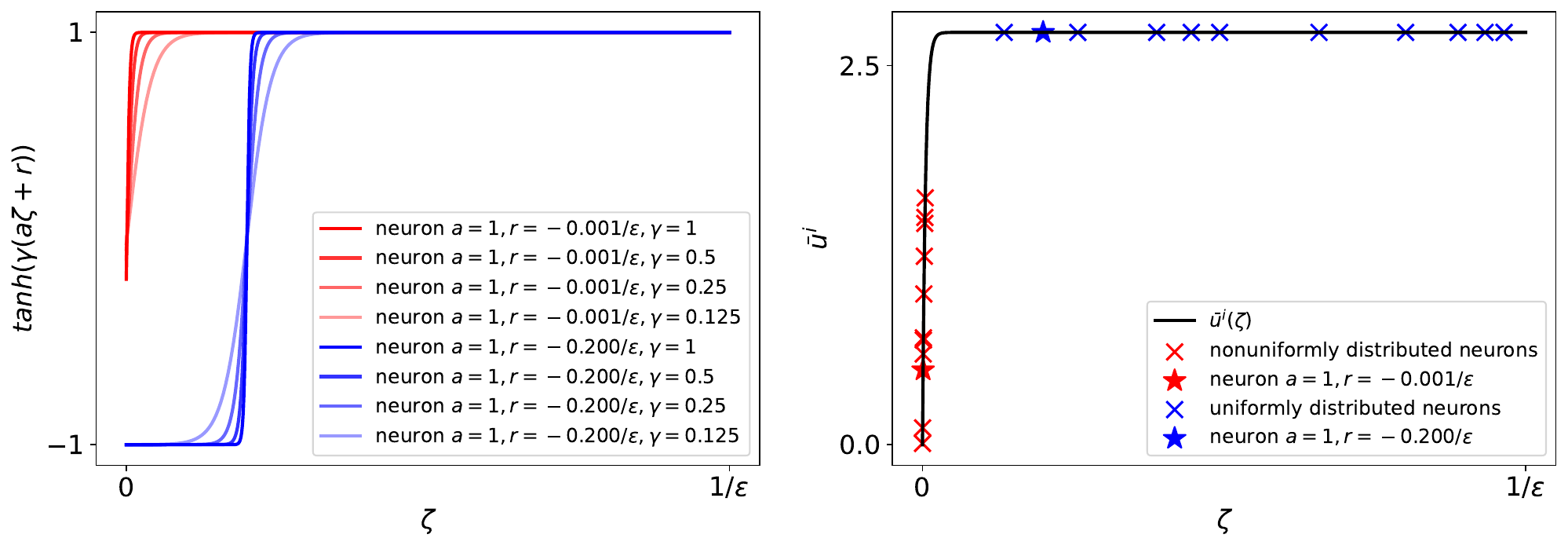}
	\caption{Left: Schematic illustration of the  neuron geometry structure under two location parameters $(a,r)$ and various shape parameter $\gamma$ in the computational domain $\Omega^{\zeta}$. Right: Comparison of two distinct hidden-layer neuron distributions across $\Omega^{\zeta}$.}
	\label{fig: inner_uni_vs_nonuni}
\end{figure*}

As illustrated in Figure \ref{fig: inner_uni_vs_nonuni}, the left panel presents two sets of hidden-layer neurons under different  values of location parameters $(a,r)$ in the scaled domain $\Omega^{\zeta}$, where  the  width of the transition region gets larger along the  decrease of the shape parameter $\gamma$. The right panel compares two distinct hidden-layer neuron distributions across  $\Omega^{\zeta}$ for solving the inner solution $\bar{u}^i(\zeta)$. 
A uniform neuron distribution in $\Omega^{\zeta}$ fails to capture the large gradient information within the scaled boundary layer due to the mild variation of most neurons in this region. In contrast, a nonuniform distribution concentrates neurons inside the scaled boundary layer region of $\Omega^{\zeta}$, enabling effective resolution of rapid variations. Traditional methods, such as FEM, typically rely on local basis functions. When confined strictly within the boundary layer, these functions fail to capture information outside this region. However, since the inner solution remains approximately constant outside the boundary layer, and  the neurons act as global basis functions that are nearly constant beyond the transition layer, neurons distributed within the scaled boundary layer can effectively capture the solution characteristics outside this region. 

\begin{remark}
	Van Dyke's matching principle  states that the $m$-term inner expansion of the $n$-term outer expansion equals the $n$-term outer expansion of the $m$-term inner expansion \cite{nayfeh2024perturbation}. By incorporating Van Dyke's matching rule as a soft constraint into the loss function, even higher-order MAE-TransNet methods potentially could  be constructed. 
\end{remark}

\subsubsection{The Case of Multiple Boundary Layers}
\label{sec3.1.2}
\begin{figure*}[!htb]
	\centering
	\includegraphics[width=.9\textwidth]{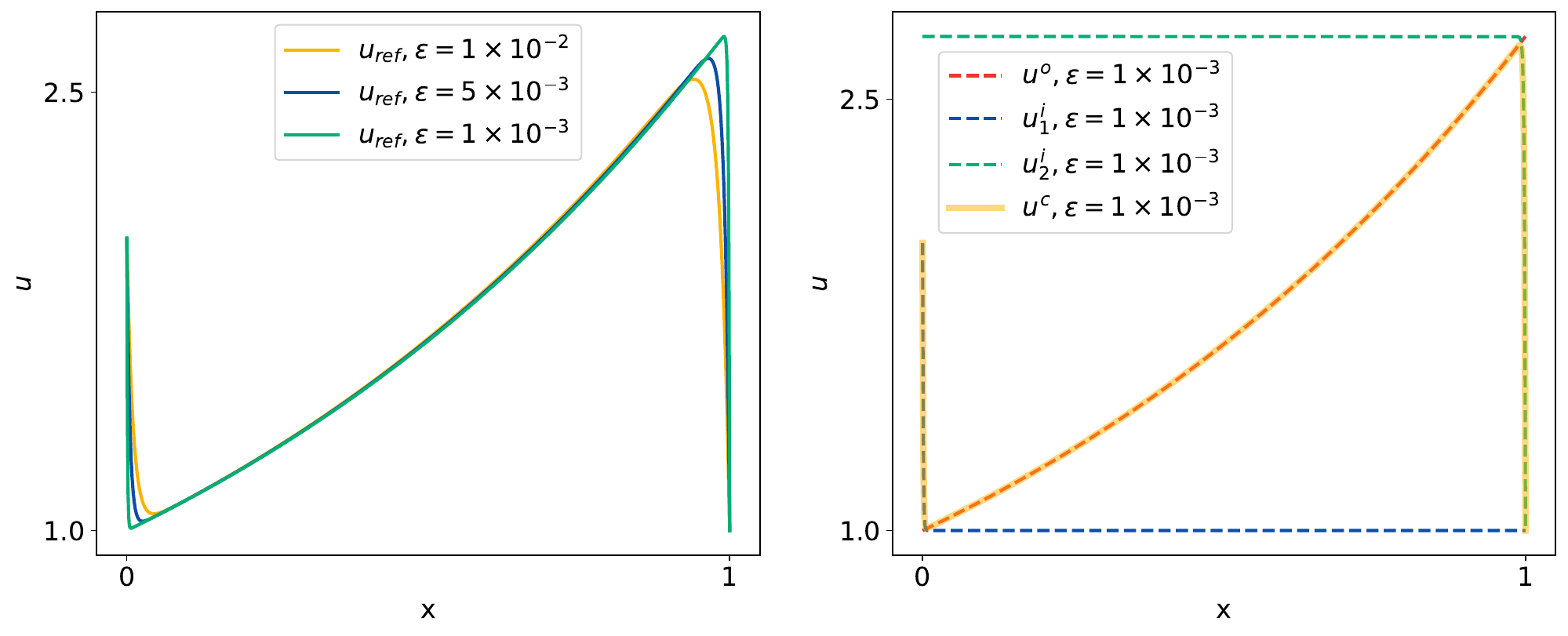}
	\caption{Left: The reference solutions of the 1D singular perturbation problem (\ref{equ: case2_original}) with different values of $\varepsilon$. Right: The outer, inner and composite solutions for the problem (\ref{equ: case2_original}) with $\varepsilon=1\times10^{-3}$.}
	\label{fig: case2_example}
\end{figure*}

Since the MAE approach is amenable to singular perturbation problems with multiple boundary layers, the proposed MAE-TransNet method  can be easily generalized  to solve such problems. 
The sole modification involves reformulating (\ref{equ: MAE_composite}) into the following equation: 
\begin{equation}\label{equ: MAE_composite_plus}
	u^c(x) = u^o(x) + \sum_{k=1}^{K}(u^i_k(x) - (u^o)^{in}_k(x)),\quad x\in \Omega,
\end{equation}
where $K\geq 2$ represents the number of boundary layers, $u^i_k$ denotes the inner solution of the $k$-th boundary layer, and $(u^o)^{in}_k$ signifies the inner limit of $u^o$ at the $k$-th boundary layer. For illustration, let us consider the following 1D singular perturbation problem in the domain $\Omega =(0,1)$ \cite{holmes2009introduction}: 
\begin{equation}\label{equ: case2_original}
	\left\{
	\begin{aligned}
		&\varepsilon^2 \frac{\mathrm{d}^2 u(x)}{\mathrm{d} x^2} + \varepsilon x\frac{\mathrm{d} u(x)}{\mathrm{d} x} - u(x) = -e^x, \quad x\in (0, 1), \\
		&u(0)=2, \quad u(1)=1.
	\end{aligned}
	\right.
\end{equation}
This problem involves two boundary layers, one is at  the point $x = 0$ and the other is at  $x=1$  with the same thickness $\varepsilon$.
As shown in Figure \ref{fig: case2_example}, the left plot presents the reference solutions $u_{ref}$ of the problem (\ref{equ: case2_original}) for different values of $\varepsilon$, while the right plot presents the outer, inner, and composite solutions for the problem (\ref{equ: case2_original}) with $\varepsilon=1\times10^{-3}$.

As $\varepsilon \rightarrow 0$, the problem (\ref{equ: case2_original}) reduces to 
\begin{equation}
	-u^o(x) = -e^x,\quad x\in (0,1),
\end{equation}
and this directly gives  the outer solution $u^o(x)=e^x$, which cannot satisfy either boundary conditions $u(0) = 2$ and $u(1)=1$. To establish the inner solution $u^i_1(x)$ at $x=0$, the scaling transformation $\zeta=\frac{x}{\delta_1(\varepsilon)}:=\frac{x}{\varepsilon}$, $\varepsilon \rightarrow 0$ and the matching principle are used to obtain the following boundary value problem:
\begin{equation}\label{equ: case2_inner1}
	\left\{
	\begin{aligned}
		&\frac{\mathrm{d}^2 \bar{u}^i_1(\zeta)}{\mathrm{d} \zeta^2} - \bar{u}^i_1(\zeta) = -1, \quad \zeta\in(0,\frac{1}{\varepsilon}), \\ 
		&\bar{u}^i_1(0)=2, \quad \bar{u}^i_1(\frac{1}{\varepsilon}) = u^o(0).
	\end{aligned}
	\right.
\end{equation}
Similarly, for the inner solution $u^i_2(x)$ at $x=1$, $\zeta=\frac{1-x}{\delta_2(\varepsilon)}:=\frac{1-x}{\varepsilon}$, $\varepsilon \rightarrow 0$ and the matching principle are used to obtain the following boundary value problem:
\begin{equation}\label{equ: case2_inner2}
	\left\{
	\begin{aligned}
		&\frac{\mathrm{d}^2 \bar{u}^i_2(\zeta)}{\mathrm{d} \zeta^2} - \frac{\mathrm{d} \bar{u}^i_2(\zeta)}{\mathrm{d} \zeta} - \bar{u}^i_2(\zeta) = -e, \quad \zeta \in (0, \frac{1}{\varepsilon}), \\ 
		&\bar{u}^i_2(0)=1, \quad \bar{u}^i_2(\frac{1}{\varepsilon}) = u^o(1).
	\end{aligned}
	\right.
\end{equation}
Note the computational domain for $\bar{u}^i_1(\zeta)$ and  $\bar{u}^i_2(\zeta)$ is $\Omega^{\zeta}_1= \Omega^{\zeta}_2= (0, \frac{1}{\varepsilon})$.
According to (\ref{equ: MAE_composite_plus}) with $K=2$, the composite solution of problem (\ref{equ: case2_original}) is then given by $$u^c=u^o + u^i_1 - (u^o)^{in}_1 + u^i_2 - (u^o)^{in}_2.$$

\begin{figure*}[!htb]
	\centering
	\includegraphics[width=1.\textwidth]{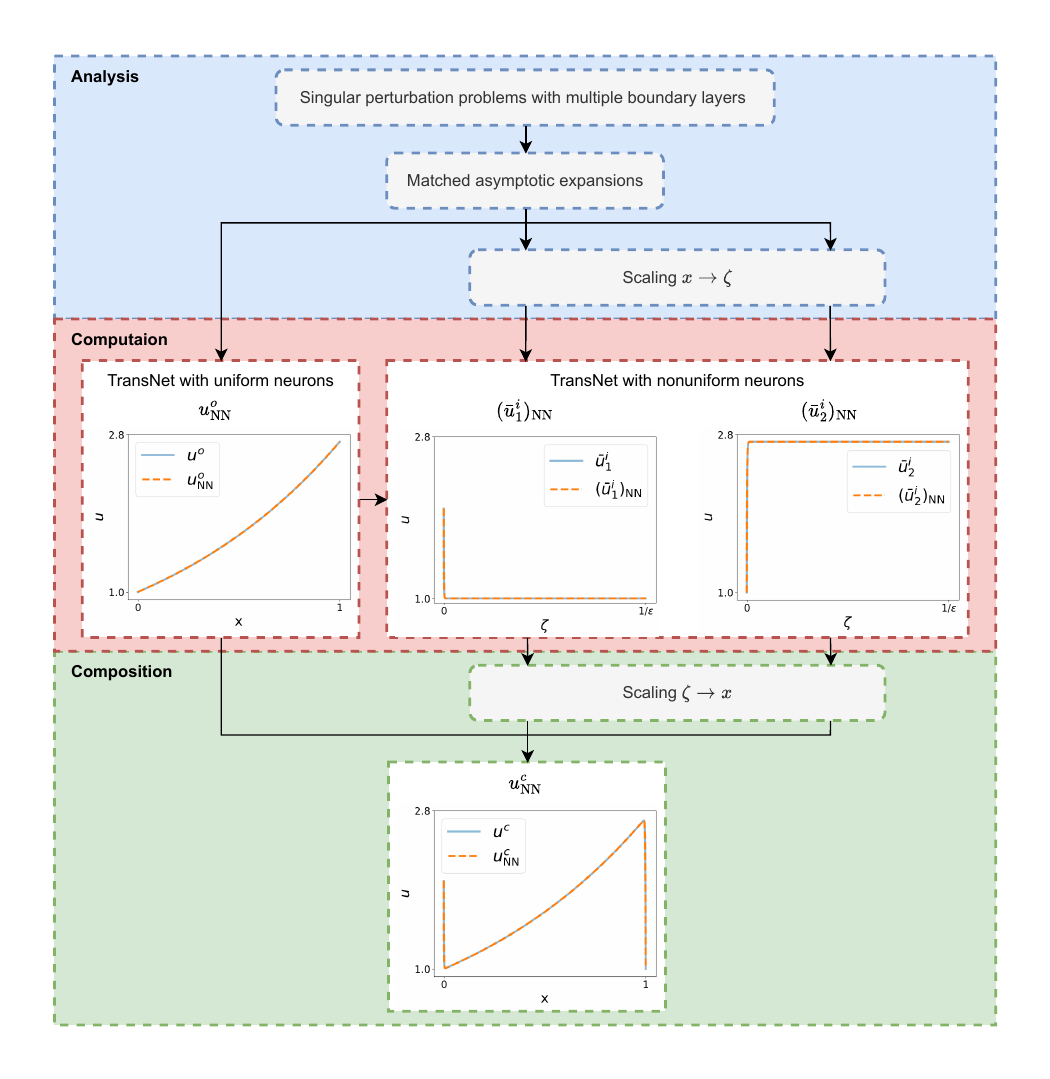}
	\vspace{-0.2cm}
	\caption{A schematic diagram of MAE-TransNet for 1D singular perturbation problems with $K=2$ boundary layers.}
	\label{fig: MAE-TransNet_multiple}
\end{figure*}

As shown in Figure \ref{fig: MAE-TransNet_multiple}, in the analysis module, $u^o(x)$ is directly obtained for  $x\in\Omega$ and the scaling transformations are introduced to magnify the two boundary layers. 
In the computation module, the network inner solutions $(\bar{u}^i_1)_{\mathrm{NN}}(\zeta)$ from (\ref{equ: case2_inner1}) and $(\bar{u}^i_2)_{\mathrm{NN}}(\zeta)$ from (\ref{equ: case2_inner2}) are computed in the domain $\Omega^{\zeta}$.
In the composite module, $u_{\mathrm{NN}}^c(x)$ is finally determined through (\ref{equ: MAE_composite_plus}).

Finally, we summarize in Algorithm \ref{alg: MAE-TransNet_multiple} the implementation details of the proposed MAE-TransNet for 1D singular perturbation problems with $K\geq 1$ boundary layers.

\begin{algorithm}[!htb]
	\caption{\enskip MAE-TransNet method for 1D singular perturbation problems.}
	\label{alg: MAE-TransNet_multiple}
	\begin{algorithmic}[1]
		\Require{The number of boundary layers $K$, the shape parameters $\{\gamma^o, \gamma^i_1, \cdots, \gamma^i_K\}$ and the number of hidden-layer neurons $\{M^o, M^i_1,\cdots, M^i_K\}$ for outer solution and inner solutions respectively, the scaling transformation factor $\{\delta_1(\varepsilon),\cdots,\delta_K(\varepsilon)\}$ for inner solutions.
		}
		\Ensure{The  MAE-TransNet composite solution $u_{\mathrm{NN}}^c$.}
		
		\State Form the outer solution problem  defined over $\Omega$ and compute the network outer solution $u_{\mathrm{NN}}^o$ by feeding $\gamma^o$ and $M^o$ into Algorithm \ref{alg: TransNet}.
		
		\For{$j = 1:K$}
			\State Find  the scaled domain $\Omega^{\zeta}_j$ with  the scaling factor $1/\delta_j(\varepsilon)$ and correspondingly form the $j$-inner solution problem  defined over $\Omega^{\zeta}_j$.
			
			\State Construct the set of hidden-layer  neurons $\left\{\sigma\left(\gamma^i_j\left({a}_m^{\top} {\zeta}+r_m\right)\right)\right\}_{m=1}^{M^i_j}$ which are uniformly distributed only within the scaled $j$-boundary layer region of $\Omega^{\zeta}_j$.
			
			\State Generate collocation points in $\Omega^{\zeta}_j$ and on $\partial \Omega^{\zeta}_j$ by uniform sampling.
			
			\State Optimize the parameters $\boldsymbol{\alpha}^i_j$ by minimizing the loss function (\ref{equ: TransNet_loss}) associated with the $j$-inner solution problem 
			over the collocation points  using a least squares solver to obtain  the $j$-inner network solution $(\bar{u}^i_{j})_{\mathrm{NN}}$.
		\EndFor
		
		\State Compute and return $u_{\mathrm{NN}}^c$ based on (\ref{equ: MAE_composite_plus}).
	\end{algorithmic}
\end{algorithm}

\subsection{Multidimensional Singular Perturbation Problems With Boundary Layers}
\label{sec3.2}
In this subsection, we extend the proposed MAE-TransNet method from one-dimensional cases to multidimensional cases. For simplicity, we illustrate  the MAE-TransNet method for multidimensional cases through 2D problems.

\subsubsection{The Case of a Single Boundary Layer Case}
\label{sec3.2.1}
\begin{figure*}[!htb]
	\centering
	\includegraphics[width=0.95\textwidth]{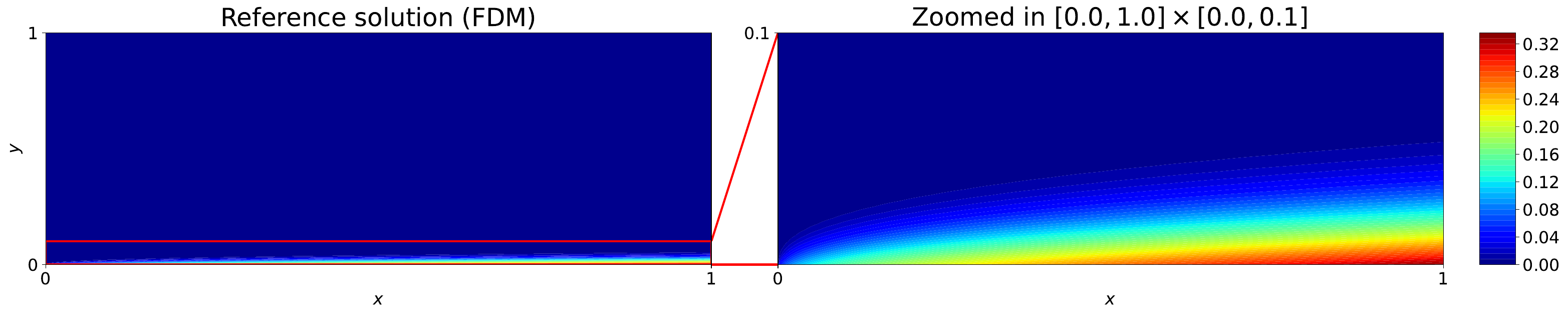}
	\caption{The reference solution for the  2D Couette flow problem (\ref{equ: case5_original}) with $\varepsilon=1 \times 10^{-4}$.
		Left: Over the whole domain $\Omega$. Right: Zoomed in $[0, 1] \times [0, 0.1]$.}
	\label{fig: case5_example}
\end{figure*}

Let us consider the 2D Couette flow problem (advection-diffusion transport) in the domain $\Omega=(0,1)^2$ \cite{arzani2023theory}: 
\begin{equation}\label{equ: case5_original}
	\left\{
	\begin{aligned}
		&10y \frac{\partial u(x, y)}{\partial x}-\varepsilon\left(\frac{\partial^2 u(x, y)}{\partial x^2}+\frac{\partial^2 u(x, y)}{\partial y^2}\right) = 0, \quad (x,y) \in (0, 1)^2,\\
		&u(x=0,y)=0, \quad \frac{\partial u}{\partial y}(x,y=0)=-10 ,\quad  \frac{\partial u}{\partial x}(x=1,y)=0, \quad u(x,y=1)=0,
	\end{aligned}
	\right.
\end{equation}
which is a typical 2D singular perturbation problem with a single boundary layer at $y=0$. The reference solution for $\varepsilon=1 \times 10^{-4}$ is shown in Figure \ref{fig: case5_example}.

As $\varepsilon\rightarrow0$, the problem (\ref{equ: case5_original}) reduces to the following problem for the outer solution: 
\begin{equation}
	\frac{\partial u(x, y)}{\partial x}= 0,\quad (x,y)\in (0,1)^2,
\end{equation}
with the boundary condition $u(x=0,y)=0$,  this directly gives  the outer solution $u^o(x,y)=0$.
To establish the inner solution $u^i$, the scaling transformation $\eta = \frac{y}{\delta(\varepsilon)} := \frac{y}{\sqrt{\varepsilon}}$ (i.e., asymptotic expansion in $y$) and the matching principle are used to obtain the following boundary value problem: 
\begin{equation}\label{equ: case5_inner}
	\left\{
	\begin{aligned}
		&10\sqrt{\varepsilon}\eta\frac{\partial \bar{u}^i(x, \eta)}{\partial x}=\frac{\partial^2 \bar{u}^i(x, \eta)}{\partial \eta^2}, \quad (x,\eta) \in (0, 1)\times (0, \frac{1}{\sqrt{\varepsilon}}),\\ 
		&\bar{u}^i(x=0,\eta)=0, \quad \frac{\partial \bar{u}^i}{\partial \eta}(x, \eta=0)=-10\sqrt{\varepsilon}, \quad \bar{u}^i(x,\eta=\frac{1}{\sqrt{\varepsilon}})=0,
	\end{aligned}
	\right.
\end{equation}
where $\bar{u}^i(x,\eta) = {u}^i(x,\sqrt{\varepsilon}\eta)= {u}^i(x,y)$.
Subsequently, the composite solution $u_{\mathrm{NN}}^c$ is constructed through (\ref{equ: MAE_composite}). The implementation details can be similarly  referred to Algorithm \ref{alg: MAE-TransNet_multiple} with $K=1$ in Section \ref{sec3.1.2}, thus are skipped here.

\subsubsection{The Case of Coupled Boundary Layers}
\label{sec3.2.2}

Due to the absence of MAE theory for coupled boundary layer problems, the direct deployment of MAE-TransNet could result in significant errors in the coupled region. 
To address this issue, an auxiliary TransNet in the coupled region is combined with MAE-TransNet over the whole domain, thereby improving its capability to handle multidimensional coupled boundary layer problems.

\begin{figure*}[!htb]
	\centering
	\includegraphics[width=.8\textwidth]{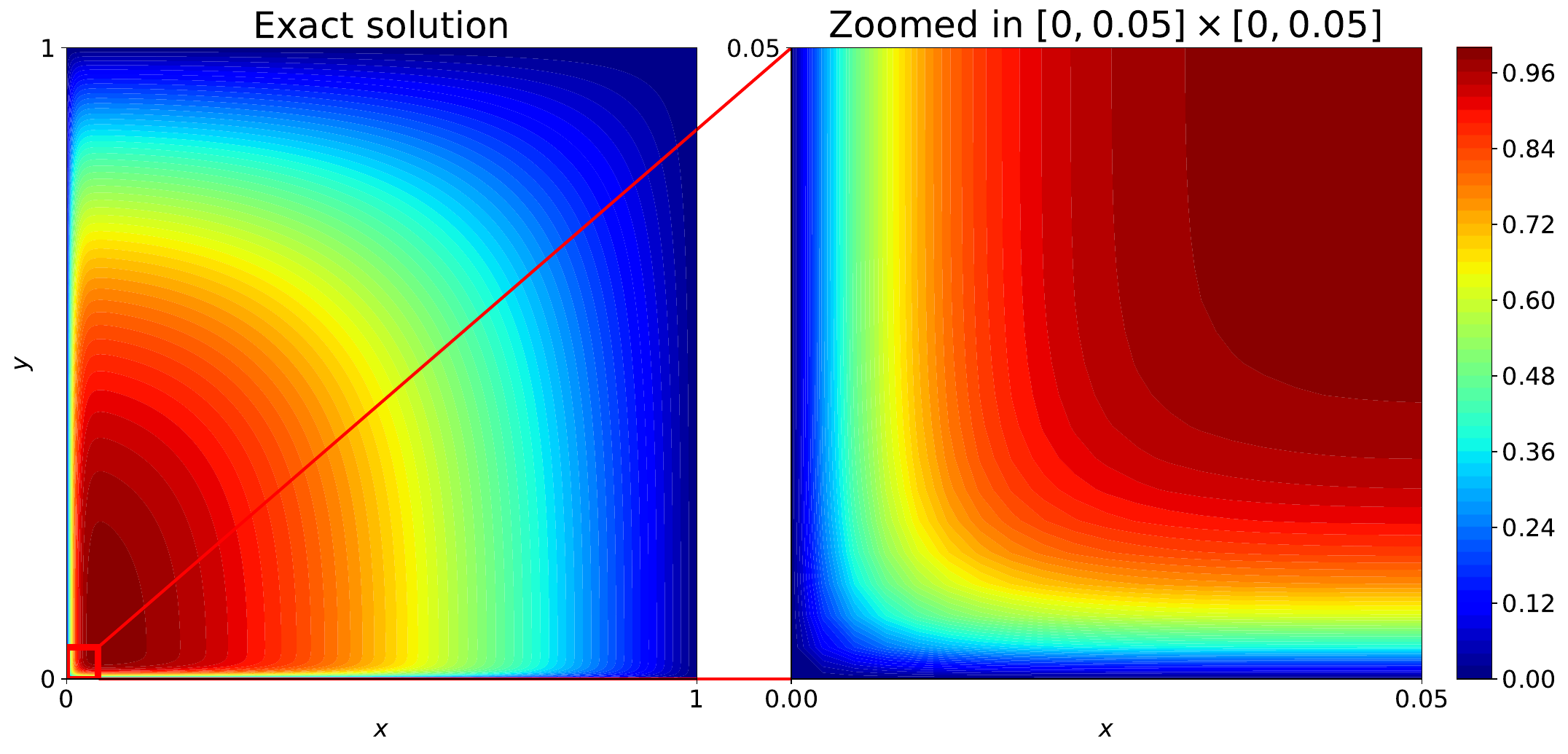}
	\caption{The exact solution of the 2D singular perturbation problem (\ref{equ: case6_original})  of coupled boundary layers with $\varepsilon=2^{-6}$.
		Left: Over the whole domain $\Omega$. Right: Zoomed in $[0, 0.05] \times [0, 0.05]$.}
	\label{fig: case6_example}
\end{figure*}

\begin{figure*}[!htb]
	\centering
	\includegraphics[width=1.\textwidth]{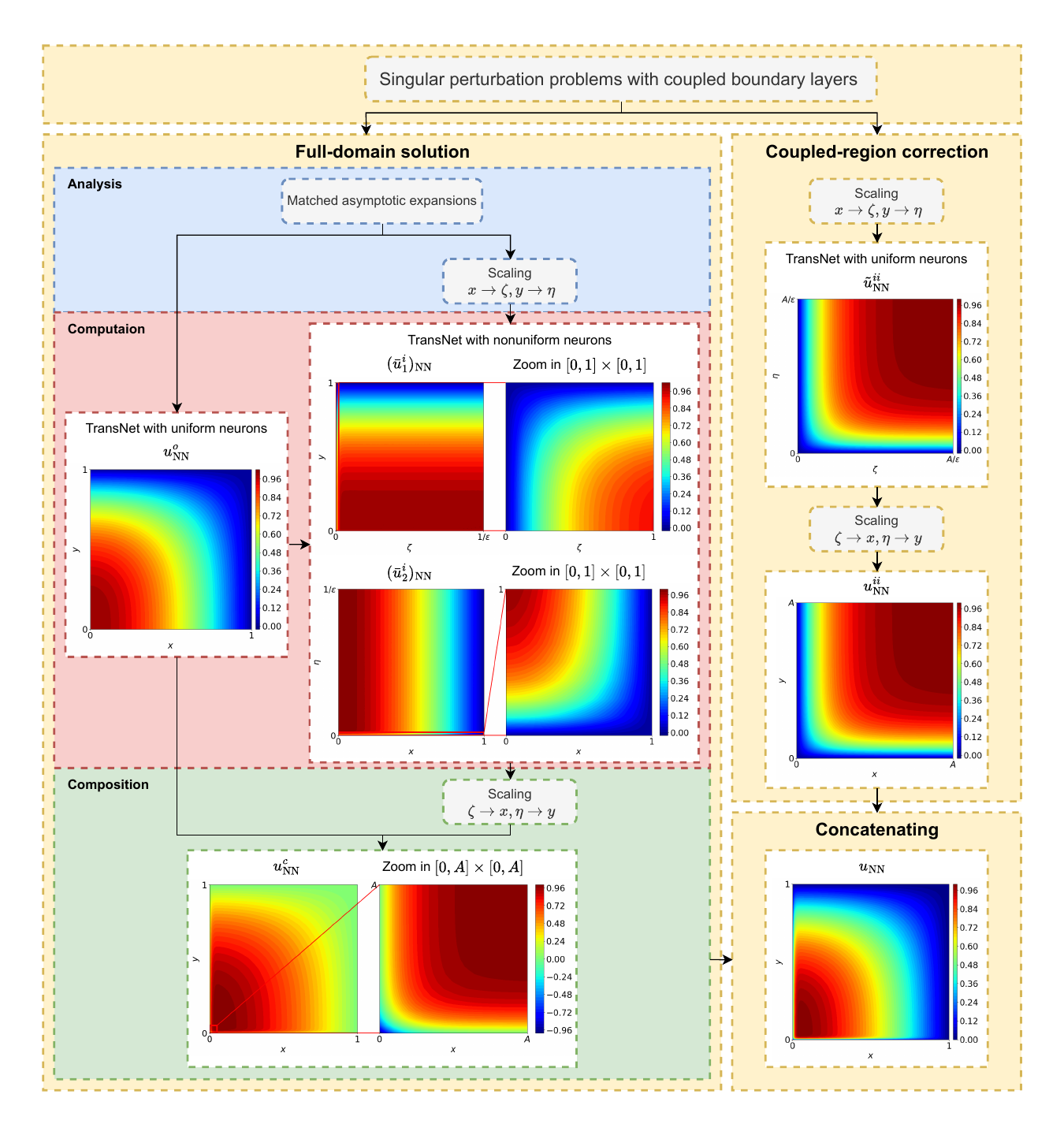}
	\vspace{-0.5cm}
	\caption{A schematic diagram of MAE-TransNet for 2D singular perturbation problems with coupled boundary layers. During the concatenating stage, $u_{\mathrm{NN}}$ is synthesized through $u_{\mathrm{NN}}^c$ in $[0,1]^2 \setminus [0,A]^2$ and $u_{\mathrm{NN}}^{ii}$ in $[0,A]^2$. }
	\label{fig: MAE-TransNet_coupled}
\end{figure*}

To illustrate how MAE-TransNet can be employed in this case, let us consider the following 2D coupled boundary layer problem \cite{podila2024wavelet} in the domain $\Omega=(0,1)^2$: 
\begin{equation}\label{equ: case6_original}
	\left\{
	\begin{aligned}
		&-\varepsilon\left(\frac{\partial^2 u(x, y)}{\partial x^2}+\frac{\partial^2 u(x, y)}{\partial y^2}\right)-(x+2) \frac{\partial u(x, y)}{\partial x}\\
		&\hspace{4cm}-\left(y^3+3\right) \frac{\partial u(x, y)}{\partial y}+u(x,y)=f(x, y), \quad (x,y)\in(0,1)^2,\\[2pt]
		&u(x=0,y)=0, \quad u(x=1,y)=0, \quad u(x,y=0)=0, \quad u(x,y=1)=0,
	\end{aligned}
	\right.
\end{equation}
where $f(x, y)$ is chosen appropriately so that the exact solution is  
\begin{equation}\label{equ: case6_exact}
	u(x, y)=\cos \left(\frac{\pi x}{2}\right)\left(1-e^{-2 x / \varepsilon}\right)\left(1-y^3\right)\left(1-e^{-3 y / \varepsilon}\right). \nonumber
\end{equation}
The exact solution for $\varepsilon=2^{-6}$ is shown in Figure \ref{fig: case6_example}. 
Figure \ref{fig: MAE-TransNet_coupled} presents  our proposed computational framework, in which a MAE-TransNet is implemented for finding the full-domain solution in $\Omega$, and an extra auxiliary TransNet is used for correction in the coupled region $\Omega^{ii}=(0,A)^2$ with $\varepsilon\le A \ll 1$.

Over the whole domain, as $\varepsilon\rightarrow0$, the problem \eqref{equ: case6_original} is simplified and the outer solution $u^o$ satisfies the following boundary value problem:
\begin{equation}\label{equ: case6_outer}
	\left\{
	\begin{aligned}
		&-(x+2) \frac{\partial u^o(x, y)}{\partial x}-\left(y^3+3\right) \frac{\partial u^o(x, y)}{\partial y}+u^o(x, y)=f^o(x,y), \quad (x,y) \in (0, 1)^2,\\ 
		&u^o(x=1, y)=0, \quad u^o(x, y=1)=0,
	\end{aligned}
	\right.
\end{equation}
where 
\begin{equation}\label{equ: case6_outer_f}
	f^o(x,y) = \left(1-y^3\right)\cos \frac{\pi x}{2} + (x+2)\frac{\pi}{2}\left(1-y^3\right)\sin \frac{\pi x}{2} + \left(y^3+3\right)3y^2\cos \frac{\pi x}{2}. \nonumber
\end{equation}
In order to capture the two inner solutions along the $x$-axis (i.e., $u^i_1$) and the $y$-axis (i.e., $u^i_2$) respectively, the scaling transformation $\zeta = \frac{x}{\delta_1(\varepsilon)} := \frac{x}{\varepsilon}$ (i.e., asymptotic expansion in $x$) and $\eta = \frac{y}{\delta_2(\varepsilon)} := \frac{y}{\varepsilon}$ (i.e, asymptotic expansion in $y$), and the matching principle are used to obtain the following two boundary value problems: 
\begin{equation}\label{equ: case6_innerx}
	\left\{
	\begin{aligned}
		&-\frac{\partial^2 \bar{u}^i_1(\zeta, y)}{\partial \zeta^2} - 2\frac{\partial \bar{u}^i_1(\zeta, y)}{\partial \zeta}=0, \quad (\zeta,y) \in (0, \frac{1}{\varepsilon})\times(0, 1),\\ 
		&\bar{u}^i_1(\zeta=0,y)=0, \quad \bar{u}^i_1(\zeta=\frac{1}{\varepsilon}, y)=u^o(x=0, y), \quad \bar{u}^i_1(\zeta, y=1)=0, 
	\end{aligned}
	\right.
\end{equation}
for solving $\bar{u}^i_1(\zeta, y)$ in  $\Omega^{\zeta}_1= (0, \frac{1}{\varepsilon})\times(0, 1)$,
and 
\begin{equation}\label{equ: case6_innery}
	\left\{
	\begin{aligned}
		&-\frac{\partial^2 \bar{u}^i_2(x, \eta)}{\partial \eta^2} - 3\frac{\partial \bar{u}^i_2(x, \eta)}{\partial \eta}=0, \quad (x,\eta) \in (0, 1)\times (0, \frac{1}{\varepsilon}), \\ 
		&\bar{u}^i_2(x,\eta=0)=0, \quad \bar{u}^i_2(x, \eta=\frac{1}{\varepsilon})=u^o(x, y=0), \quad \bar{u}^i_2(x=1, \eta)=0. 
	\end{aligned}
	\right.
\end{equation}
for solving $\bar{u}^i_2(x,\eta)$ in   $\Omega^{\eta}_2= (0, 1)\times(0, \frac{1}{\varepsilon})$.
The MAE-TransNet method described in Algorithm \ref{alg: MAE-TransNet_multiple} with $K=2$ in Section \ref{sec3.1.2} can be similarly  employed to obtain  $u_{\mathrm{NN}}^o$ from (\ref{equ: case6_outer}), $(\bar{u}_1^i)_{\mathrm{NN}}$ from (\ref{equ: case6_innerx}), $(\bar{u}_2^i)_{\mathrm{NN}}$ from (\ref{equ: case6_innery}), and  $u^c_{\mathrm{NN}}$ from (\ref{equ: MAE_composite_plus}). 

In the coupled region $\Omega^{ii}=(0,A)^2$, the two scaling transformations mentioned above are utilized simultaneously,  as special solution in $\Omega^{\zeta,\eta}=(0,\frac{A}{\varepsilon})^2$ is denoted as $\tilde{u} ^{ii}(\zeta,\eta) = u^{ii}(x,y)$ which satisfies the following boundary value problem: 
\begin{equation}\label{equ: case6_inner}
	\left\{
	\begin{aligned}
		&-\left(\frac{\partial^2 \tilde{u}^{ii}(\zeta, \eta)}{\partial \zeta^2}+\frac{\partial^2 \tilde{u}^{ii}(\zeta, \eta)}{\partial \eta^2}\right) - 2\frac{\partial \tilde{u}^{ii}(\zeta, \eta)}{\partial \zeta} - 3\frac{\partial \tilde{u}^{ii}(\zeta, \eta)}{\partial \eta}=0, \quad (\zeta, \eta)\in(0, \frac{A}{\varepsilon})^2,\\ 
		&\tilde{u}^{ii}(\zeta=0, \eta)=0, \quad \tilde{u}^{ii}(\zeta=\frac{A}{\varepsilon}, \eta)=u^c(x=A, \varepsilon\eta), \\
		&\tilde{u}^{ii}(\zeta, \eta=0)=0, \quad \tilde{u}^{ii}(\zeta, \eta=\frac{A}{\varepsilon})=u^c(\varepsilon\zeta, y=A), 
	\end{aligned}
	\right.
\end{equation}
where the top and right boundary conditions are supplied by $u^c_{\mathrm{NN}}$ found in the previous step. Subsequently, a TransNet with uniformly distributed hidden-layer neurons is employed  to solve the problem \eqref{equ: case6_inner} for the network  solution in the coupled region, denoted as $\tilde{u}_{\mathrm{NN}}^{ii}$.

Finally, by concatenating $u_{\mathrm{NN}}^c$ in $\Omega \setminus \Omega^{ii}$ and $u_{\mathrm{NN}}^{ii}$ in $\Omega^{ii}$, a consistently valid solution $u_{\mathrm{NN}}$ across the entire computational domain is obtained.
Algorithm \ref{alg: MAE-TransNet_coupled} presents the implementation details of the proposed MAE-TransNet for multidimensional singular perturbation problems with coupled boundary layers.

\begin{algorithm}[!htb]
	\caption{\enskip MAE-TransNet method for multidimensional singular perturbation problems with coupled boundary layers.}
	\label{alg: MAE-TransNet_coupled}
	\begin{algorithmic}[1]
		\Require{The number of boundary layers $K$, the shape parameters $\{\gamma^o, \gamma^i_1, \gamma^i_2, \dots, \gamma^i_K, \gamma^{ii}\}$ and the number of hidden-layer neurons $\{M^o, M^i_1, M^i_2, \dots, M^i_K, M^{ii}\}$ for outer solution, inner solution and solution in the coupled region respectively, the scaling transformation factor $\{\delta_{1}(\varepsilon), \delta_{2}(\varepsilon), \dots, \delta_{K}(\varepsilon)\}$ for inner solutions.
			}
		\Ensure{The MAE-TransNet solution $u_{\mathrm{NN}}$.}
		
		\State Form the outer solution problem defined over $\Omega$ and the inner solution problems defined over $\{\Omega^{\zeta}_1, \Omega^{\zeta}_2, \cdots \Omega^{\zeta}_K\}$ with the scaling factors $\{1/\delta_{1}(\varepsilon), 1/\delta_{2}(\varepsilon), \dots, 1/\delta_{K}(\varepsilon)\}$ respectively, then compute the network composite solution $u_{\mathrm{NN}}^c$ by feeding $\{\gamma^o, \gamma^i_1, \gamma^i_2, \dots, \gamma^i_K\}$ and $\{M^o, M^i_1, M^i_2, \dots, M^i_K\}$ into Algorithm \ref{alg: MAE-TransNet_multiple}.
		
		\State Find the coupled region $\Omega^{\zeta,\eta}$ with the scaling factors $1/\delta_m(\varepsilon)$ and $1/\delta_n(\varepsilon)$ (the coupled region $\Omega^{ii}$ is assumed to arise from the interaction between the $m$-th and $n$-th boundary layers), then correspondingly form the problem defined over $\Omega^{\zeta,\eta}$.
		
		\State Compute the network solution $\tilde{u}^{ii}$ over the coupled region $\Omega^{\zeta,\eta}$ by feeding $\gamma^{ii}$ and $M^{ii}$ into Algorithm \ref{alg: TransNet}.
		
		\State Return $u_{\mathrm{NN}}$ by concatenating $u_{\mathrm{NN}}^c$ in $\Omega \setminus \Omega^{ii}$ and $u_{\mathrm{NN}}^{ii}$ in $\Omega^{ii}$.
	\end{algorithmic}
\end{algorithm}

\section{Numerical Experiments}
\label{sec4}
In this section, we conduct numerical experiments for the proposed MAE-TransNet method on a series of singular perturbation problems in various dimensions, including linear problems with single or multiple boundary layers  (test cases 1, 2, and 3), a nonlinear problem (test case 4), the 2D Couette flow problem (test case 5), a 2D coupled boundary layers problem (test case 6), and the 3D Burgers vortex problem (test case 7).

The collocation points (i.e., the training points) and the test points are sampled uniformly in the computational domain (including interior domain and boundaries), and the number of test points is set to $2^d$ times as large as that of collocation points, where $d$ is the problem dimension.
For the singular perturbation problem with a specific $\varepsilon$, the number of collocation points is maintained consistent across the computation of the outer, inner and coupled-region solutions, if needed.

The main neural network parameters used in the experiments are presented below: 
\begin{itemize}
	\item $M^o$ denotes the number of hidden-layer neurons for the outer solution. $M^i, M_1^i$ and $M_2^i$ denote the numbers of hidden-layer neurons for the inner solution. $M^{ii}$ denotes the number of hidden-layer neurons for the coupled-region solution.
	\item $(\boldsymbol{a}_m, r_m)$ are named as the location parameters, where $\left\{\boldsymbol{a}_m\right\}_{m=1}^M$ are i.i.d., uniformly distributed on the $d$-dimensional unit sphere and $\left\{r_m\right\}_{m=1}^M$ are i.i.d., uniformly distributed in $[0,1]$.
	\item $\gamma^o$ denotes the shape parameter for the outer solution. $\gamma^i, \gamma_1^i$ and $\gamma_2^i$ denote the shape parameters for the inner solution. $\gamma^{ii}$ denotes the shape parameter for the coupled-region solution. These shape parameters are determined by the golden-section search as done in \cite{lu2025multiple}.
\end{itemize}

As in \cite{zhang2024transferable}, the discrete $L_2$ error and the discrete $L_{\infty}$ error are used to evaluate the network solution $u_{\mathrm{NN}}$ obtained by different neural network method
\begin{equation}\label{equ: error}
	\|u-u_{\mathrm{NN}}\|_2 = \sqrt{\frac{1}{K}\sum_{k=1}^K(u(\boldsymbol{x}^{k})-u_{\mathrm{NN}}(\boldsymbol{x}^{k}))^2}, \quad
	\|u-u_{\mathrm{NN}}\|_\infty = \max\left\{\left|u(\boldsymbol{x}^{k})-u_{\mathrm{NN}}(\boldsymbol{x}^{k})\right|\right\}_{k=1}^{K}, \nonumber
\end{equation}
where $\boldsymbol{x}^{k}$ is the $k$-th test point and $K$ is the total number of test points. Our code is implemented using PyTorch on a workstation with an NVIDIA GeForce RTX 4090.

\subsection{Test Case 1: 1D Linear Problem With a Single Boundary Layer}
\label{sec4.1}
We consider the 1D advection-diffusion-reaction problem in the domain $\Omega=(0,1)$ \cite{arzani2023theory, bender2013advanced}: 
\begin{equation}\label{equ: case1_original}
	\left\{
	\begin{aligned}
		&\varepsilon \frac{\mathrm{d}^2 u(x)}{\mathrm{d} x^2} + (1+\varepsilon)\frac{\mathrm{d} u(x)}{\mathrm{d} x} + u(x)=0, \quad x\in(0, 1), \\
		&u(0)=0, \quad u(1)=1. 
	\end{aligned}
	\right.
\end{equation}
It is well known that one boundary layer occurs at $x=0$  with the thickness $\varepsilon$ for this problem. In fact the exact solution is given by $u(x)=(e^{-x}-e^{-x/{\varepsilon}})/(e^{-1}-e^{-1/\varepsilon})$.

As $\varepsilon \rightarrow 0$, the outer solution $u^o$ satisfies the following boundary value problem in the domain $\Omega=(0,1)$:
\begin{equation}\label{equ: case1_outer}
	\left\{
	\begin{aligned}
		&\frac{\mathrm{d} u^o(x)}{\mathrm{d} x} + u^o(x) = 0,  \quad x\in(0, 1), \\ 
		&u^o(1)=1.
	\end{aligned}
	\right.
\end{equation}
To establish the inner solution $u^i$ at $x=0$, the scaling transformation $\zeta = \frac{x}{\delta(\varepsilon)}:=\frac{x}{\varepsilon}$ is introduced to magnify the boundary layer and the second boundary condition is obtained through  the matching principle, then as $\varepsilon\rightarrow0$, the problem (\ref{equ: case1_original}) reduces to the following boundary value problem in the domain $\Omega^\zeta=(0, \frac{1}{\varepsilon})$: 
\begin{equation}\label{equ: case1_inner}
	\left\{
	\begin{aligned}
		&\frac{\mathrm{d}^2 \bar{u}^i(\zeta)}{\mathrm{d} \zeta^2} + \frac{\mathrm{d} \bar{u}^i(\zeta)}{\mathrm{d} \zeta} = 0, \quad \zeta\in(0, \frac{1}{\varepsilon}),\\ 
		&\bar{u}^i(0)=0, \quad \bar{u}^i(\frac{1}{\varepsilon}) = u^o(0).
	\end{aligned}
	\right.
\end{equation}
It is easy to verify $u^o(x) = e^{1-x}$ and $\bar{u}^i(\zeta) = e - e^{1-\zeta}$,  Finally, the composite solution to the problem (\ref{equ: case1_original}) $u^c(x) = e^{1-x}-e^{1-x/\varepsilon}$ is obtained from (\ref{equ: MAE_composite}). 

\begin{figure*}[!htb]
	\centering
	\subfigure{
		\begin{minipage}[t]{.9\textwidth}
			\centering
			\includegraphics[width=\linewidth]{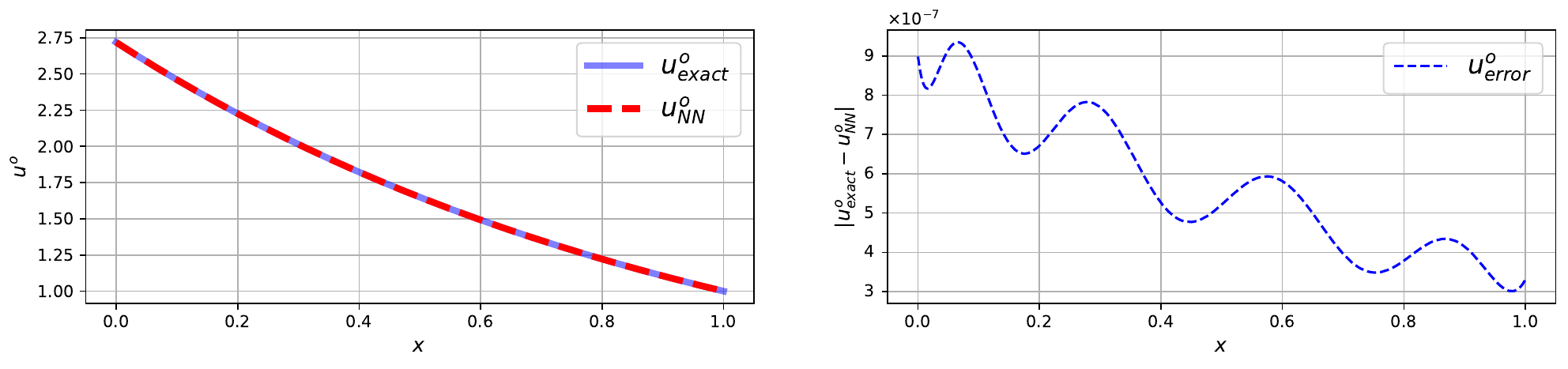}
		\end{minipage}
	} \\
	\vspace{-0.4cm}
	\subfigure{
		\begin{minipage}[t]{.9\textwidth}
			\centering
			\includegraphics[width=\linewidth]{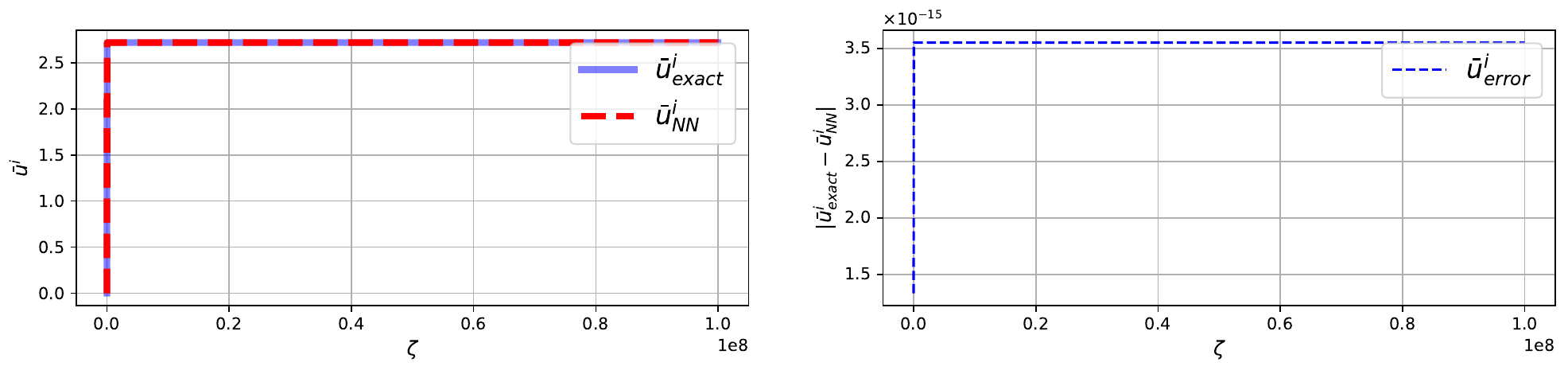}
		\end{minipage}
	} \\
	\vspace{-0.4cm}
	\subfigure{
		\begin{minipage}[t]{.9\textwidth}
			\centering
			\includegraphics[width=\linewidth]{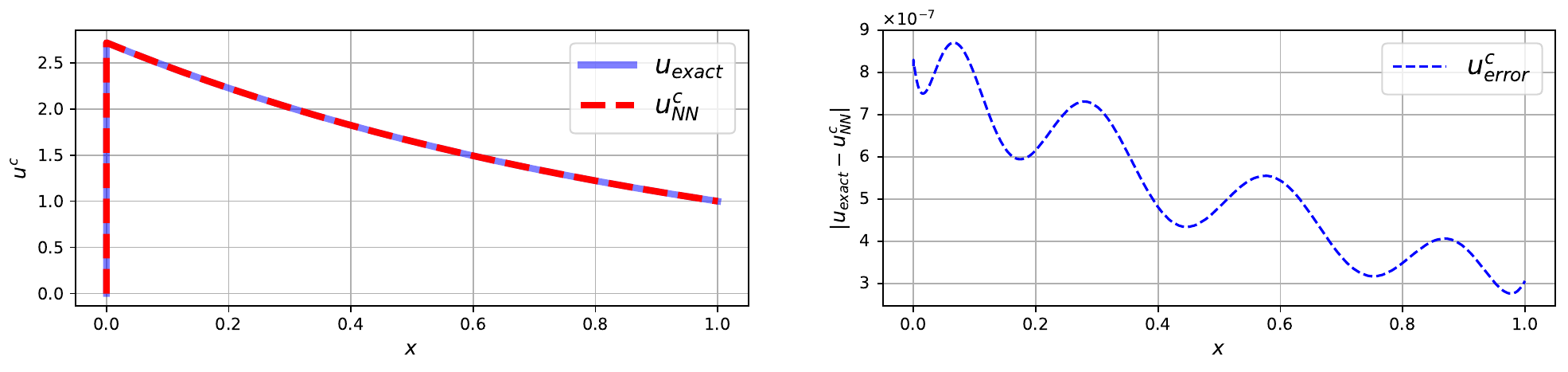}
		\end{minipage}
	} \\
	\caption{Comparisons between the exact solutions $u^o$, $u^i$ and $u$  and corresponding MAE-TransNet solutions   $u_{\mathrm{NN}}^o$, $u_{\mathrm{NN}}^i$ and $u^c_{\mathrm{NN}}$ (from top to bottom) for the test case 1 with $\varepsilon=1\times10^{-8}$ in Subsection \ref{sec4.1}.}
	\label{fig: case1_result}
\end{figure*}

\begin{figure*}[!htb]
	\centering
	\includegraphics[width=.9\textwidth]{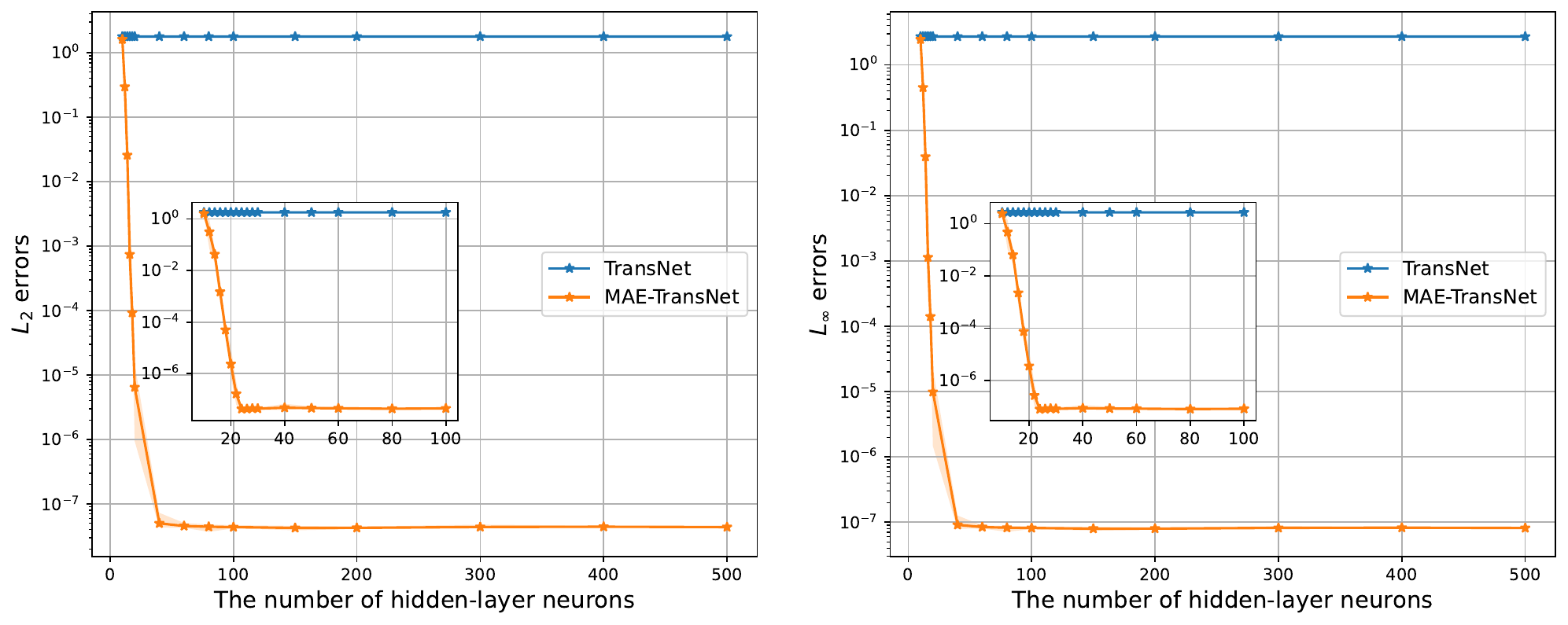}
	\caption{Comparison of the performance between MAE-TransNet solution $u_{\mathrm{NN}}^c$ and TransNet solution $u_{\mathrm{NN}}$ for the test case 1 with $\varepsilon=1\times10^{-8}$ in Subsection \ref{sec4.1}.
		Left: $L_2$ errors $\| u_{\mathrm{NN}}^c - u\|_2$ and $\| u_{\mathrm{NN}} - u\|_2$. Right: $L_{\infty}$ errors $\| u_{\mathrm{NN}}^c - u\|_{\infty}$ and $\| u_{\mathrm{NN}} - u\|_{\infty}$. The confidence band represents the range between the 30-th and 70-th percentiles across 50 repeated runs.}
	\label{fig: case1_result_ablation}
\end{figure*}

The  thickness parameter $\varepsilon$ is first set to be $1 \times 10^{-8}$. For our MAE-TransNet method, the shape parameter for the outer solution is set to $\gamma^o = 1$, and for the inner solution, it is set to $\gamma^i = 0.5$. 
The number of hidden-layer neurons for the outer solution and the inner solution are set to $M^o =M^i = 10$. 
The number of collocation points in $\Omega$ and $\Omega^\zeta$ is uniformly set to $N_\Omega = N_{\Omega^\zeta} = 99999$ and that on $\partial\Omega$ and $\partial\Omega^\zeta$ is uniformly set to $N_{\partial\Omega} = N_{\partial\Omega^\zeta} = 2$, then the total number of collocation points is denoted as $N = N_\Omega + N_{\partial\Omega} = N_{\Omega^\zeta} + N_{\partial\Omega^\zeta} = 100001$.
The left column of Figure \ref{fig: case1_result} presents the comparisons of the exact solutions $u^o$, $u^i$ and $u$  and corresponding MAE-TransNet solutions   $u_{\mathrm{NN}}^o$, $u_{\mathrm{NN}}^i$ and $u^c_{\mathrm{NN}}$, while the right column shows corresponding pointwise absolute errors. Using MAE-TransNet with only 10 neurons for each of two boundary value problems, the produced numerical results exhibit errors with magnitudes $10^{-7}$ in $u_{\mathrm{NN}}^o$, $10^{-15}$ in $\bar{u}_{\mathrm{NN}}^i$, and $10^{-7}$ in $u_{\mathrm{NN}}^c$. 
Figure \ref{fig: case1_result_ablation} compares the performance of MAE-TransNet and TransNet methods with different numbers of neurons under identical collocation and test points. 
The shape parameters for MAE-TransNet are mentioned above, and those for TransNet are determined via the golden section search.
It is observed that the MAE-TransNet solution $u_{\mathrm{NN}}^{c}$ achieve the best accuracy with just a total of $M^i$ + $M^o = 24$ neurons, and the errors
$\|u_{\mathrm{NN}}^{c}-u\|_2$ and $\|u_{\mathrm{NN}}^{c}-u\|_{\infty}$ show no further improvements along the increase of neurons. This is mainly because the MAE 
error $u^{c}-u$ already  dominates the network approximation  error  $u_{\mathrm{NN}}^{c}-u^c$ in this case. In addition, we easily find that the solutions exhibit the MAE-TransNet method significantly and consistently outperforms the original TransNet method.

\begin{figure*}[!htb]
	\centering
	\includegraphics[width=.45\textwidth]{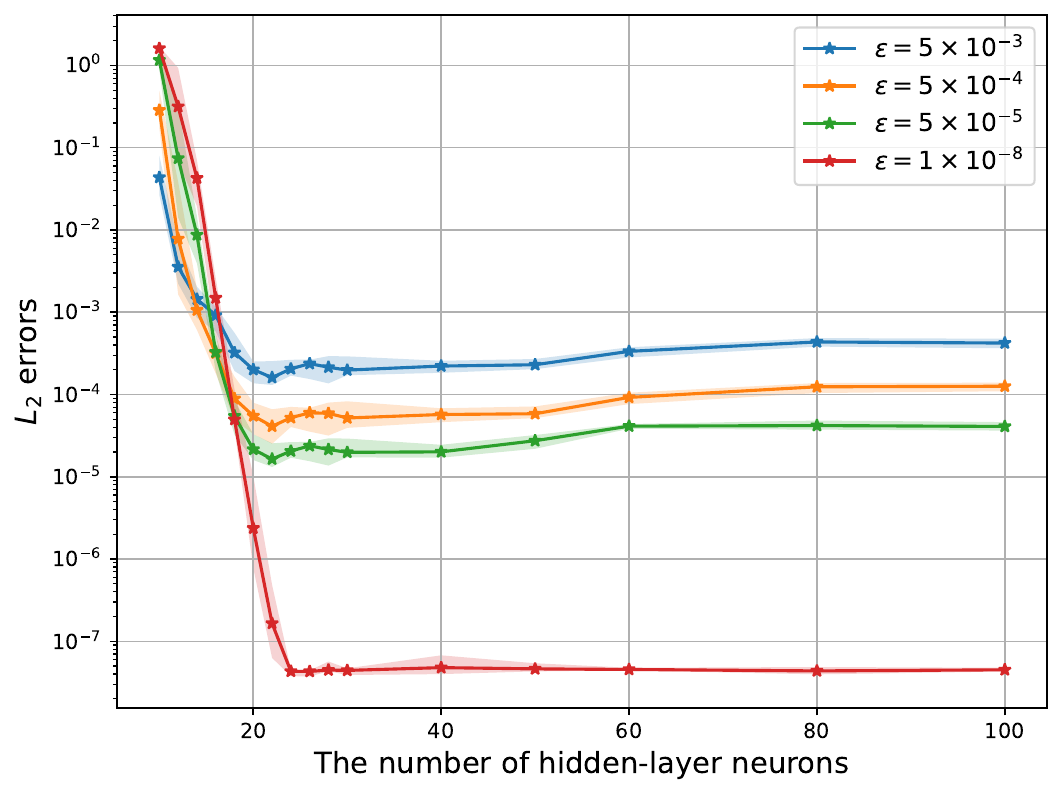}
		\vspace{-0.2cm}
	\caption{Comparison of the convergence of MAE-TransNet solution $u_{\mathrm{NN}}^c$ for the test case 1 with $\varepsilon=5\times10^{-3}, 5\times10^{-4}, 5\times10^{-5}$ and $1\times10^{-8}$ respectively in Subsection \ref{sec4.1}.
	The confidence band represents the range between the 30-th and 70-th percentiles across 50 repeated runs.}
	\label{fig: case1_convergence}
\end{figure*}

Figure \ref{fig: case1_convergence} compares the convergence of MAE-TransNet solution $u_{\mathrm{NN}}^c$ for different values of $\varepsilon$ ($5 \times 10^{-3}$, $5 \times 10^{-4}$, $5 \times 10^{-5}$ and $1 \times 10^{-8}$ respectively). As $\varepsilon \rightarrow 0$, convergence is achieved with a small number of hidden-layer neurons, demonstrating the robustness of MAE-TransNet.
Table \ref{tab: case1_table} reports the performance comparisons of the four neural network methods (PINN, BL-PINN, TransNet and MAE-TransNet) using the same collocation and test points for different values of $\varepsilon$. 
For PINN, the network consists of 5 hidden layers with 120 neurons per layer, and the learning rate is set to $1\times10^{-4}$ with $2000$ epochs. For BL-PINN, all parameters follow the code in \cite{arzani2023theory}. For instance, each network corresponds to the inner solution and the outer solution has $5$ hidden layers with $60$ neurons per layer, and the learning rate is set to $1\times10^{-4}$ with $2000$ epochs. 
The shape parameters for TransNet and MAE-TransNet are mentioned above. 
It is observed that: first, PINN and TransNet fails to solve this singular perturbation problem; second, as BL-PINN does, the solution accuracy of MAE-TransNet gets improved when $\varepsilon$ decreases, which is consistent with the theory of MAEs; third, compared to BL-PINN, our MAE-TransNet requires significantly fewer neurons  and less running time while achieving more accurate solutions.

\begin{table}[!htb]
	\centering
	\caption{Performance comparison of the four neural network methods (PINN, BL-PINN, TransNet and MAE-TransNet) for the test case 1  in Subsection \ref{sec4.1}.}\vspace{0.2cm}
	\label{tab: case1_table}
	\begin{tabular}{llrrrll}
		\toprule
		\multicolumn{1}{l}{Method} & \multicolumn{1}{l}{${\varepsilon}$} & 
		\multicolumn{1}{l}{Points} &\multicolumn{1}{l}{Neurons} & \multicolumn{1}{l}{{Time(s)}} & \multicolumn{1}{l}{${L_2}$}Error &
		\multicolumn{1}{l}{${L_{\infty}}$ Error}\\
		\midrule
		\multirow{4}{*}{PINN} & $5 \times 10^{-3}$ & 201 & 600 & 16.9 & 1.49e$+$0 & 2.38e$+$0 \\
		& $5 \times 10^{-4}$ & 2,001 & 600 & 18.9 & 1.53e$+$0 & 2.37e$+$0 \\
		& $5 \times 10^{-5}$ & 20,001 & 600 & 19.7 & 1.55e$+$0 & 2.35e$+$0 \\ 
		& $1 \times 10^{-8}$ & 100,001 & 600 & 260.3 & 1.53e$+$0 & 2.37e$+$0 \\ 
		\midrule
		\multirow{4}{*}{BL-PINN}& $5 \times 10^{-3}$ & 201 & 600 & 34.4 & 4.80e$-$3 & 4.92e$-$2 \\       
		& $5 \times 10^{-4}$ & 2,001 & 600 & 38.5 & 2.27e$-$3 & 6.88e$-$3 \\
		& $5 \times 10^{-5}$ & 20,001 & 600 & 40.2 & 1.90e$-$3 & 5.88e$-$3 \\
		& $1 \times 10^{-8}$ & 100,001 & 600 & 185.6 & 1.25e$-$3 & 3.03e$-$3 \\
		\midrule
		\multirow{4}{*}{TransNet}   & $5 \times 10^{-3}$ & 201 & 20 & 0.0059& 1.79e$+$0 & 2.65e$+$0 \\
		& $5 \times 10^{-4}$ & 2,001 & 20 & 0.0116 & 1.78e$+$0 & 2.70e$+$0 \\
		& $5 \times 10^{-5}$ & 20,001 & 20 & 0.0671 & 1.78e$+$0 & 2.70e$+$0 \\
		& $1 \times 10^{-8}$ & 100,001 & 20 & 0.1133 & 1.78e$+$0 & 2.71e$+$0 \\
		\midrule
		\multirow{4}{*}{MAE-TransNet}   & $5 \times 10^{-3}$ & 201 & 20 & 0.0044& 1.02e$-$4 & 2.62e$-$3\\
		& $5 \times 10^{-4}$ & 2,001 & 20 & 0.0078 & 6.11e$-$5 & 2.14e$-$3\\
		& $5 \times 10^{-5}$ & 20,001 & 20 & 0.0241 & 1.06e$-$5 & 1.58e$-$3\\
		& $1 \times 10^{-8}$ & 100,001 & 20 & 0.0841 & 3.00e$-$6 & 4.56e$-$6 \\
		\bottomrule
	\end{tabular}
\end{table}

\subsection{Test Case 2: 1D Linear Problem With Two Boundary Layers of the Same Thickness}
\label{sec4.2}
\begin{figure*}[!htb]
	\centering
	\subfigure{
		\begin{minipage}[t]{.9\textwidth}
			\centering
			\includegraphics[width=\linewidth]{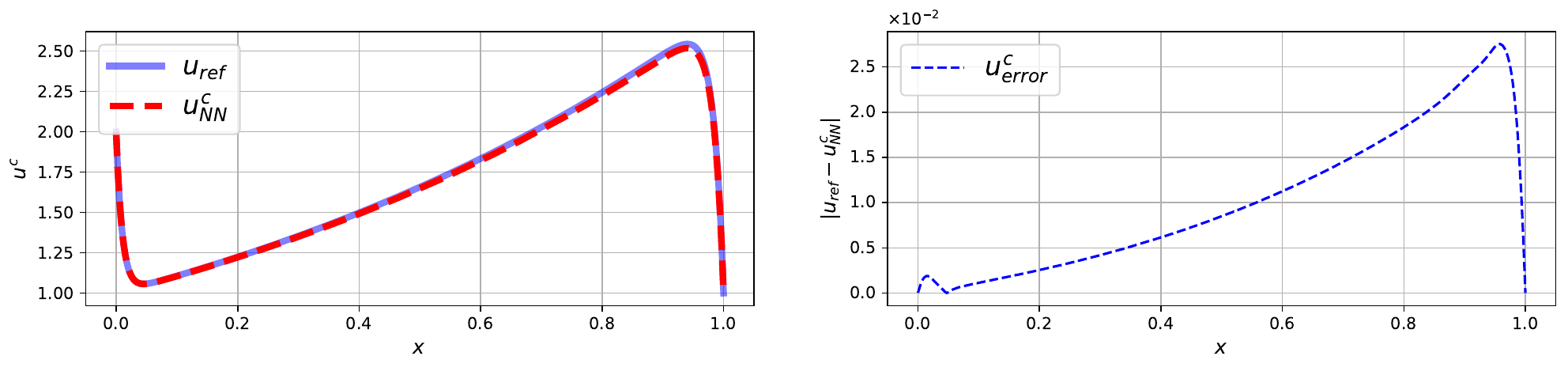}
		\end{minipage}
	} \\
	\vspace{-0.4cm}
	\subfigure{
		\begin{minipage}[t]{.9\textwidth}
			\centering
			\includegraphics[width=\linewidth]{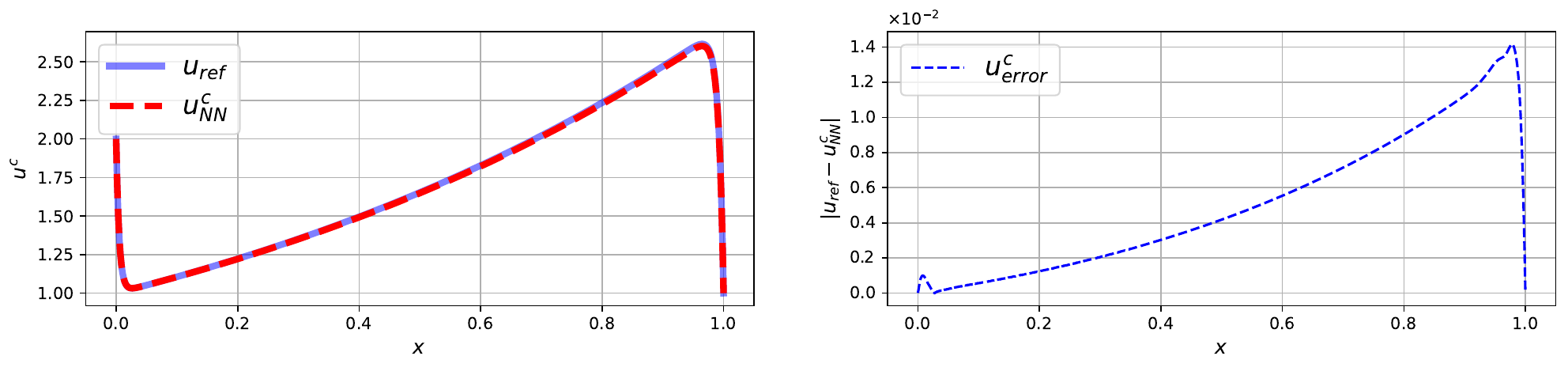}
		\end{minipage}
	} \\
	\vspace{-0.4cm}
	\subfigure{
		\begin{minipage}[t]{.9\textwidth}
			\centering
			\includegraphics[width=\linewidth]{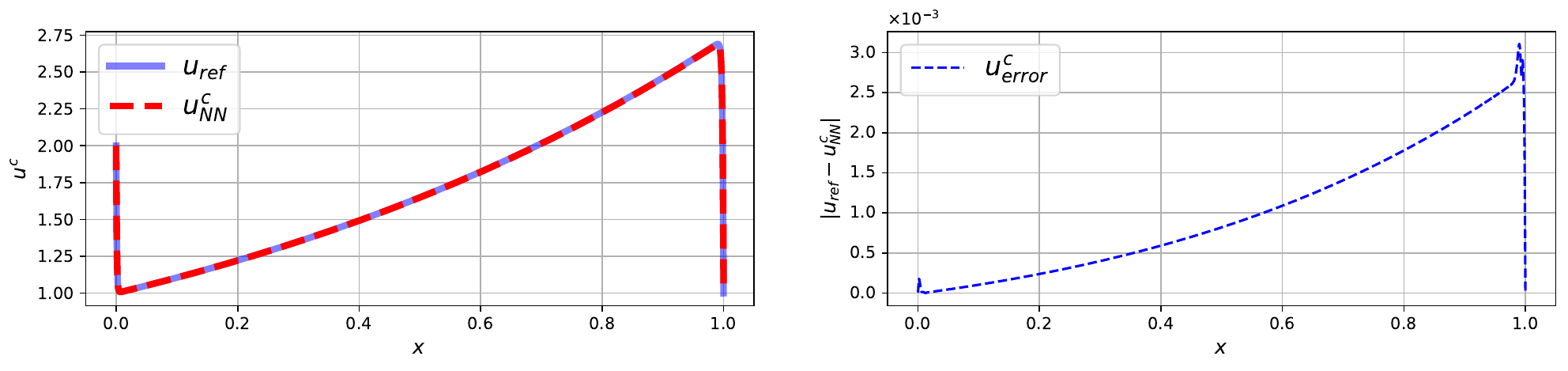}
		\end{minipage}
	} \\
	\vspace{-0.4cm}
	\subfigure{
		\begin{minipage}[t]{.9\textwidth}
			\centering
			\includegraphics[width=\linewidth]{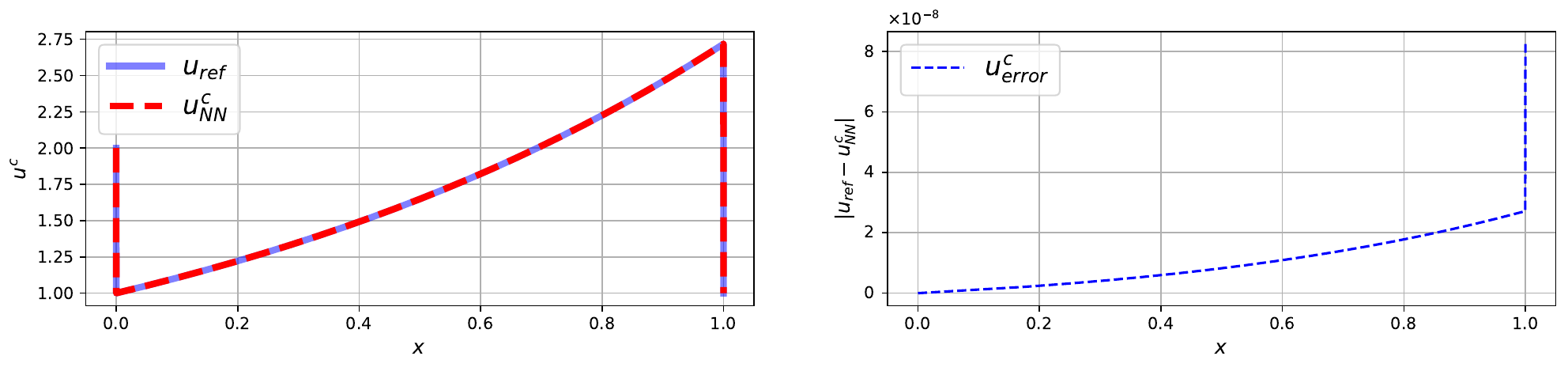}
		\end{minipage}
	} \\
	\caption{Comparison between the reference solution $u_{ref}$ and corresponding MAE-TransNet solution $u_{\mathrm{NN}}^c$ for the test case 2 with $\varepsilon=1\times10^{-2}$, $\varepsilon=5\times10^{-3}$, $\varepsilon=1\times10^{-3}$ and $\varepsilon=1\times10^{-8}$ (from top to bottom) in Subsection \ref{sec4.2}.}
	\label{fig: case2_result}
\end{figure*}

\begin{figure*}[!htb]
	\centering
	\includegraphics[width=.9\textwidth]{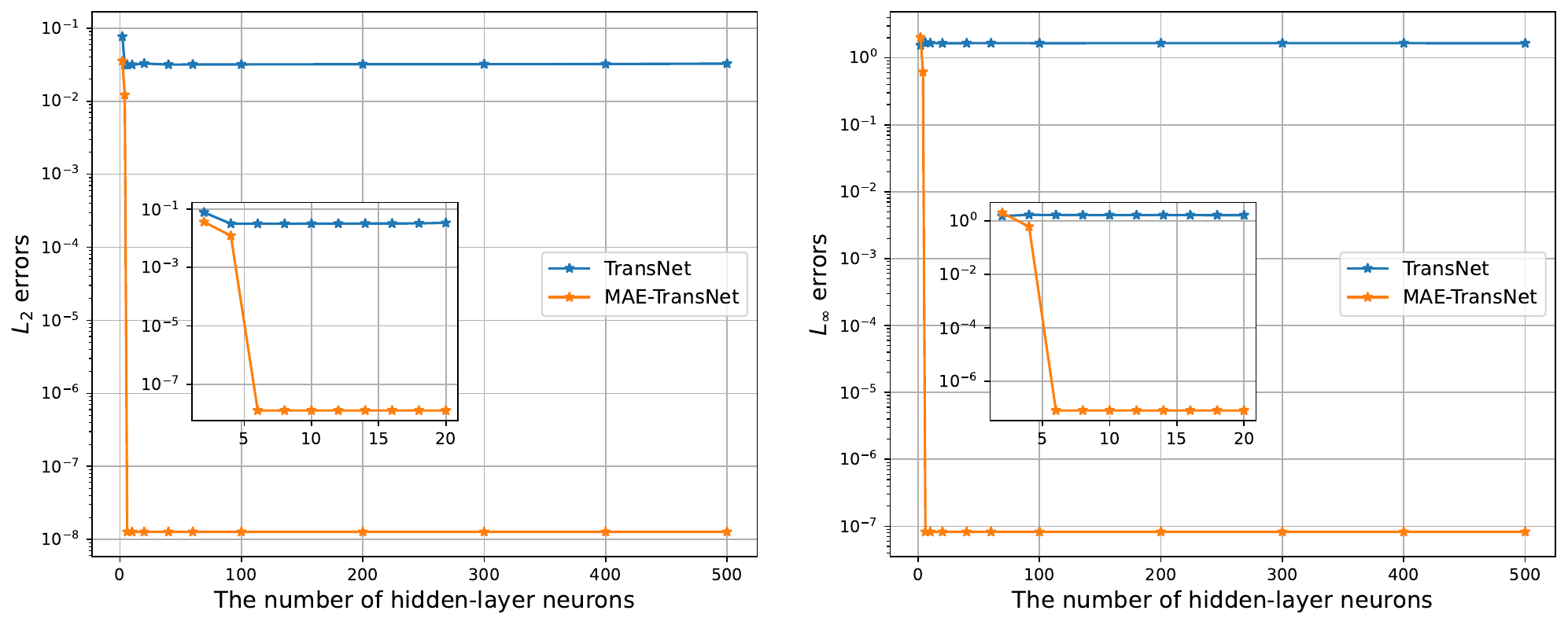}
		\vspace{-0.2cm}
	\caption{Comparison of the performance between MAE-TransNet solution $u_{\mathrm{NN}}^c$ and TransNet solution $u_{\mathrm{NN}}$ for the test case 2 with $\varepsilon=1\times10^{-8}$ in Subsection \ref{sec4.2}.
		Left: $L_2$ errors $\| u_{\mathrm{NN}}^c - u\|_2$ and $\| u_{\mathrm{NN}} - u\|_2$. Right: $L_{\infty}$ errors $\| u_{\mathrm{NN}}^c - u\|_{\infty}$ and $\| u_{\mathrm{NN}} - u\|_{\infty}$. The confidence band represents the range between the 30-th and 70-th percentiles across 50 repeated runs.}
	\label{fig: case2_result_ablation}
\end{figure*}

We consider the linear boundary value problem (\ref{equ: case2_original}) presented in Section \ref{sec3.1.2}. The shape parameters of our MAE-TransNet for the two inner solutions are set to $\gamma^i_1 = \gamma^i_2 = 0.25$, and the numbers of hidden-layer neurons are set to $M^i_1 = M^i_2 = 10$. A classic  finite difference method (FDM) with a sampling interval length of $5\times10^{-5}$ is used to generate the high-quality reference solution $u_{ref}$. 
The left column of Figure \ref{fig: case2_result} presents the comparisons of the reference solution $u_{ref}$ and corresponding MAE-TransNet solution $u_{\mathrm{NN}}^c$ with different values of $\varepsilon$, while the right column shows corresponding pointwise absolute errors.
Our MAE-TransNet method achieves high accuracy with different values of $\varepsilon$ while using the same hidden layer parameters, highlighting its transferability. 
Figure \ref{fig: case2_result_ablation} compares the convergence of MAE-TransNet and TransNet methods for $\varepsilon=1\times10^{-8}$. 
It is observed that the MAE-TransNet solution $u_{\mathrm{NN}}^{c}$ achieves the best accuracy with just a total of $M^i$ + $M^o = 6$ neurons, while significantly and consistently outperforming the original TransNet method.
Table \ref{tab: case2_table} reports the performance comparisons of the two neural network methods (TransNet and MAE-TransNet) using the same collocation and test points for different values of $\varepsilon$ ($1 \times 10^{-2}$, $5 \times 10^{-3}$, $1 \times 10^{-3}$ and $1 \times 10^{-8}$, respectively). 
It is observed that TransNet again fails to accurately solve the singular perturbation problem with two boundary layers of the same thickness, and as $\varepsilon$ decreases, the error of the MAE-TransNet solution also decreases, demonstrating strong agreement with the predictions of MAE theory.

\begin{table}[!htb]
	\centering
	\caption{Performance comparison of the two neural network methods (TransNet and MAE-TransNet) for the test case 2 in Subsection \ref{sec4.2}.}\vspace{0.2cm}
	\label{tab: case2_table}
	\begin{tabular}{llrrll}
		\toprule
		\multicolumn{1}{l}{Method} & \multicolumn{1}{l}{$\varepsilon$} & 
		\multicolumn{1}{l}{Points} &\multicolumn{1}{l}{Neurons} & \multicolumn{1}{l}{$L_2$ Error} &
		\multicolumn{1}{l}{$L_{\infty}$ Error}\\
		\midrule
		\multirow{4}{*}{TransNet}   & $1 \times 10^{-2}$ & 2,001 & 20 & 1.44e$-$1 & 1.62e$+$0 \\
		& $5 \times 10^{-3}$ & 2,001 & 20 & 5.25e$-$1 & 1.59e$+$0 \\
		& $1 \times 10^{-3}$ & 2,001 & 20 & 5.72e$-$1 & 1.74e$+$0 \\
		& $1 \times 10^{-8}$ & 2,001 & 20 & 3.26e$-$2 & 1.64e$+$0 \\
		\midrule
		\multirow{4}{*}{MAE-TransNet}   & $1 \times 10^{-2}$ & 2,001 & 20 & 1.27e$-$2 & 2.74e$-$2\\
		& $5 \times 10^{-3}$ & 2,001 & 20 & 6.36e$-$3 & 1.40e$-$2\\
		& $1 \times 10^{-3}$ & 2,001 & 20 & 1.19e$-$3 & 3.13e$-$3\\
		& $1 \times 10^{-8}$ & 2,001 & 20 & 1.27e$-$8 & 8.25e$-$8\\
		\bottomrule
	\end{tabular}
\end{table}

\subsection{Test Case 3: 1D Linear Problem With Two Boundary Layers of Different Thicknesses}
\label{sec4.3}
We consider the following linear problem in the domain $\Omega=(0,1)$ \cite{bender2013advanced}: 
\begin{equation}\label{equ: case3_original}
	\left\{
	\begin{aligned}
		&\varepsilon \frac{\mathrm{d}^2 u(x)}{\mathrm{d} x^2} - x^2\frac{\mathrm{d} u(x)}{\mathrm{d} x} - u(x) = 0, \quad x\in(0, 1), \\
		&u(0)=1, \quad u(1)=1.
	\end{aligned}
	\right.
\end{equation}
This problem involves two boundary layers, one is at the point $x=0$ with the thickness $\sqrt{\varepsilon}$ and the other is at $x=1$ with the thickness $\varepsilon$. 

As $\varepsilon \rightarrow 0$, the problem (\ref{equ: case3_original}) reduces to the following outer solution problem:
\begin{equation}\label{equ: case3_outer}
	- x^2\frac{\mathrm{d} u^o(x)}{\mathrm{d} x} - u^o(x) = 0, \quad x\in(0,1),
\end{equation}
which gives the outer solution $u^o(x)=C_0 e^{1/x}$. Moreover, as $x \rightarrow 0^+$, $u^o(x)$ becomes infinite unless $C_0=0$, that is, $u^o(x)=0$.
For the inner solution $u^i_1(x)$ at $x=0$, the scaling transformation $\zeta=\frac{x}{\delta_1(\varepsilon)}:=\frac{x}{\sqrt{\varepsilon}}$ and the matching principle are used to obtain the following boundary value problem: 
\begin{equation}\label{equ: case3_inner1}
	\left\{
	\begin{aligned}
		&\frac{\mathrm{d}^2 \bar{u}^i_1(\zeta)}{\mathrm{d} \zeta^2} - \bar{u}^i_1(\zeta) = 0, \quad \zeta\in(0,\frac{1}{\sqrt{\varepsilon}}), \\ 
		&\bar{u}^i_1(0)=1, \quad \bar{u}^i_1(\frac{1}{\sqrt{\varepsilon}}) = u^o(0).
	\end{aligned}
	\right.
\end{equation}
Similarly, for the inner solution $u^i_2(x)$ at $x=1$, $\zeta=\frac{1-x}{\delta_2(\varepsilon)}:=\frac{1-x}{\varepsilon}$, $\varepsilon \rightarrow 0$ and the matching principle are used to obtain the following problem:
\begin{equation}\label{equ: case3_inner2}
	\left\{
	\begin{aligned}
		&\frac{\mathrm{d}^2 \bar{u}^i_2(\zeta)}{\mathrm{d} \zeta^2} + \frac{\mathrm{d} \bar{u}^i_2(\zeta)}{\mathrm{d} \zeta} = 0, \quad \zeta \in (0, \frac{1}{\varepsilon}),\\ 
		&\bar{u}^i_2(0)=1, \quad \bar{u}^i_2(\frac{1}{\varepsilon}) = u^o(1).
	\end{aligned}
	\right.
\end{equation}
Then, the composite solution of problem (\ref{equ: case3_original}) is obtained from (\ref{equ: MAE_composite_plus}) with $K=2$. 

\begin{figure*}[!htb]
	\centering
	\subfigure{
		\begin{minipage}[t]{.9\textwidth}
			\centering
			\includegraphics[width=\linewidth]{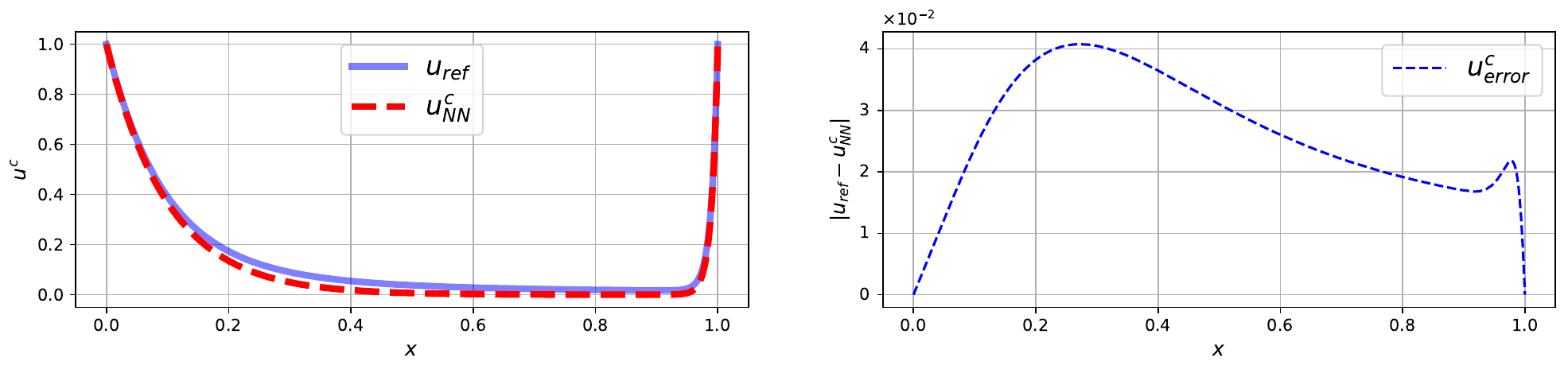}
		\end{minipage}
	} \\
	\vspace{-0.4cm}
	\subfigure{
		\begin{minipage}[t]{.9\textwidth}
			\centering
			\includegraphics[width=\linewidth]{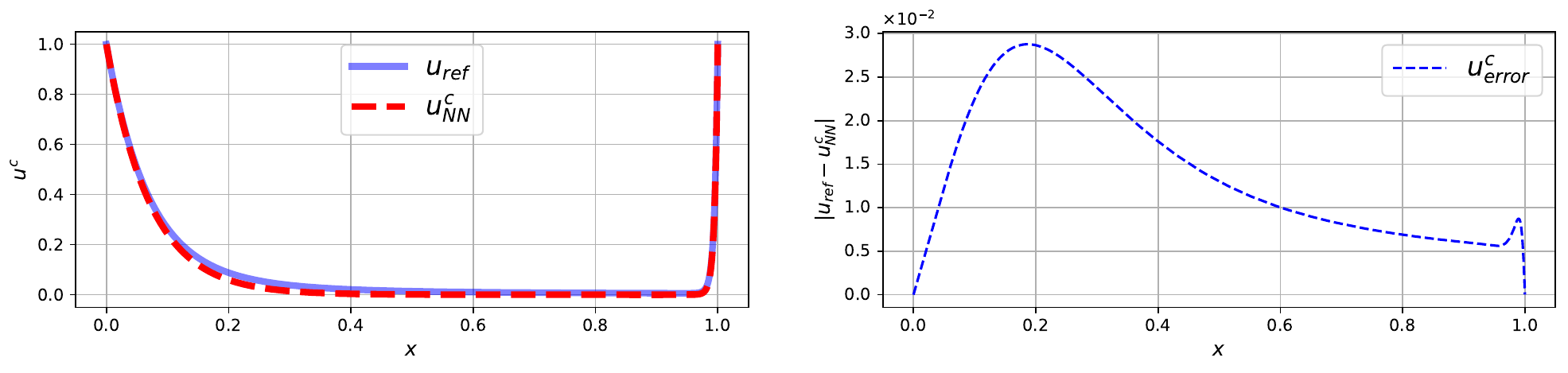}
		\end{minipage}
	} \\
	\vspace{-0.4cm}
	\subfigure{
		\begin{minipage}[t]{.9\textwidth}
			\centering
			\includegraphics[width=\linewidth]{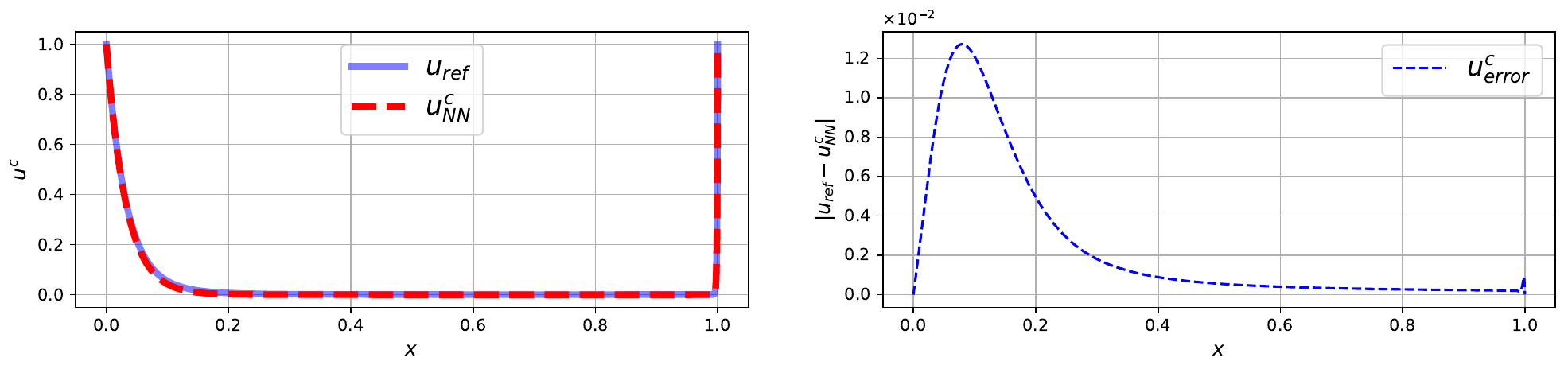}
		\end{minipage}
	} \\
	\caption{Comparison between the reference solution $u_{ref}$ and corresponding MAE-TransNet solution $u_{{\mathrm{NN}}}^c$ for the test case 3 with $\varepsilon=1\times10^{-2}$, $\varepsilon=5\times10^{-3}$ and $\varepsilon=1\times10^{-3}$ (from top to bottom) in Subsection \ref{sec4.3}.}
	\label{fig: case3_result}
\end{figure*}

\begin{figure*}[!htb]
	\centering
	\includegraphics[width=.9\textwidth]{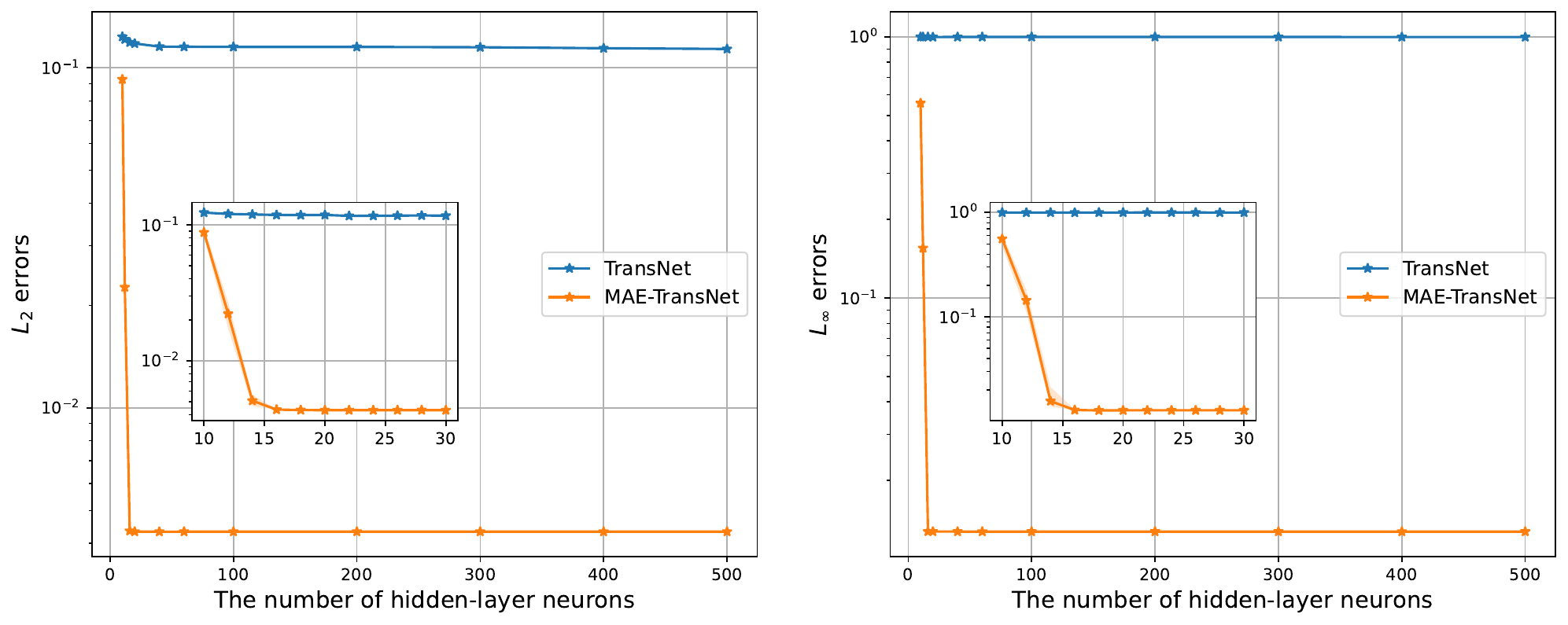}
		\vspace{-0.2cm}
	\caption{Comparison of the performance between MAE-TransNet solution $u_{\mathrm{NN}}^c$ and TransNet solution $u_{\mathrm{NN}}$ for the test case 3 with $\varepsilon=1\times10^{-3}$ in Subsection \ref{sec4.3}.
		Left: $L_2$ errors $\| u_{\mathrm{NN}}^c - u\|_2$ and $\| u_{\mathrm{NN}} - u\|_2$. Right: $L_{\infty}$ errors $\| u_{\mathrm{NN}}^c - u\|_{\infty}$ and $\| u_{\mathrm{NN}} - u\|_{\infty}$. The confidence band represents the range between the 30-th and 70-th percentiles across 50 repeated runs.}
	\label{fig: case3_result_ablation}
\end{figure*}

The shape parameters for the two inner solutions are set to $\gamma^i_1 = \gamma^i_2 = 0.25$, and the numbers of hidden-layer neurons are set to $M^i_1 = M^i_2 = 10$. A FDM with a sampling interval length of $5\times10^{-5}$ is again implemented to provide the reference solution $u_{ref}$. 
The left column of Figure \ref{fig: case3_result} presents the comparisons of the reference solution $u_{ref}$ and corresponding MAE-TransNet solution $u_{\mathrm{NN}}^c$ with different values of $\varepsilon$, while the right column shows corresponding pointwise absolute errors.
MAE-TransNet effectively captures boundary layers of varying thicknesses for different values of $\varepsilon$, while maintaining the same hidden layer parameters, demonstrating the transferability of MAE-TransNet.
Figure \ref{fig: case3_result_ablation} compares the convergence of MAE-TransNet and TransNet methods for $\varepsilon=1\times10^{-3}$. 
Table \ref{tab: case3_table} reports the performance comparisons of the two neural network methods (TransNet and MAE-TransNet) using the same collocation and test points for different values of $\varepsilon$ ($1 \times 10^{-2}$, $5 \times 10^{-3}$ and $1 \times 10^{-3}$ respectively). 

\begin{table}[!htb]
	\centering
	\caption{Performance comparison of the two neural network methods (TransNet and MAE-TransNet) for the test case 3 in Subsection \ref{sec4.3}.}\vspace{0.2cm}
	\label{tab: case3_table}
	\begin{tabular}{llrrll}
		\toprule
		\multicolumn{1}{l}{Method} & \multicolumn{1}{l}{$\varepsilon$} & 
		\multicolumn{1}{l}{Points} &\multicolumn{1}{l}{Neurons} & \multicolumn{1}{l}{$L_2$ Error} &
		\multicolumn{1}{l}{$L_{\infty}$ Error}\\
		\midrule
		\multirow{3}{*}{TransNet}   & $1 \times 10^{-2}$ & 2,001 & 20 & 2.42e$-$1 & 9.98e$-$1 \\
		& $5 \times 10^{-3}$ & 2,001 & 20 & 1.74e$-$1 & 9.68e$-$1 \\
		& $1 \times 10^{-3}$ & 2,001 & 20 & 1.24e$-$1 & 9.37e$-$1 \\
		\midrule
		\multirow{3}{*}{MAE-TransNet}   & $1 \times 10^{-2}$ & 2,001 & 20 & 2.88e$-$2 & 4.15e$-$2\\
		& $5 \times 10^{-3}$ & 2,001 & 20 & 1.62e$-$2 & 2.87e$-$2\\
		& $1 \times 10^{-3}$ & 2,001 & 20 & 4.32e$-$3 & 1.27e$-$2\\
		\bottomrule
	\end{tabular}
\end{table}

\subsection{Test Case 4: 1D Nonlinear Problem With a Single Boundary Layer}
\label{sec4.4}
We consider the following nonlinear advection-diffusion-reaction problem in the domain $\Omega=(0,1)$ \cite{arzani2023theory, bender2013advanced}: 
\begin{equation}\label{equ: case4_original}
	\left\{
	\begin{aligned}
		&\varepsilon \frac{\mathrm{d}^2 u(x)}{\mathrm{d} x^2} + 2\frac{\mathrm{d} u(x)}{\mathrm{d} x} + e^{u(x)}=0, \quad x\in(0, 1), \\
		&u(0)=0, \quad u(1)=0.
	\end{aligned}
	\right.
\end{equation}
It is well known that one boundary layer occurs at $x=0$ with the thickness $\varepsilon$. 

As $\varepsilon \rightarrow 0$, the problem (\ref{equ: case4_original}) reduces to the following outer solution problem:
\begin{equation}\label{equ: case4_outer}
	\left\{
	\begin{aligned}
		&2\frac{\mathrm{d} u^o(x)}{\mathrm{d} x} + e^{u^o(x)} = 0, \quad x\in(0, 1), \\ 
		&u^o(1)=0.
	\end{aligned}
	\right.
\end{equation}
For the inner solution $u^i(x)$ at $x=0$, the scaling transformation $\zeta = \frac{x}{\delta(\varepsilon)}:=\frac{x}{\varepsilon}$ and the matching principle are used to obtain the following boundary value problem: 
\begin{equation}\label{equ: case4_inner}
	\left\{
	\begin{aligned}
		&\frac{\mathrm{d}^2 \bar{u}^i(\zeta)}{\mathrm{d} \zeta^2} + 2\frac{\mathrm{d} \bar{u}^i(\zeta)}{\mathrm{d} \zeta} = 0, \quad \zeta\in(0, \frac{1}{\varepsilon}), \\ 
		&\bar{u}^i(0)=0, \quad \bar{u}^i(\frac{1}{\varepsilon}) = u^o(0).
	\end{aligned}
	\right.
\end{equation}
Finally, the composite solution of problem (\ref{equ: case4_original}) is determined through (\ref{equ: MAE_composite}). 

\begin{figure*}[!htb]
	\centering
	\subfigure{
		\begin{minipage}[t]{.9\textwidth}
			\centering
			\includegraphics[width=\linewidth]{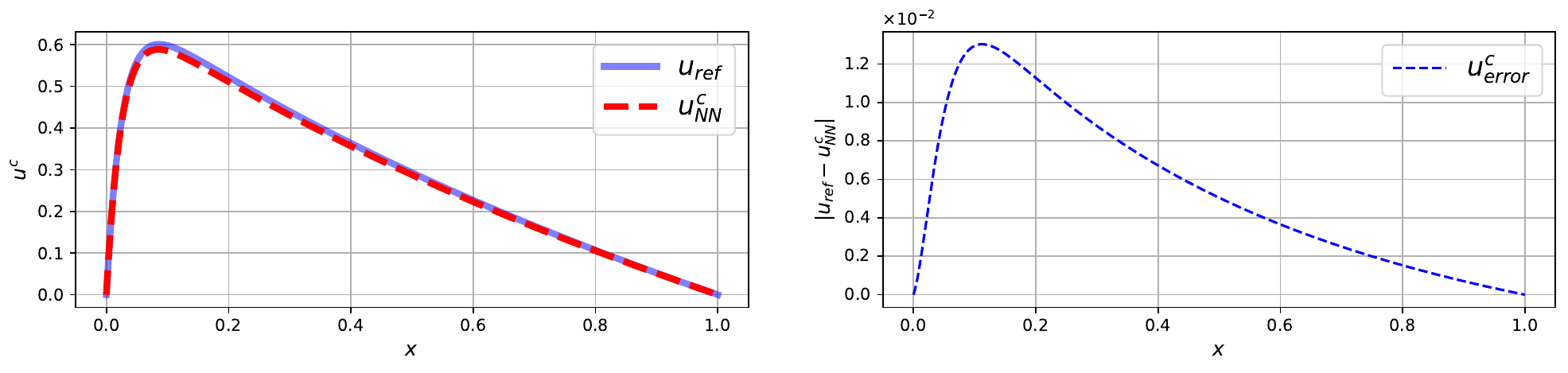}
		\end{minipage}
	} \\
	\vspace{-0.4cm}
	\subfigure{
		\begin{minipage}[t]{.9\textwidth}
			\centering
			\includegraphics[width=\linewidth]{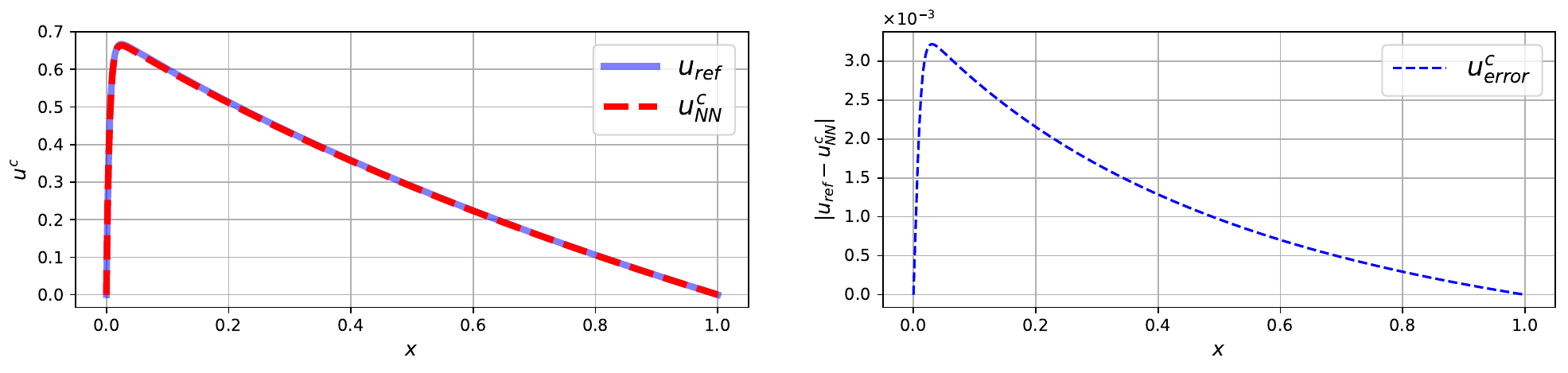}
		\end{minipage}
	} \\
	\vspace{-0.4cm}
	\subfigure{
		\begin{minipage}[t]{.9\textwidth}
			\centering
			\includegraphics[width=\linewidth]{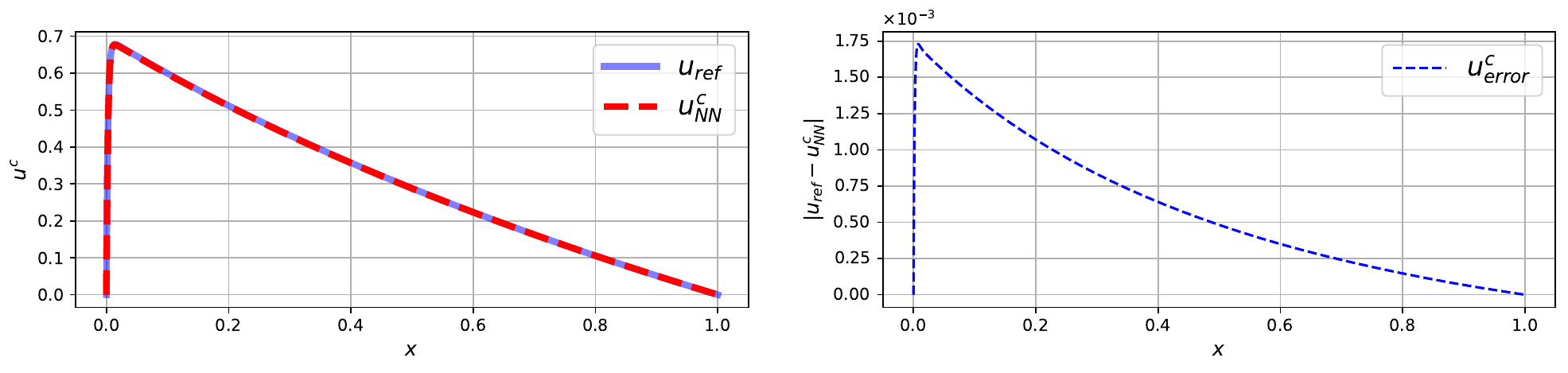}
		\end{minipage}
	} \\
	\caption{Comparison between the reference solution $u_{ref}$ and corresponding MAE-TransNet solution $u_{{\mathrm{NN}}}^c$ for the test case 4 with $\varepsilon=5\times10^{-2}$, $\varepsilon=1\times10^{-2}$ and $\varepsilon=5\times10^{-3}$ (from top to bottom) in Subsection \ref{sec4.4}.}
	\label{fig: case4_result}
\end{figure*}

\begin{figure*}[!htb]
	\centering
	\includegraphics[width=.9\textwidth]{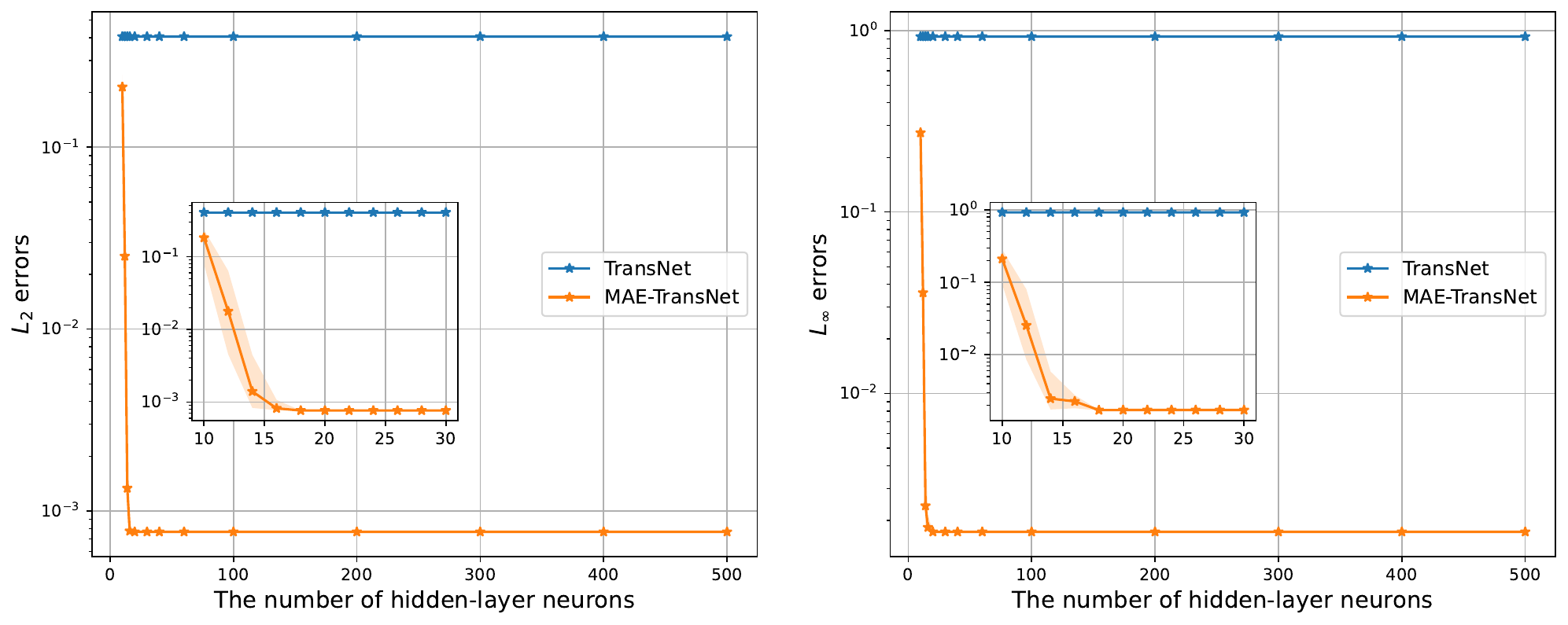}
		\vspace{-0.2cm}
	\caption{Comparison of the performance between MAE-TransNet solution $u_{\mathrm{NN}}^c$ and TransNet solution $u_{\mathrm{NN}}$ for the test case 4 with $\varepsilon=5\times10^{-3}$ in Subsection \ref{sec4.4}.
		Left: $L_2$ errors $\| u_{\mathrm{NN}}^c - u\|_2$ and $\| u_{\mathrm{NN}} - u\|_2$. Right: $L_{\infty}$ errors $\| u_{\mathrm{NN}}^c - u\|_{\infty}$ and $\| u_{\mathrm{NN}} - u\|_{\infty}$. The confidence band represents the range between the 30-th and 70-th percentiles across 50 repeated runs.}
	\label{fig: case4_result_ablation}
\end{figure*}

The shape parameters for the outer and inner solutions are set to $\gamma^o = \gamma^i = 1$, and the numbers of hidden-layer neurons are set to $M^o = M^i = 10$. The reference solution $u_{ref}$ is obtained from 20 Picard iterations, in each iteration the FDM with a sampling interval length of $5\times10^{-4}$ is used for solution of the resulting linear system.
The left column of Figure \ref{fig: case4_result} presents the comparisons of the reference solution $u_{ref}$ and corresponding MAE-TransNet solution $u_{{\mathrm{NN}}}^c$ with different values of $\varepsilon$, while the right column shows corresponding pointwise absolute errors.
For the nonlinear singular perturbation problem with different values of $\varepsilon$, MAE-TransNet can still achieve high accuracy while using the same hidden layer parameters. 
Figure \ref{fig: case4_result_ablation} compares the convergence of MAE-TransNet and TransNet methods for $\varepsilon=5\times10^{-3}$. 
Table \ref{tab: case4_table} reports the performance comparisons of the two neural network methods (TransNet and MAE-TransNet) using the same collocation and test points for different values of $\varepsilon$ ($5 \times 10^{-2}$, $1 \times 10^{-2}$ and $5 \times 10^{-3}$ respectively). 
We observed  that TransNet fails to accurately solve  the nonlinear singular perturbation problem, and as $\varepsilon$ decreases, the error of MAE-TransNet solution decreases as well, remaining consistent with the MAEs theory. 

\begin{table}[!h]
	\centering
	\caption{Performance comparison of the two neural network methods (TransNet and MAE-TransNet) for the test case 4 in Subsection \ref{sec4.4}.}\vspace{0.2cm}
	\label{tab: case4_table}
	\begin{tabular}{llrrll}
		\toprule
		\multicolumn{1}{l}{Method} & \multicolumn{1}{l}{$\varepsilon$} & 
		\multicolumn{1}{l}{Points} &\multicolumn{1}{l}{Neurons} & \multicolumn{1}{l}{$L_2$ Error} &
		\multicolumn{1}{l}{$L_{\infty}$ Error}\\
		\midrule
		\multirow{3}{*}{TransNet}   & $5 \times 10^{-2}$ & 1,001 & 20 & 2.59e$-$1 & 6.00e$-$1 \\
		& $1 \times 10^{-2}$ & 1,001 & 20 & 2.75e$-$1 & 6.66e$-$1 \\
		& $5 \times 10^{-3}$ & 1,001 & 20 & 2.77e$-$1 & 6.78e$-$1 \\
		\midrule
		\multirow{3}{*}{MAE-TransNet}   & $5 \times 10^{-2}$ & 1,001 & 20 & 6.83e$-$2 & 1.30e$-$2\\
		& $1 \times 10^{-2}$ & 1,001 & 20 & 1.51e$-$3 & 3.29e$-$3\\
		& $5 \times 10^{-3}$ & 1,001 & 20 & 7.67e$-$4 & 1.73e$-$3\\
		\bottomrule
	\end{tabular}
\end{table}

\subsection{Test Case 5: 2D  Couette Flow Problem}
\label{sec4.5}
We consider the 2D advection-diffusion transport in the Couette flow problem (\ref{equ: case5_original}), as detailed in Section \ref{sec3.2.1} and compare the performance of our MAE-TransNet with BL-PINN.  Note  that the outer solution $u^o(x,y)=0$ is known for this example, so we don't need to solve it.
For BL-PINN, all parameters are adopted from the code in \cite{arzani2023theory}. 
For MAE-TransNet, the  shape parameter is set to $\gamma^i = 0.2$ and the number of hidden-layer neurons is set to $M^i=128$ to solve for the inner solution. 
The number of collocation points in $\Omega$ and $\Omega^\zeta$ is uniformly set to $N_{\Omega} = N_{\Omega^\zeta} = 199\times199$ and that on $\partial\Omega$ and $\partial\Omega^\zeta$ is uniformly set to $N_{\partial\Omega} = N_{\partial\Omega^\zeta} = 4000$, then the total number of collocation points is set to $N = N_\Omega + N_{\partial\Omega} = N_{\Omega^\zeta} + N_{\partial\Omega^\zeta} = 43601$.
The reference solution $u_{ref}$ is obtained by FDM with a sampling interval length of $1.25 \times 10^{-3}$. 
Table \ref{tab: case5_table} reports the performance comparisons of the two neural network methods (BL-PINN and MAE-TransNet) using the same collocation and test points for different values of $\varepsilon$ ($1 \times 10^{-2}$, $1 \times 10^{-3}$ and $1 \times 10^{-4}$ respectively).
We find that the solution accuracy of MAE-TransNet still gets improved when $\varepsilon$ decreases, which is consistent with the theory of MAEs.  Compared to BL-PINN, our MAE-TransNet requires much fewer neurons and less running time while achieving even better accuracy.

\begin{figure*}[!htb]
	\centering
	\includegraphics[width=0.95\textwidth]{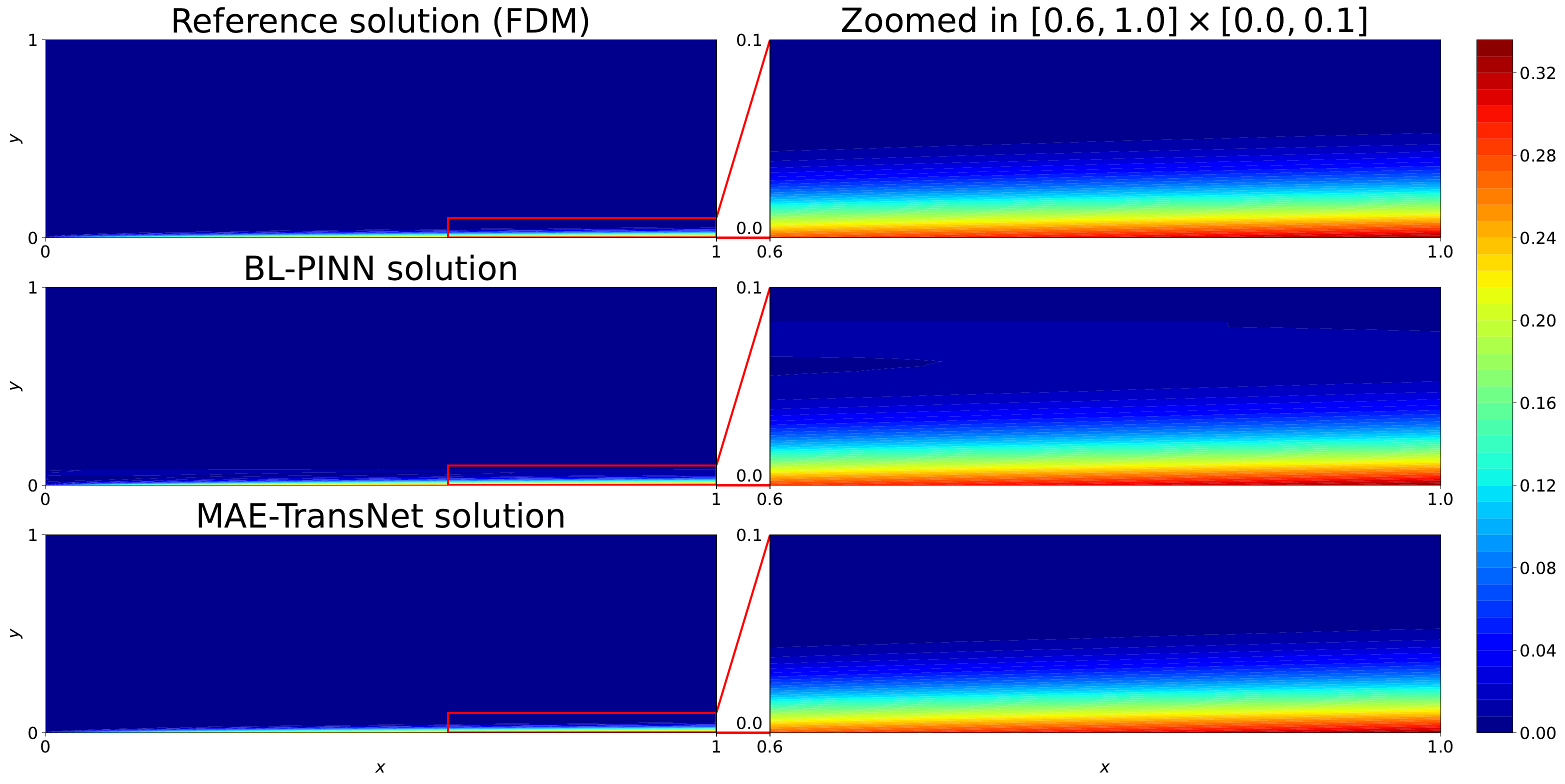}
	\caption{Comparison of the performance between MAE-TransNet and BL-PINN for the test case 5 with $\varepsilon = 1 \times 10^{-4}$ in Subsection \ref{sec4.5}. 
		Left: Over the whole domain $\Omega$. Right: Zoomed in $[0.6, 1] \times [0, 0.1]$.}
	\label{fig: case5_result}
\end{figure*}

\begin{figure*}[!htb]
	\centering
	\includegraphics[width=.85\textwidth]{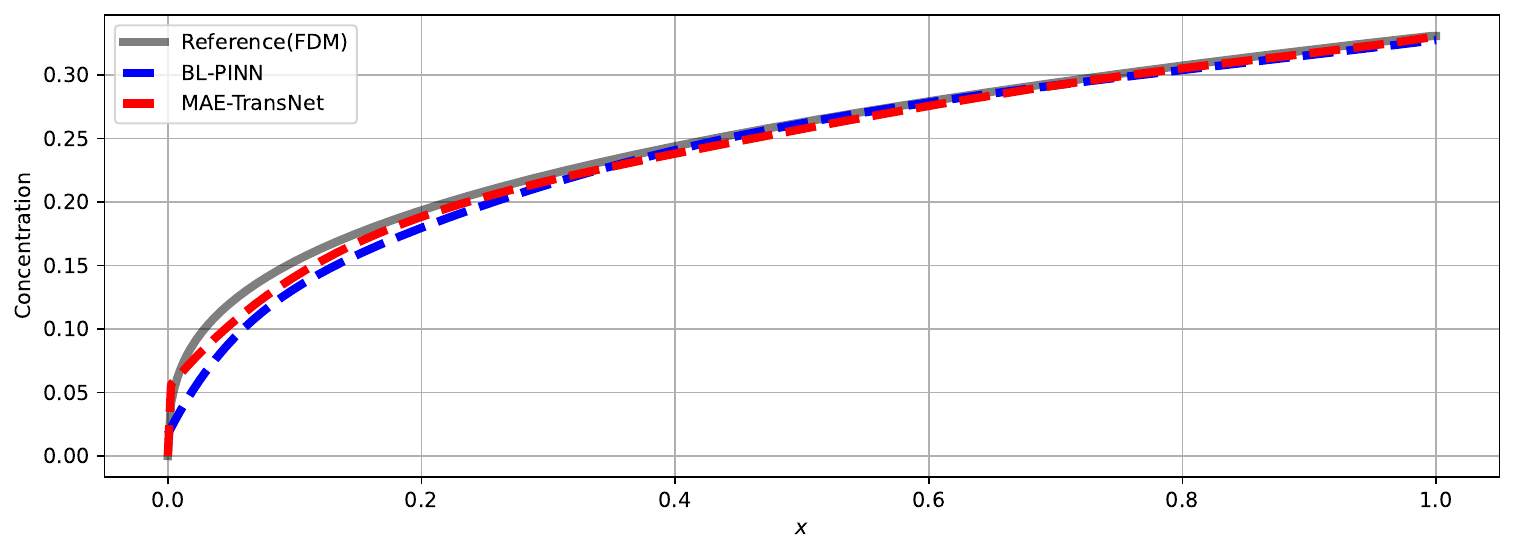}
	\caption{Comparison of the performance at the bottom between MAE-TransNet and BL-PINN for the test case 5 with $\varepsilon = 1 \times 10^{-4}$ in Subsection \ref{sec4.5}.}
	\label{fig: case5_result_bottom}
\end{figure*}

Table \ref{tab: case5_table1} reports the performance of MAE-TransNet with different numbers of hidden-layer neurons in the case of $\varepsilon = 1 \times 10^{-4}$.
It is easy to see  that  MAE-TransNet gets improved accuracy when the number of neurons increases.
The results of BL-PINN with 896 neurons and MAE-TransNet with 128 neurons in the case of  $\varepsilon = 1 \times 10^{-4}$ are presented in Figure \ref{fig: case5_result} and Figure \ref{fig: case5_result_bottom}. 
We observe  that both the MAE-TransNet and BL-PINN solutions closely match the reference solution, 
and the MAE-TransNet solution converges noticeably closer to the reference solution on the bottom wall within the boundary layer (i.e., $y=0$). 

\begin{table}[!h]
	\centering
	\caption{Performance comparison of the two neural network methods (BL-PINN and MAE-TransNet) for the test case 5 in Subsection \ref{sec4.5}.}\vspace{0.2cm}
	\label{tab: case5_table}
	\begin{tabular}{llrrrll}
		\toprule
		\multicolumn{1}{l}{Method} & \multicolumn{1}{l}{$\varepsilon$} & 
		\multicolumn{1}{l}{Points} &\multicolumn{1}{l}{Neurons} & \multicolumn{1}{l}{Time(s)} & \multicolumn{1}{l}{$L_2$ Error} &
		\multicolumn{1}{l}{$L_{\infty}$ Error}\\
		\midrule
		\multirow{3}{*}{BL-PINN}& $1 \times 10^{-2}$ & 43,601 & 896 & 20,228 & 1.42e$-$1 & 6.86e$-$1\\       
		& $1 \times 10^{-3}$ & 43,601 & 896 & 20,236 & 1.10e$-$2 & 9.18e$-$2\\
		& $1 \times 10^{-4}$ & 43,601 & 896 & 20,200 & 1.34e$-$3 & 3.79e$-$2\\
		\midrule
		\multirow{3}{*}{MAE-TransNet}   & $1 \times 10^{-2}$ & 43,601 & 128 & 0.717 & 9.07e$-$3 & 2.54e$-$1\\
		& $1 \times 10^{-3}$ & 43,601 & 128 & 0.721 & 3.23e$-$3 & 8.40e$-$2\\
		& $1 \times 10^{-4}$ & 43,601 & 128 & 0.713 & 1.18e$-$3 & 2.04e$-$2\\
		\bottomrule
	\end{tabular}
\end{table}

\begin{table}[!h]
	\centering
	\caption{Performance of MAE-TransNet with different numbers of hidden-layer neurons for the test case 5 with $\varepsilon = 1 \times 10^{-4}$ in Subsection \ref{sec4.5}.}\vspace{0.2cm}
	\label{tab: case5_table1}
	\begin{tabular}{llrrrll}
		\toprule
		\multicolumn{1}{l}{Method} & \multicolumn{1}{l}{$\varepsilon$} & 
		\multicolumn{1}{l}{Points} & \multicolumn{1}{l}{Neurons} & \multicolumn{1}{l}{Time(s)} & \multicolumn{1}{l}{$L_2$ Error} &
		\multicolumn{1}{l}{$L_{\infty}$ Error}\\
		\midrule
		\multirow{8}{*}{MAE-TransNet}  & \multirow{8}{*}{$1 \times 10^{-4}$} & \multirow{8}{*}{43,601}
		& 16  & 0.101 & 5.54e$-$3 & 6.66e$-$2\\
		& & & 32  & 0.111 & 3.89e$-$3 & 5.31e$-$2\\
		& & & 48  & 0.156 & 2.39e$-$3 & 3.59e$-$2\\
		& & & 64  & 0.318 & 1.78e$-$3 & 3.05e$-$2\\
		& & & 80  & 0.464 & 1.48e$-$3 & 2.64e$-$2\\
		& & & 96  & 0.555 & 1.16e$-$3 & 2.13e$-$2\\
		& & & 112 & 0.655 & 1.13e$-$3 & 2.08e$-$2\\
		& & & 128 & 0.713 & 1.18e$-$3 & 2.04e$-$2\\
		\bottomrule
	\end{tabular}
\end{table}

\subsection{Test Case 6: 2D Coupled Boundary Layers Problem}
\label{sec4.6}
\begin{figure*}[!htb]
	\centering
	\includegraphics[width=.7\textwidth]{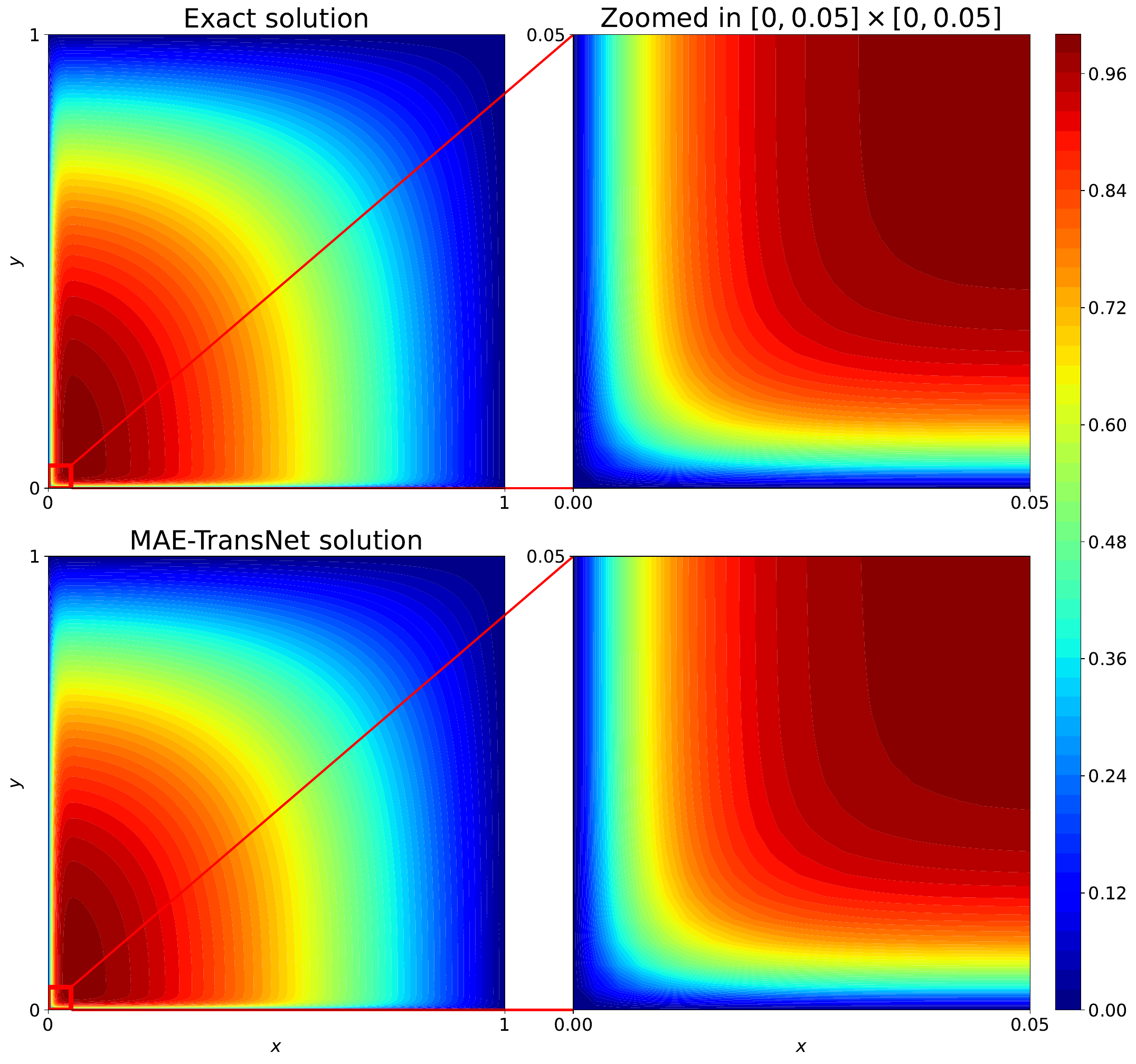}
	\caption{Comparison of the exact solution and the MAE-TransNet solution for the test case 6 with $\varepsilon=2^{-6}$ in Subsection \ref{sec4.6}. 
		Left: over the whole domain $\Omega$. Right: zoomed in the coupled region $\Omega^{ii}=[0,0.05] \times [0,0.05]$.}
	\label{fig: case6_result}
\end{figure*}

\begin{figure*}[!htb]
	\centering
	\includegraphics[width=0.95\textwidth]{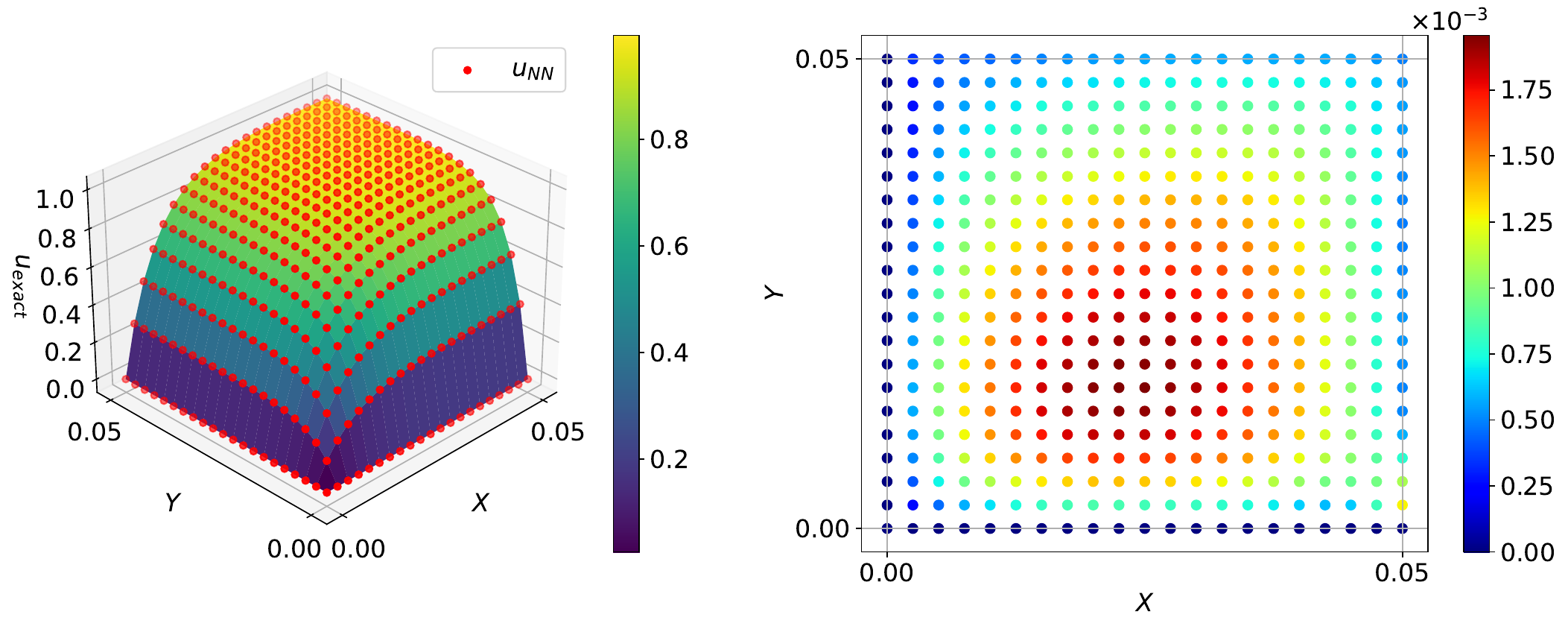}
	\caption{Comparison of the exact solution and the MAE-TransNet solution in the coupled region $\Omega^{ii}=[0,0.05] \times [0,0.05]$ for the test case 6 with $\varepsilon=2^{-6}$ in Subsection \ref{sec4.6}.
		Left: the surface represents the exact solution, while the red points denote the results from MAE-TransNet. Right: pointwise absolute error.}
	\label{fig: case6_result_inner}
\end{figure*}

We consider the 2D singular perturbation convection-diffusion problem (\ref{equ: case6_original}), as detailed in Section \ref{sec3.2.2}. 
Taking $\varepsilon=2^{-6}$ as an example, the parameter associated with the coupled region is set to $A=0.05$ (i.e., the coupled region $\Omega^{ii}=[0,0.05] \times [0,0.05]$).
For our MAE-TransNet method, the shape parameters for the outer, the two inner, and the coupled-region solutions are set to $\gamma^o=1$, $\gamma^i_1=\gamma^i_2=0.2$, and $\gamma^{ii}=1$.
The numbers of hidden-layer neurons for the outer, the two inner, and the coupled-region solutions are set to $M^o=M^i_1=M^i_2=M^{ii}=200$. 
The total number of collocation points in $\Omega$, $\Omega_1^\zeta$, $\Omega_2^\eta$ and $\Omega^{\zeta,\eta}$ is uniformly set to $N = 399 \times399 + 16000 = 175201$.
Figure \ref{fig: case6_result} compares the MAE-TransNet solution and the exact solution with $\varepsilon=2^{-6}$.
It is observed that the two solutions exhibit almost no differences either across the entire domain $\Omega$ or in the coupled region $\Omega^{ii}$.
The left plot of Figure \ref{fig: case6_result_inner} presents the comparison of the exact solution and the corresponding MAE-TransNet solution in the coupled region $\Omega^{ii}$, while the right plot shows the corresponding pointwise absolute errors.
The produced numerical results exhibit errors with magnitudes $O(10^{-3})$ for the MAE-TransNet solution, demonstrating its capability to effectively resolve mutual interactions between boundary layers.
Table \ref{tab: case6_table1} reports the performance of MAE-TransNet with different numbers of hidden-layer neurons in the case of $\varepsilon = 2^{-6}$.

\begin{table}[!h]
	\centering
	\caption{Performance of MAE-TransNet with different numbers of hidden-layer neurons for the test case 6 with $\varepsilon = 2^{-6}$ in Subsection \ref{sec4.6}.}\vspace{0.2cm}
	\label{tab: case6_table1}
	\begin{tabular}{llrrrll}
		\toprule
		\multicolumn{1}{l}{Method} & \multicolumn{1}{l}{$\varepsilon$} & 
		\multicolumn{1}{l}{Points} & \multicolumn{1}{l}{Neurons} & \multicolumn{1}{l}{Time(s)} & \multicolumn{1}{l}{$L_2$ Error} &
		\multicolumn{1}{l}{$L_{\infty}$ Error}\\
		\midrule
		\multirow{5}{*}{MAE-TransNet}  & \multirow{5}{*}{$2^{-6}$}  & \multirow{5}{*}{175,201}
		& 200 & 3.29 & 3.27e$-$1 & 5.69e$-$1\\
		& & & 400  & 6.27 & 8.00e$-$3 & 2.42e$-$2\\
		& & & 800  & 13.32 & 3.40e$-$4 & 1.95e$-$3\\
		& & & 1600 & 26.60 & 1.00e$-$4 & 1.45e$-$3\\
		& & & 3200 & 52.40 & 7.81e$-$5 & 1.41e$-$3\\
		\bottomrule
	\end{tabular}
\end{table}

\subsection{Test Case 7: 3D Burgers Vortex Problem}
\label{sec4.7}
The  Burgers vortex describes a stationary, self-similar flow based on the balance among radial flow, axial stretching and viscous diffusion. We consider axisymmetric transport in the 3D Burgers vortex problem as discussed in \cite{arzani2023theory, panton2024incompressible}. In cylindrical coordinates $(r,\theta, x)$, the velocity field can be written as:
\begin{equation}\label{equ: case7_velocity_field}
	\begin{aligned}
		v_r=-\frac{a}{2}r, \quad
		v_\theta = \frac{\Gamma_0}{2\pi r}\left(1-e^{-br^2}\right), \quad
		v_x = ax, 
	\end{aligned}
	\nonumber
\end{equation}
where $a=0.2$, $\Gamma_0=2\pi$ and $b=1$. The diffusion coefficient is set to $\varepsilon=1\times10^{-4}$, and the cylindrical domain has a radius of $r=0.5$ and a height of $x=0.3$. Due to its axisymmetric nature, the original 3D problem can be simplified into a 2D problem in the domain $\Omega=(0.002, 0.5)\times(0, 0.3)$ via cylindrical coordinate parametrization: 
\begin{equation}\label{equ: case7_original}
	\left\{
	\begin{aligned}
		&\varepsilon\frac{\partial^2 c(r, x)}{\partial r^2} + \frac{\varepsilon}{r}\frac{\partial c(r, x)}{\partial r}+ \varepsilon\frac{\partial^2 c(r, x)}{\partial x^2}\\
		&\hspace{2cm}=v_r\frac{\partial c(r, x)}{\partial r} + v_x\frac{\partial c(r, x)}{\partial x}, \quad (r, x) \in (0.002, 0.5)\times(0, 0.3),\\
		&\frac{\partial c}{\partial r}(r=0.002,x)=0, \quad c(r=0.5, x)=0, \\
		&\frac{\partial c}{\partial x}(r,x=0)=-5, \quad \frac{\partial c}{\partial x}(r, x=0.3)=0.
	\end{aligned}
	\right.
\end{equation}

\begin{figure*}[!htb]
	\centering
	\includegraphics[width=.70\textwidth]{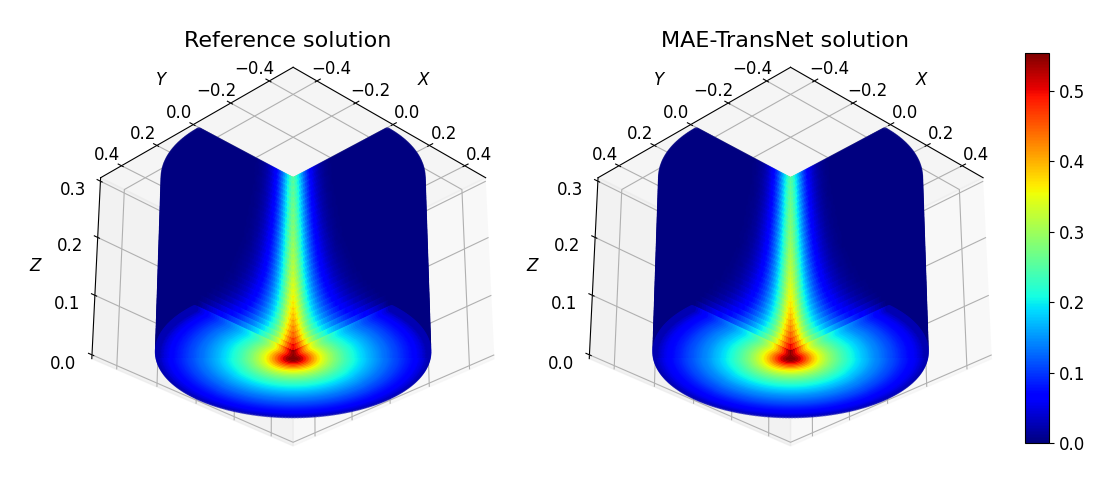}
	\includegraphics[width=.65\textwidth]{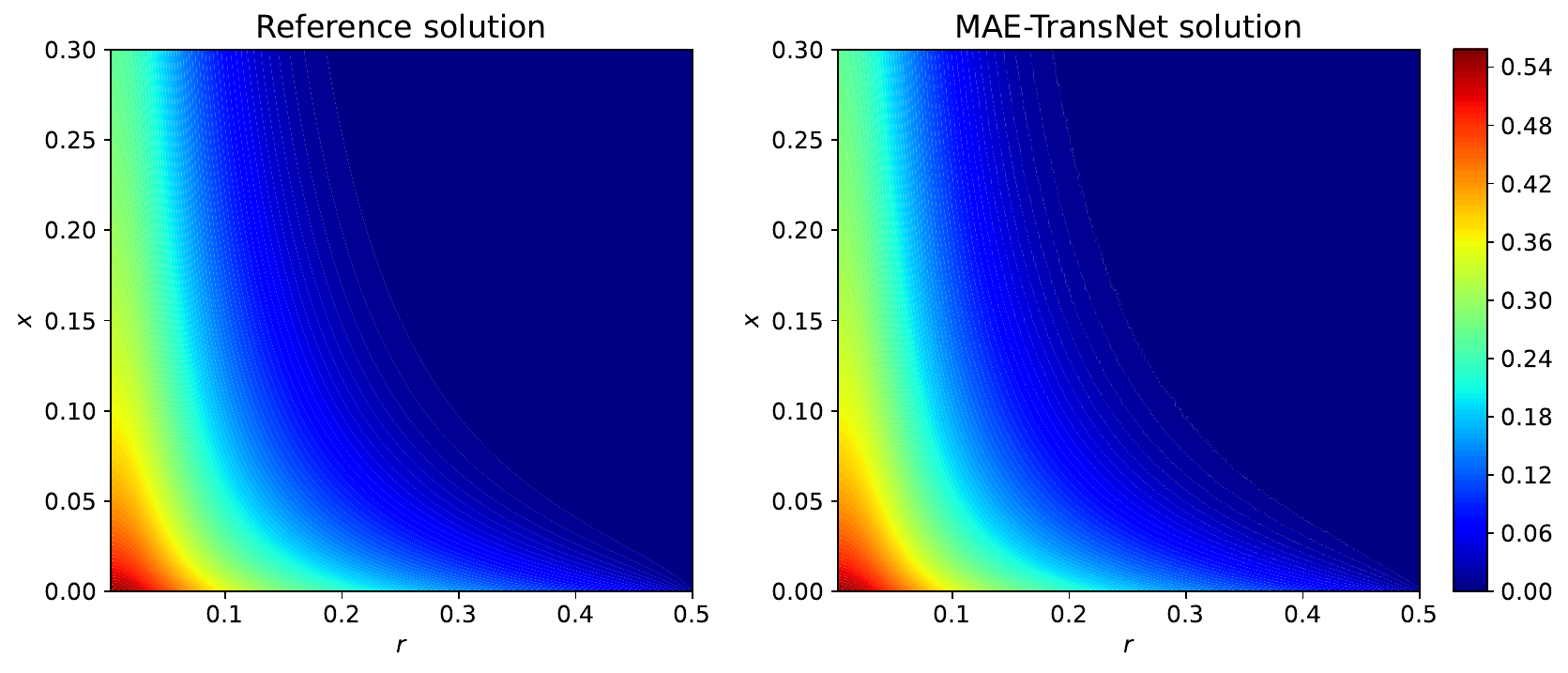}
	\caption{Comparison of the reference solution and the MAE-TransNet solution for the test case 7 with $\varepsilon=1 \times 10^{-4}$ in Subsection \ref{sec4.7}.
		Top: the half section views. Bottom: the cross-section in cylindrical coordinates.}
	\label{fig: case7_result}
\end{figure*}

\begin{figure*}[!htb]
	\centering
	\includegraphics[width=.70\textwidth]{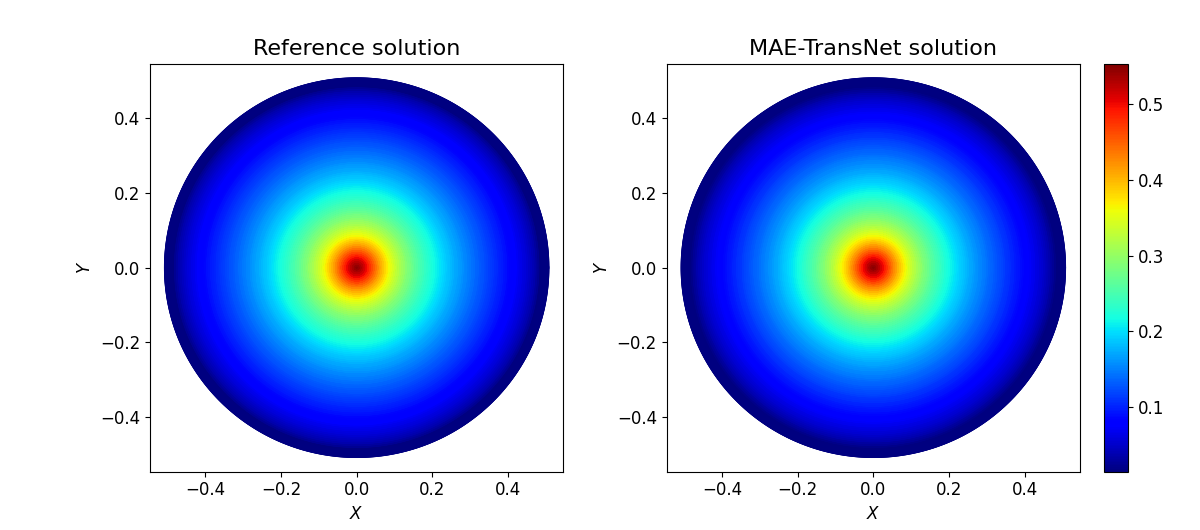}
	\includegraphics[width=.60\textwidth]{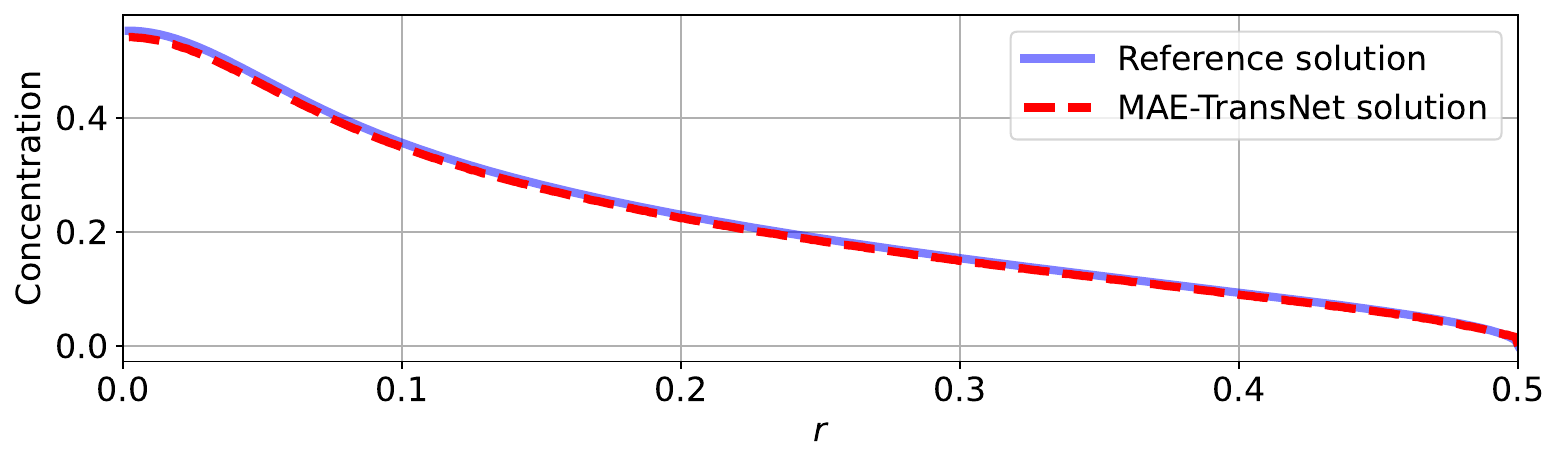}
	\caption{Comparison of the reference solution and the MAE-TransNet solution at the bottom (i.e., $z=0$)  for the test case 7 with $\varepsilon=1 \times 10^{-4}$ in Subsection \ref{sec4.7}.
		Top: the section views. Bottom: in cylindrical coordinates.}
	\label{fig: case7_result_bottom_circle}
\end{figure*}

Notably, when the boundary layer is sufficiently thick such that $\Omega^{ii}=\Omega$, the solution can be exclusively computed within the coupled region.
The scaling transformations $\zeta =\frac{r}{\delta_1(\varepsilon)}:= \frac{r}{\sqrt{\varepsilon}}$, $\eta =\frac{x}{\delta_2(\varepsilon)}:= \frac{x}{\sqrt{\varepsilon}}$ are used to obtain the following boundary value problem:
\begin{equation}\label{equ: case7_inner}
	\left\{
	\begin{aligned}
		&\frac{\partial^2 \tilde{c}^{ii}(\zeta, \eta)}{\partial \zeta^2} + \frac{1}{\zeta}\frac{\partial \tilde{c}^{ii}(\zeta, \eta)}{\partial \zeta} + \frac{\partial^2 \tilde{c}^{ii}(\zeta, \eta)}{\partial \eta^2} \\
		&\hspace{1.3cm} = -0.1\zeta\frac{\partial \tilde{c}^{ii}(\zeta, \eta)}{\partial \zeta} + 0.2\eta\frac{\partial \tilde{c}^{ii}(\zeta, \eta)}{\partial \eta} , \quad (\zeta, \eta)\in(\frac{0.002}{\sqrt{\varepsilon}}, \frac{0.5}{\sqrt{\varepsilon}}) \times (0, \frac{0.3}{\sqrt{\varepsilon}}),\\ 
		&\frac{\partial \tilde{c}^{ii}}{\partial \zeta}(\zeta=\frac{0.002}{\sqrt{\varepsilon}},\eta)=0, \quad \tilde{c}^{ii}(\zeta=\frac{0.5}{\sqrt{\varepsilon}}, \eta)=0, \\
		&\frac{\partial \tilde{c}^{ii}}{\partial \eta}(\zeta,\eta=0)=-5\sqrt{\varepsilon}, \quad \frac{\partial \tilde{c}^{ii}}{\partial \eta}(\zeta, \eta=\frac{0.3}{\sqrt{\varepsilon}})=0,
	\end{aligned}
	\right.
\end{equation}
where the boundary conditions are supplied by the problem (\ref{equ: case7_original}). Subsequently, a TransNet with uniformly distributed hidden-layer neurons is employed to solve the problem (\ref{equ: case7_inner}). The shape parameter is set to $\gamma^{ii}=0.1$ and the number of hidden-layer neurons is set to $M^{ii}=500$. 
The total number of collocation points in $\Omega^{\zeta,\eta}$ is set to $N = 497 \times 299 + 1600 = 150203$.
The reference solution $u_{ref}$ is obtained from the FDM with a sampling interval length of $1 \times 10^{-3}$.
Figure \ref{fig: case7_result} presents the half-section views and the cross-sections in cylindrical coordinates of the MAE-TransNet solution and the reference solution.
It is observed that the two solutions exhibit visually no differences, demonstrating that MAE-TransNet effectively captures the global features of the Burgers vortex. 
Figure \ref{fig: case7_result_bottom_circle} compares the MAE-TransNet solution and the reference solution at the bottom (i.e., $z=0$), and
it shows that MAE-TransNet also effectively captures the intricate details in the boundary layers. 
Table \ref{tab: case7_table1} reports the performance of MAE-TransNet with different numbers of hidden-layer neurons in the case of $\varepsilon=1\times10^{-4}$.

\begin{table}[!h]
	\centering
	\caption{Performance of MAE-TransNet with different numbers of hidden-layer neurons for the test case 7 with $\varepsilon=1 \times 10^{-4}$ in Subsection \ref{sec4.7}.}\vspace{0.2cm}
	\label{tab: case7_table1}
	\begin{tabular}{llrrrll}
		\toprule
		\multicolumn{1}{l}{Method} & \multicolumn{1}{l}{$\varepsilon$} & 
		\multicolumn{1}{l}{Points} & \multicolumn{1}{l}{Neurons} & \multicolumn{1}{l}{Time(s)} & \multicolumn{1}{l}{$L_2$ Error} &
		\multicolumn{1}{l}{$L_{\infty}$ Error}\\
		\midrule
		\multirow{6}{*}{MAE-TransNet} & \multirow{6}{*}{$1 \times 10^{-4}$} & \multirow{6}{*}{150,203}
		& 20 & 0.45 & 5.66e$-$2 & 1.74e$-$1\\
		& & & 50  & 0.92 & 6.61e$-$3 & 5.37e$-$2\\
		& & & 100 & 1.63 & 5.37e$-$3 & 2.97e$-$2\\
		& & & 200 & 3.00 & 3.55e$-$3 & 1.35e$-$2\\
		& & & 300 & 4.51 & 3.36e$-$3 & 1.19e$-$2\\
		& & & 500 & 7.36 & 3.13e$-$3 & 1.20e$-$2\\
		\bottomrule
	\end{tabular}
\end{table}

\section{Conclusions}
\label{sec5}
In this work, we present the MAEs-based transferable neural networks
for solving singular perturbation problems characterized by solutions exhibiting sharp gradients within narrow boundary layers. 
Using MAEs, the original problem is decomposed into inner and outer boundary value problems. The outer solution, valid outside the boundary layer, is approximated by the TransNet with uniform hidden-layer neurons, while the inner solution, valid within the boundary layer, is approximated by the improved TransNet with nonuniform 
hidden-layer neurons. A matching term is then applied to combine these solutions into a uniformly valid solution across the computational domain.
To address the challenging multidimensional coupled boundary layer system in the absence of theoretical guidance, MAE-TransNet computes solutions over the entire domain, followed by an auxiliary TransNet that effectively corrects the solution in the coupled region.
Numerical experiments confirm that our MAE-TransNet method inherits both the accuracy of MAEs and the efficiency of TransNets, thereby significantly improving computational precision and efficiency compared to other neural network methods, such as BL-PINN in \cite{arzani2023theory}. 
Additionally, the experiments demonstrate that the same hidden-layer parameters can be employed to boundary layer problems with different values of $\varepsilon$, further highlighting its transferability and substantially reducing the cost of parameter tuning.
The MAE-TransNet method provides a promising tool for solving singular perturbation problems with boundary layers. 
There are several potential related topics that will be studied in future work. Although various numerical results illustrate that the proposed method converges well and has good accuracy, exploring rigorous theoretical convergence analysis and error estimates is still an interesting topic. It is also worthy of further investigation into more challenging singular perturbation problems involving moving thin layers \cite{wang2024general, kim2022fast} or turning points \cite{SHARMA201310575}, as well as multiscale systems discussed in \cite{XU2025114129}.

%

\bmsection*{Acknowledgments}
L. Zhu’s work was partially supported by the National Natural Science Foundation of China under grant
number 11871091, 12571580.

%

\bmsection*{Conflict of interest}

The authors declare no conflicts of interests.

\bmsection*{Data Availability Statement}

The data that support the findings of this study are openly available inMAE-TransNet at \url{https://github.com/ZhequanShen/MAE-TransNet}.

\bibliography{MAE-TransNet}

@book{nayfeh2024perturbation,
  title={Perturbation methods},
  author={Nayfeh, Ali H},
  year={2024},
  publisher={John Wiley \& Sons}
}

@book{bender2013advanced,
  title={Advanced mathematical methods for scientists and engineers I: Asymptotic methods and perturbation theory},
  author={Bender, Carl M and Orszag, Steven A},
  year={2013},
  publisher={Springer Science \& Business Media}
}

@article{raissi2019physics,
  title={Physics-informed neural networks: A deep learning framework for solving forward and inverse problems involving nonlinear partial differential equations},
  author={Raissi, Maziar and Perdikaris, Paris and Karniadakis, George E},
  journal={Journal of Computational Physics},
  volume={378},
  pages={686--707},
  year={2019},
  publisher={Elsevier}
}

@article{arzani2023theory,
  title={Theory-guided physics-informed neural networks for boundary layer problems with singular perturbation},
  author={Arzani, Amirhossein and Cassel, Kevin W and D'Souza, Roshan M},
  journal={Journal of Computational Physics},
  volume={473},
  pages={111768},
  year={2023},
  publisher={Elsevier}
}

@article{faria2017equation,
  title={Equation level matching: An extension of the method of matched asymptotic expansion for problems of wave propagation},
  author={Faria, Luiz M and Rosales, Rodolfo R},
  journal={Studies in Applied Mathematics},
  volume={139},
  number={2},
  pages={265--287},
  year={2017},
  publisher={Wiley Online Library}
}

@article{zhang2024transferable,
  title={Transferable Neural Networks for Partial Differential Equations},
  author={Zhang, Zezhong and Bao, Feng and Ju, Lili and Zhang, Guannan},
  journal={Journal of Scientific Computing},
  volume={99},
  number={1},
  pages={2},
  year={2024},
  publisher={Springer}
}

@article{pieper2024nonuniform,
  title={Nonuniform random feature models using derivative information},
  author={Pieper, Konstantin and Zhang, Zezhong and Zhang, Guannan},
  journal={arXiv preprint arXiv:2410.02132},
  year={2024}
}

@book{holmes2009introduction,
  title={Introduction to the foundations of applied mathematics},
  author={Holmes, Mark H},
  volume={56},
  year={2009},
  publisher={Springer}
}

@article{podila2024wavelet,
  title={Wavelet-based approximation for two-dimensional singularly perturbed elliptic problems},
  author={Podila, Pramod Chakravarthy and Sundrani, Vishwas and Ramos, Higinio and Vigo-Aguiar, Jes{\'u}s},
  journal={Journal of Computational and Applied Mathematics},
  pages={116069},
  year={2024},
  publisher={Elsevier}
}

@book{panton2024incompressible,
  title={Incompressible flow},
  author={Panton, Ronald L},
  year={2024},
  publisher={John Wiley \& Sons}
}

@article{COUPEZ201365,
title = {Solution of high-Reynolds incompressible flow with stabilized finite element and adaptive anisotropic meshing},
journal = {Computer Methods in Applied Mechanics and Engineering},
volume = {267},
pages = {65-85},
year = {2013},
issn = {0045-7825},
author = {T. Coupez and E. Hachem},
}

@article{RIOU2012302,
title = {A new numerical strategy for the resolution of high-Péclet advection–diffusion problems},
journal = {Computer Methods in Applied Mechanics and Engineering},
volume = {241-244},
pages = {302-310},
year = {2012},
issn = {0045-7825},
author = {H. Riou and P. Ladevèze},
}

@article{steinruck2012asymptotic,
  title={Asymptotic methods in fluid mechanics: survey and recent advances},
  author={Steinr{\"u}ck, Herbert},
  year={2012},
  publisher={Springer Science \& Business Media}
}

@article{LI2023116107,
title = {Fast heat transfer simulation for laser powder bed fusion},
journal = {Computer Methods in Applied Mechanics and Engineering},
volume = {412},
pages = {116107},
year = {2023},
issn = {0045-7825},
author = {Xiaohan Li and Nick Polydorides},
}

@article{cao2023physics,
  title={Physics-informed neural networks with parameter asymptotic strategy for learning singularly perturbed convection-dominated problem},
  author={Cao, Fujun and Gao, Fei and Guo, Xiaobin and Yuan, Dongfang},
  journal={Computers \& Mathematics with Applications},
  volume={150},
  pages={229--242},
  year={2023},
  publisher={Elsevier}
}

@book{roos2008robust,
  title={Robust numerical methods for singularly perturbed differential equations},
  author={Roos, Hans-G{\"o}rg},
  year={2008},
  publisher={Springer}
}

@article{clavero2003uniformly,
  title={A uniformly convergent scheme on a nonuniform mesh for convection--diffusion parabolic problems},
  author={Clavero, C and Jorge, JC and Lisbona, F},
  journal={Journal of Computational and Applied Mathematics},
  volume={154},
  number={2},
  pages={415--429},
  year={2003},
  publisher={Elsevier}
}

@article{ge2011multigrid,
  title={Multigrid method based on the transformation-free HOC scheme on nonuniform grids for 2D convection diffusion problems},
  author={Ge, Yongbin and Cao, Fujun},
  journal={Journal of Computational Physics},
  volume={230},
  number={10},
  pages={4051--4070},
  year={2011},
  publisher={Elsevier}
}

@article{kumar2015adaptive,
  title={An adaptive mesh strategy for singularly perturbed convection diffusion problems},
  author={Kumar, Vivek and Srinivasan, Balaji},
  journal={Applied Mathematical Modelling},
  volume={39},
  number={7},
  pages={2081--2091},
  year={2015},
  publisher={Elsevier}
}

@article{du2018adaptive,
  title={An adaptive staggered discontinuous Galerkin method for the steady state convection--diffusion equation},
  author={Du, Jie and Chung, Eric},
  journal={Journal of Scientific Computing},
  volume={77},
  number={3},
  pages={1490--1518},
  year={2018},
  publisher={Springer}
}

@article{singh2020parameter,
  title={A parameter-uniform hybrid finite difference scheme for singularly perturbed system of parabolic convection-diffusion problems},
  author={Singh, Maneesh Kumar and Natesan, Srinivasan},
  journal={International Journal of Computer Mathematics},
  volume={97},
  number={4},
  pages={875--905},
  year={2020},
  publisher={Taylor \& Francis}
}

@article{hsieh2018robust,
  title={A robust finite difference scheme for strongly coupled systems of singularly perturbed convection-diffusion equations},
  author={Hsieh, Po-Wen and Yang, Suh-Yuh and You, Cheng-Shu},
  journal={Numerical Methods for Partial Differential Equations},
  volume={34},
  number={1},
  pages={121--144},
  year={2018},
  publisher={Wiley Online Library}
}

@article{kumar2022new,
  title={A new stable finite difference scheme and its error analysis for two-dimensional singularly perturbed convection--diffusion equations},
  author={Kumar, Kamalesh and Podila, Pramod Chakravarthy},
  journal={Numerical Methods for Partial Differential Equations},
  volume={38},
  number={5},
  pages={1215--1231},
  year={2022},
  publisher={Wiley Online Library}
}

@article{gharibi2021convergence,
  title={Convergence analysis of weak Galerkin flux-based mixed finite element method for solving singularly perturbed convection-diffusion-reaction problem},
  author={Gharibi, Zeinab and Dehghan, Mehdi},
  journal={Applied Numerical Mathematics},
  volume={163},
  pages={303--316},
  year={2021},
  publisher={Elsevier}
}

@article{lin2018weak,
  title={A Weak Galerkin Finite Element Method for Singularly Perturbed Convection-Diffusion--Reaction Problems},
  author={Lin, Runchang and Ye, Xiu and Zhang, Shangyou and Zhu, Peng},
  journal={SIAM Journal on Numerical Analysis},
  volume={56},
  number={3},
  pages={1482--1497},
  year={2018},
  publisher={SIAM}
}

@article{zhang2021high,
  title={High-order finite element method on a Bakhvalov-type mesh for a singularly perturbed convection--diffusion problem with two parameters},
  author={Zhang, Jin and Lv, Yanhui},
  journal={Applied Mathematics and Computation},
  volume={397},
  pages={125953},
  year={2021},
  publisher={Elsevier}
}

@article{TOPRAKSEVEN2024130,
title = {An efficient weak Galerkin FEM for third-order singularly perturbed convection-diffusion differential equations on layer-adapted meshes},
journal = {Applied Numerical Mathematics},
volume = {204},
pages = {130-146},
year = {2024},
issn = {0168-9274},
author = {Suayip Toprakseven and Natesan Srinivasan},
}

@article{ZHANG2017549,
title = {Supercloseness of continuous interior penalty method for convection–diffusion problems with characteristic layers},
journal = {Computer Methods in Applied Mechanics and Engineering},
volume = {319},
pages = {549-566},
year = {2017},
issn = {0045-7825},
author = {Jin Zhang and Martin Stynes},
}

@article{yu2018deep,
  title={The deep Ritz method: a deep learning-based numerical algorithm for solving variational problems},
  author={Yu, Bing and others},
  journal={Communications in Mathematics and Statistics},
  volume={6},
  number={1},
  pages={1--12},
  year={2018},
  publisher={Springer}
}

@article{sirignano2018dgm,
  title={DGM: A deep learning algorithm for solving partial differential equations},
  author={Sirignano, Justin and Spiliopoulos, Konstantinos},
  journal={Journal of Computational Physics},
  volume={375},
  pages={1339--1364},
  year={2018},
  publisher={Elsevier}
}

@article{19M1274067,
author = {Lu, Lu and Meng, Xuhui and Mao, Zhiping and Karniadakis, George Em},
title = {DeepXDE: A Deep Learning Library for Solving Differential Equations},
journal = {SIAM Review},
volume = {63},
number = {1},
pages = {208-228},
year = {2021},
}

@article{CAO2024117222,
title = {Multistep asymptotic pre-training strategy based on PINNs for solving steep boundary singular perturbation problems},
journal = {Computer Methods in Applied Mechanics and Engineering},
volume = {431},
pages = {117222},
year = {2024},
issn = {0045-7825},
author = {Fujun Cao and Fei Gao and Dongfang Yuan and Junmin Liu},
}

@article{HUANG2024100496,
title = {Multi-scale physics-informed neural networks for solving high Reynolds number boundary layer flows based on matched asymptotic expansions},
journal = {Theoretical and Applied Mechanics Letters},
volume = {14},
number = {2},
pages = {100496},
year = {2024},
issn = {2095-0349},
author = {Jianlin Huang and Rundi Qiu and Jingzhu Wang and Yiwei Wang},
}

@article{aldirany2024multi,
  title={Multi-level neural networks for accurate solutions of boundary-value problems},
  author={Aldirany, Ziad and Cottereau, R{\'e}gis and Laforest, Marc and Prudhomme, Serge},
  journal={Computer Methods in Applied Mechanics and Engineering},
  volume={419},
  pages={116666},
  year={2024},
  publisher={Elsevier}
}

@article{dong2021local,
  title={Local extreme learning machines and domain decomposition for solving linear and nonlinear partial differential equations},
  author={Dong, Suchuan and Li, Zongwei},
  journal={Computer Methods in Applied Mechanics and Engineering},
  volume={387},
  pages={114129},
  year={2021},
  publisher={Elsevier}
}

@article{chen2022bridging,
  title={Bridging traditional and machine learning-based algorithms for solving PDEs: the random feature method},
  author={Chen, Jingrun and Chi, Xurong and Yang, Zhouwang and others},
  journal={J Mach Learn},
  volume={1},
  pages={268-298},
  year={2022}
}

@article{cai2020deep,
  title={Deep least-squares methods: An unsupervised learning-based numerical method for solving elliptic PDEs},
  author={Cai, Zhiqiang and Chen, Jingshuang and Liu, Min and Liu, Xinyu},
  journal={Journal of Computational Physics},
  volume={420},
  pages={109707},
  year={2020},
  publisher={Elsevier}
}

@article{calabro2021extreme,
  title={Extreme learning machine collocation for the numerical solution of elliptic PDEs with sharp gradients},
  author={Calabr{\`o}, Francesco and Fabiani, Gianluca and Siettos, Constantinos},
  journal={Computer Methods in Applied Mechanics and Engineering},
  volume={387},
  pages={114188},
  year={2021},
  publisher={Elsevier}
}

@article{dong2022computing,
  title={On computing the hyperparameter of extreme learning machines: Algorithm and application to computational PDEs, and comparison with classical and high-order finite elements},
  author={Dong, Suchuan and Yang, Jielin},
  journal={Journal of Computational Physics},
  volume={463},
  pages={111290},
  year={2022},
  publisher={Elsevier}
}

@article{chen2023random,
  title={The random feature method for time-dependent problems},
  author={Chen, Jingrun and Luo, Yixin and others},
  journal={arXiv preprint arXiv:2304.06913},
  year={2023}
}

@article{21M1436087,
author = {Clark, William A. and Gomes, Mario W. and Rodriguez-Gonzalez, Arnaldo and Stein, Leo C. and Strogatz, Steven H.},
title = {Surprises in a Classic Boundary-Layer Problem},
journal = {SIAM Review},
volume = {65},
number = {1},
pages = {291-315},
year = {2023},
}

@article{1017007,
author = {Hoppensteadt, Frank},
title = {Analysis of Some Problems Having Matched Asymptotic Expansion Solutions},
journal = {SIAM Review},
volume = {17},
number = {1},
pages = {123-135},
year = {1975},
}

@article{kim2022fast,
  title={A fast and accurate physics-informed neural network reduced order model with shallow masked autoencoder},
  author={Kim, Youngkyu and Choi, Youngsoo and Widemann, David and Zohdi, Tarek},
  journal={Journal of Computational Physics},
  volume={451},
  pages={110841},
  year={2022},
  publisher={Elsevier}
}

@article{tang2022asymptotic,
  title={Asymptotic Analysis on the Sharp Interface Limit of the Time-Fractional Cahn--Hilliard Equation},
  author={Tang, Tao and Wang, Boyi and Yang, Jiang},
  journal={SIAM Journal on Applied Mathematics},
  volume={82},
  number={3},
  pages={773--792},
  year={2022},
  publisher={SIAM}
}

@article{bressloff2021asymptotic,
  title={Asymptotic analysis of target fluxes in the three-dimensional narrow capture problem},
  author={Bressloff, Paul C},
  journal={Multiscale Modeling \& Simulation},
  volume={19},
  number={2},
  pages={612--632},
  year={2021},
  publisher={SIAM}
}

@article{SHARMA201310575,
title = {A review on singularly perturbed differential equations with turning points and interior layers},
journal = {Applied Mathematics and Computation},
volume = {219},
number = {22},
pages = {10575-10609},
year = {2013},
issn = {0096-3003},
author = {Kapil K. Sharma and Pratima Rai and Kailash C. Patidar},
}

@article{lu2025multiple,
  title={A multiple transferable neural network method with domain decomposition for elliptic interface problems},
  author={Lu, Tianzheng and Ju, Lili and Zhu, Liyong},
  journal={Journal of Computational Physics},
  volume={530},
  pages={113902},
  year={2025},
  publisher={Elsevier}
}

@article{wang2024general,
  title={General-Kindred physics-informed neural network to the solutions of singularly perturbed differential equations},
  author={Wang, Sen and Zhao, Peizhi and Ma, Qinglong and Song, Tao},
  journal={Physics of Fluids},
  volume={36},
  number={11},
  pages = {113604},
  year={2024},
  publisher={AIP Publishing}
}

@article{fitzsimons1985petrov,
  title={Petrov--Galerkin finite element methods with a hinged test space for singularly perturbed problems},
  author={Fitzsimons, C and Miller, JJH and O'riordan, E},
  journal={International journal for numerical methods in engineering},
  volume={21},
  number={10},
  pages={1803--1812},
  year={1985},
  publisher={Wiley Online Library}
}

@article{patidar2006uniformly,
  title={Uniformly convergent non-standard finite difference methods for singularly perturbed differential-difference equations with delay and advance},
  author={Patidar, Kailash C and Sharma, Kapil K},
  journal={International Journal for Numerical Methods in Engineering},
  volume={66},
  number={2},
  pages={272--296},
  year={2006},
  publisher={Wiley Online Library}
}

@article{XU2025114129,
  title = {Network scaling and scale-driven loss balancing for intelligent characterization of poroelastic systems},
  journal = {Journal of Computational Physics},
  volume = {537},
  pages = {114129},
  year = {2025},
  issn = {0021-9991},
  author = {Yang Xu and Fatemeh Pourahmadian},
}

%

\end{document}